\documentclass[a4paper,10pt,twoside]{cpc-hepnp}

\usepackage{multicol}
\usepackage{graphicx}
\usepackage{booktabs}
\usepackage{amssymb,bm,mathrsfs,bbm,amscd}
\usepackage[tbtags]{amsmath}
\usepackage{lastpage}
\usepackage{CJK}

\usepackage{graphicx,color,dcolumn,booktabs,bm}
\usepackage{longtable,lscape,comment,braket}
\usepackage{txfonts}
\usepackage{overpic}
\usepackage{amssymb}
\usepackage{indentfirst}
\usepackage{feynmf}   
\usepackage{slashed}  
\usepackage{cases}
\usepackage{epstopdf}
\usepackage{psfrag}
\usepackage{subfigure}
\usepackage[colorlinks,
            citecolor=blue,
            anchorcolor=red,
            menucolor=red,
            linkcolor=red,
            filecolor=red,
            runcolor=red,
            urlcolor=blue,
            frenchlinks=red]{hyperref}

\makeatletter

\@addtoreset{equation}{section}

\makeatother

\begin{document}
\begin{CJK*}{GB}{gbsn}

\fancyhead[c]{\small Chinese Physics C~~~Vol. **, No. * (****)
******} \fancyfoot[C]{\small ******-\thepage}

\footnotetext[0]{Received *****}

\title{Estimating the production rates of D-wave charmed mesons via the semileptonic decays of bottom mesons
\thanks{This project is partly supported by the National Natural Science Foundation of China under Grant Nos. 11222547, 11175073, and 11647301. Xiang Liu is also supported in part by the National Program for Support of Top-notch Young Professionals and the Fundamental Research Funds for the Central Universities.
}}

\author{%
      Kan Chen (³ÂÙ©)$^{1,2;1)}$\email{chenk16@lzu.edu.cn}%
\quad Hong-Wei Ke (¿ÂºìÎÀ)$^{3;2)}$\email{khw020056@hotmail.com}%
\quad Xiang Liu (ÁõÏè)$^{1,2;3)}$\email{xiangliu@lzu.edu.cn}%
\quad Takayuki Matsuki $^{4,5;4)}$\email{matsuki@tokyo-kasei.ac.jp}%
}
\maketitle

\address{%
$^1$School of Physical Science and Technology, Lanzhou University, 73000, China\\
$^2$Research Center for Hadron and CSR Physics,
Lanzhou University and Institute of Modern Physics of CAS, Lanzhou, 730000, China\\
$^3$ School of Science, Tianjin University, Tianjin 300072, China\\
$^4$Tokyo Kasei University, 1-18-1 Kaga, Itabashi, Tokyo 173-8602, Japan\\
$^5$Theoretical Research Division, Nishina Center, RIKEN, Saitama 351-0198, Japan
}

\begin{abstract}
In this work,
using the covariant light front approach with conventional vertex functions,
we estimate the production rates of $D$-wave charmed/charmed-strange mesons via $B_{(s)}$ semileptonic decays.
Due to these calculated considerable production rates, it is possible to experimentally search for $D$-wave charmed/charmed-strange mesons via the semileptonic decays, which may provide extra approach to explore $D$-wave charmed/charmed-strange mesons.
\end{abstract}



\footnotetext[0]{\hspace*{-3mm}\raisebox{0.3ex}{$\scriptstyle\copyright$}2013
Chinese Physical Society and the Institute of High Energy Physics
of the Chinese Academy of Sciences and the Institute
of Modern Physics of the Chinese Academy of Sciences and IOP Publishing Ltd}%

\begin{multicols}{2}

\section{Introduction}

With the accumulation of the experimental data, more and more open-charm and open-bottom states were reported in experiment (see review paper \cite{Chen:2016spr} for more details).
Among the observed states, there are abundant candidates of charmed and charmed-strange mesons
including famous $D_{s0}(2317)$ and $D_{s1}(2460)$. Especially, in recent years, experimentalists have made a great progress on observing $D$-wave charmed mesons as well as $D$-wave charmed-strange mesons. For example, the observed $D^{*}(2760)$, $D(2750)$ \cite{delAmoSanchez:2010vq, Aaij:2013sza}, $D_{s1}^{*}(2860)$ and $D_{s3}^{*}(2860)$ \cite{Aaij:2014xza,Aaij:2014baa} can be good candidates of $1D$ states in charmed and charmed-strange meson families \cite{DiPierro:2001dwf,Sun:2010pg,Zhong:2010vq,Li:2010vx,Colangelo:2012xi,
Chen:2011rr,Song:2014mha,Godfrey:2014fga,Zhou:2014ytp,Wang:2014jua}. In addition, the $D_{sJ}^{*}(2860)$ \cite{Aubert:2006mh,Aubert:2009ah} can be assigned to a $1D$ state of charmed-strange meson though there exist different explanations \cite{Zhang:2006yj,Colangelo:2007ds,Li:2009qu,
Zhong:2009sk,Li:2007px,Zhong:2008kd,Chen:2009zt,vanBeveren:2009jq}. Readers can refer to Refs. \cite{Song:2015fha,Song:2015nia} for more information on $D$-wave charmed and charmed-strange mesons.

When checking the production processes involving the $D$-wave charmed and charmed-strange mesons, we notice that these states are mainly produced via the nonleptonic weak decays of bottom/bottom-strange mesons.  However, as an important decay mode, the semileptonic decays of $B/B_s$ mesons are an ideal platform to produce $D$-wave $D/D_s$ mesons because they can be estimated more accurately than nonleptonic ones. For estimating the branching ratios of these processes, we need to perform a serious theoretical study on the production of $D$-wave $D/D_s$ mesons via the semileptonic decays of $B/B_s$ mesons, which is a main task of the present work.

In this work, we adopt a light-front quark model (LFQM) \cite{Jaus:1989au,Jaus:1991cy,Ji:1992yf,Jaus:1996np,Cheng:1996if}, which is a relativistic quark model. Since the involved light-front wave function is manifestly Lorentz invariant and the hadron spin is constructed by using the Melosh-Wigner rotation \cite{Ma:1993ht,Melosh:1974cu}, the LFQM can be suitably applied to study semileptonic decays of $B/B_s$ mesons. In Refs. \cite{Cheng:2003sm,Xu:2014mqa,Li:2010bb,Cheng:2004yj,Ke:2009ed,Ke:2009mn,Lu:2007sg,Wang:2007sxa,Wang:2008ci,Shen:2008zzb,Wang:2008xt,Wang:2009mi,Chen:2009qk,Wang:2010npa}, production rates of $S$- and $P$-wave $D/D_s$ mesons have been estimated through the decay processes of $B/B_s$ in the covariant LFQM.

Until now, there has been no work on the study of the productions of $D$-wave $D/D_s$ mesons through the semileptonic decays of $B/B_s$ mesons in the covariant light-front approach, which makes the present work be the first paper on this issue. As illustrated in the following sections, the technical details of deduction relevant to the above processes are far more complicated than those of $S$- and $P$-wave mesons.
Thus, our work is not only an application of the LFQM but also development of this research field since the formula presented in this work can be helpful to the study of other processes involving $D$-wave mesons.
Because we consider it is valuable to readers, we provide more details of deduction.

Finally, we still hope that the present study can stimulate experimentalists' interest in searching for $D$-wave $D/D_s$ mesons by the semi-leptonic decays of $B/B_s$ mesons. It will open another window to explore $D$-wave $D/D_s$ mesons, on which the experimental information will become more abundant.

This paper is organized as follows. In Sec. \ref{sec2}, we introduce the covariant light-front approach for $D$-wave mesons and their corresponding form factors. In Sec. \ref{sec3}, we list our numerical results including the form factors as well as the decay branching ratios. Sec. \ref{sec4} present the relations between the obtained light front form factors and the heavy quark symmetry expectations. The final section is devoted to a summary of our work.
Appendices \ref{traceform} through \ref{toy model prove} will describe the algebraic details related to the production of $D$-wave mesons via $B_{(s)}$ semi-leptonic decay in LFQM, among which Appendix \ref{toy model prove} will be devoted to proving Lorentz invariance of matrix elements
in the toy model adopted in Ref. \cite{Jaus:1999zv} with multipole ansatz for vertex functions.

\section{Covariant light-front quark model}\label{sec2}
In the conventional light-front quark model, quark and antiquark inside a meson are required to be on their mass shells. Then, one can extract physical quantities by calculating the plus component of the corresponding matrix element. However, as discussed in Ref. \cite{Cheng:2003sm}, this treatment may result in missing the so-called Z-diagram contribution, the result of the corresponding matrix element will depend on the choice of frame. A systematic way of incorporating the zero mode effect was proposed in Ref. \cite{Jaus:1999zv} to maintain the associated current matrix elements frame independent, thus one can extract physical quantities.

In this work, we will apply the covariant light-front approach to investigate the production of $D_{(s)}^{**}$ mesons via the semileptonic decays of $B/B_s$ mesons (see Fig. \ref{diagram}), where $D_{(s)}^{**}$ denotes a general $D$-wave $D/D_s$ meson. Firstly, we briefly introduce how to deal with the transition amplitudes.
\begin{center}
 \includegraphics[width=6cm]{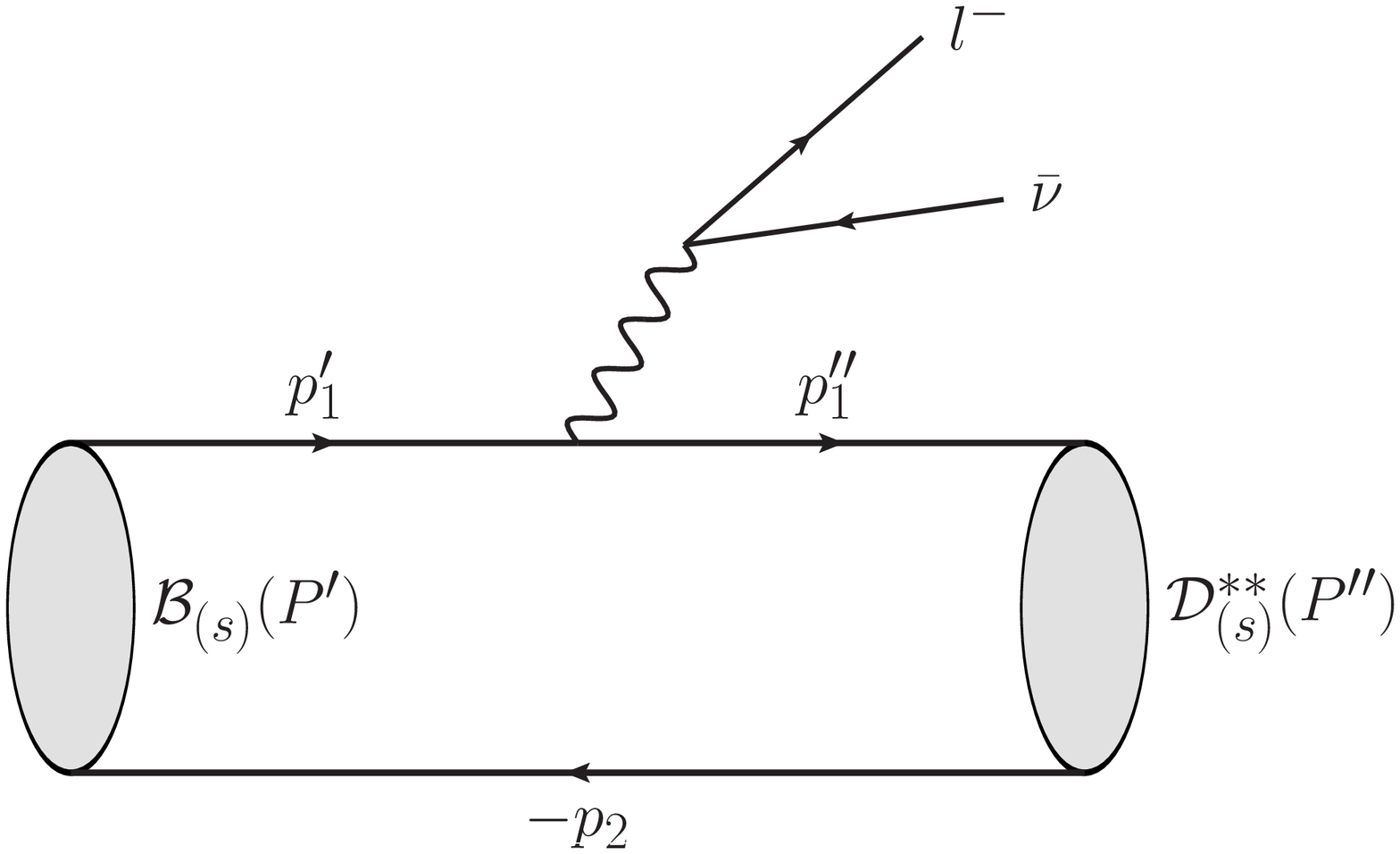}
\figcaption{\label{diagram} (color online). Diagram for the meson transition processes of $B_{(s)} \to D_{(s)}^{**}\ell^-\bar{\nu}$. Here, $P^{\prime(\prime\prime)}$ is the momentum of an incoming (outgoing) $B/B_s$ ($D$-wave $D_s$) meson. $p_1^{\prime(\prime\prime)}$ denotes the momentum carried by bottom (charm) quark, while $p_2$ is momentum of a light quark.}
\end{center}

According to Ref. \cite{Jaus:1999zv}, the relevant form factors are calculated in terms of
Feynman loop
integrals which are manifestly covariant. The constituent quarks within a hadron are off-shell, i.e., the incoming (outgoing) meson has the momentum $P^{\prime(\prime\prime)}=p_1^{\prime(\prime\prime)}+p_2$, where $p_1^{\prime(\prime\prime)}$ and $p_2$ are the off-shell momenta of quark and antiquark, respectively. These momenta can be expressed in terms of the appropriate internal variables, $(x_i, p_{\bot}^{\prime})$, defined by,
\begin{eqnarray}
p_{1}^{\prime+}&=&x_{1}P^{\prime+},\quad\quad\quad p^+_2 = x_2P^{\prime+}, \\
p_{1\bot}^{\prime}&=&x_{1}P^{\prime}_{\bot}+ p^{\prime}_{\bot},
\quad p_{2\bot}=x_{2}P^{\prime}_{\bot}- p^{\prime}_{\bot},
\end{eqnarray}
with $x_1+x_2=1$. In the light-front coordinate, $P^{\prime}=(P^{\prime-},P^{\prime+},P_{\bot}^{\prime})$ with $P^{\prime\pm}=P^{\prime0}\pm P^{\prime3}$, which has the relation $P^{\prime2}=P^{\prime+}P^{\prime-}-P^{\prime2}_{\bot}$. One needs to specify that there exist different conventions for the momentum conservation under the covariant light-front and conventional light-front approaches.
In the covariant light-front approach, four components of a momentum are conserved at each vertex, where the quark and antiquark are off-shell. In the conventional light-front approach, the plus and transverse components of a momentum are conserved quantities, where quark and antiquark are required to be on their mass shells.
Thus, it would be useful to define some internal quantities for on-shell quarks
\begin{eqnarray}
M_0^{\prime2}&=&(e_1^{\prime}+e_2)^2=\frac{p_{\bot}^{\prime2}+m_1^{\prime2}}{x_1}
+\frac{p_{\bot}^{\prime2}+m_2^{2}}{x_2},
\\
\tilde{M}_{0}^{\prime}&=&\sqrt{M_{0}^{\prime2}-(m_1^{\prime}-m_2)^2},
\\
e_{1}^{\prime}&=&\sqrt{m_1^{\prime2}+p_{\bot}^{\prime2}+p_z^{\prime2}},\quad
e_{2}=\sqrt{m_2^{2}+p_{\bot}^{\prime2}+p_z^{\prime2}},
\\
p_z^{\prime}&=&\frac{x_2 M_0^{\prime}}{2}-\frac{m_2^2+p_{\bot}^{\prime2}}{2x_2M_0^{\prime}},
\end{eqnarray}
where $M_0^{\prime2}$ is the kinetic invariant mass squared of the incoming meson. $e^{(\prime)}_i$ denotes the energy of quark $i$, while $m_1^{\prime}$ and $m_2$ are the masses of quark and antiquark, respectively.

In Ref. \cite{Cheng:2003sm}, the form factors for the semileptonic decays of bottom mesons into $S$-wave and $P$-wave charmed mesons were obtained within the framework of the covariant light-front quark model. In the following, we adopt the same approach to deduce the form factors of the production of $D$-wave charmed/charmed-strange mesons through the semileptonic decays of bottom/bottom-strange mesons. Here, the $D$-wave $D/D_s$ mesons with notations $D^{*}_{(s)1}$, $D^{*}_{(s)2}$, $D^{*\prime}_{(s)2}$, and $D^{*}_{(s)3}$ have the quantum numbers ${}^{2S+1}L_J={}^3D_1, {}^1D_2, {}^3D_2,$ and ${}^3D_3$, respectively. In the following deduction, we will use such notation for simplification.

In the heavy quark limit $m_Q\rightarrow \infty$, the heavy quark spin $s_Q$ decouples from the other degrees of freedom. Hence, a more convenient way to describe charmed/charmed-strange mesons is to use the $|J,j_{\ell}\rangle$ basis, where $J$ denotes the total spin and $j_\ell$ denotes the total angular momentum of the light quark. There exists connection between physical states $|J,j_{l}\rangle$ and the states described by $|J,S\rangle$ for $L=2$ \cite{Matsuki:2010zy,Li:2016efw}, i.e.,
\begin{eqnarray}
|D_{(s){5\over2}} \rangle&\equiv&\left|2,{5\over2}\right\rangle=\sqrt{\frac{3}{5}}\left|D^{*}_{(s)2}\right\rangle+\sqrt{\frac{2}{5}}\left|D^{*\prime}_{(s)2}\right\rangle, \label{mix1}\\
|D'_{(s){3\over2}}\rangle&\equiv&\left|2,{3\over2}\right\rangle=-\sqrt{\frac{2}{5}}\left|D^{*}_{(s)2}\right\rangle+\sqrt{\frac{3}{5}}\left|D^{*\prime}_{(s)2}\right\rangle.
\label{mix2}
\end{eqnarray}
This relation shows that two physical states $D_{(s)2}$ and $D_{(s)2}^{\prime}$ with $J^P=2^-$ are linear combinations of the $D^{*}_{(s)2}({}^1D_2)$ and $D^{*\prime}_{(s)2}({}^3D_2)$ states. When dealing with the transition amplitudes of the production of the $D_{(s)2}$ and $D_{(s)2}^{\prime}$ states, we need to consider the mixing of states shown in Eqs. (\ref{mix1}) and (\ref{mix2}).

One can write out the general definitions for the matrix elements of the production of $D$-wave $D/D_s$ mesons via the semileptonic decays of $B/B_s$ mesons, i.e.,
\end{multicols}
\begin{eqnarray}
&&\left\langle
D^{*}_{(s)1}(P'',\epsilon'')\left|V_{\mu}\right|B_{(s)}(P^\prime)\right\rangle=\epsilon_{\mu\nu\alpha\beta}\epsilon''^{*\nu}P^{\alpha}q^{\beta}g_D(q^{2}),\nonumber\\
&&\left\langle D^{*}_{(s)1}(P'',\epsilon'')\left|A_{\mu}\right |B_{(s)}(P^\prime)\right\rangle=-i\left\{\epsilon''^{*}_{\mu}f_D(q^2)+\epsilon''^{*}\cdot
P\left[P_{\mu} a_{D+}(q^2)+q_{\mu}a_{D-}(q^2)\right]\right\},\label{GD31}
\\&&\left\langle D^{*}_{(s)2}(P'',\epsilon'')\left|A_{\mu}\right|B_{(s)}(P')\right\rangle=-\epsilon_{\mu\nu\alpha\beta}\epsilon''^{*\nu\lambda}P_{\lambda}P^{\alpha}q^{\beta}n(q^2),
\nonumber\\
&&\left\langle D^{*}_{(s)2}(P'',\epsilon'')\left|V_{\mu}\right|B_{(s)}(P')\right\rangle=i\left\{m(q^2)\epsilon''^{*}_{\mu\nu}P^{\nu}+\epsilon''^{*}_{\alpha\beta}P^{\alpha}P^{\beta}
\left[P_{\mu}z_{+}(q^2)+
q_{\mu}z_{-}(q^2)\right]\right\},\label{GD120}
\label{GD12}
\\
&&\left\langle D^{*\prime}_{(s)2}(P'',\epsilon'')\left|A_{\mu}\right|B_{(s)}(P')\right\rangle=-\epsilon_{\mu\nu\alpha\beta}\epsilon''^{*\nu\lambda}P_{\lambda}P^{\alpha}q^{\beta}n'(q^2),\nonumber
\\&&\left\langle D^{*\prime}_{(s)2}(P'',\epsilon'')\left|V_{\mu}\right|B_{(s)}(P')\right\rangle
=i\left\{m'(q^2)\epsilon''^{*}_{\mu\nu}P^{\nu}+\epsilon''^{*}_{\alpha\beta}P^{\alpha}P^{\beta}\left[P_{\mu}z'_{+}(q^2)+
q_{\mu}z'_{-}(q^2)\right]\right\},
\label{GD32}
\\&&\left\langle D^{*}_{(s)3}(P'',\epsilon'')\left|V_{\mu}\right|B_{(s)}(P')\right\rangle=\epsilon_{\mu\nu\alpha\beta}\epsilon''^{*\nu\lambda\sigma}P_{\lambda}P_{\sigma}P^{\alpha}q^{\beta}y(q^2),\nonumber\\
&&\left\langle D^{*}_{(s)3}(P'',\epsilon'')\left|A_{\mu}\right|B_{(s)}(P')\right\rangle=-i\left\{w(q^2)\epsilon''^{*}_{\mu\nu\alpha}P^{\nu}P^{\alpha}+\epsilon''^*_{\alpha\beta\gamma}P^{\alpha}P^{\beta}P^{\gamma}
\left[P_{\mu}o_{+}(q^2)+q_{\mu}o_{-}(q^2)\right]\right\}.
\label{GD33}
\end{eqnarray}
\begin{multicols}{2}
Here, $P=P^{\prime}+P^{\prime\prime}$, $q=P^{\prime}-P^{\prime\prime}$ and $\epsilon_{0123}=1$. $\epsilon^{\prime*}_{\mu}$, $\epsilon^{\prime\prime*}_{\mu\nu}$ and $\epsilon^{\prime\prime*}_{\mu\nu\alpha}$ are polarization vector (tensors), to further clarify these identifies, we have presented more details in Appendix \ref{polarization tensor}. Lorentz invariance has been assumed when defining these form factors. One should notice that the $B_{(s)}\rightarrow D_{(s)}^{**}$ transition occurs through a $V-A$ current, where $D_{(s)}^{**}$ denotes the general $D$-wave charmed (charmed-strange) meson. For the semileptonic decays involving the ${}^3D_1$ and ${}^3D_3$ states, the $\epsilon_{\mu\nu\alpha\beta}$ term arises in $\left\langle D^{*}_{(s)1}\left(D^{*}_{(s)3}\right)\left|V_{\mu}\right|B_{(s)}\right\rangle$, which corresponds to the contribution of the vector current. Different from the case of the ${}^3D_1$ and ${}^3D_3$ states,
for the ${}^1D_2$ and ${}^3D_2$ states, $\epsilon_{\mu\nu\alpha\beta}$ term arises in the axial vector current. Here, a minus sign is added in front of this term so that we have $\left\langle D_{(s)2}^{*(\prime)}\left|-A_\mu\right|B_{(s)}\right\rangle=\epsilon_{\mu\nu\alpha\beta}\epsilon
^{\prime\prime*\nu\lambda}P_{\lambda}P^{\alpha}q^{\beta}n^{(\prime)}(q^2)$.
When the sign of the $\epsilon_{\mu\nu\alpha\beta}$ term is fixed, the signs of other form factors can also be determined.

Now we focus on the hadronic matrix elements given by Eqs.~(\ref{GD31})-(\ref{GD33}).
Here we show how to calculate them by taking the $B_{(s)}\rightarrow D_{(s)1}^{*}$ transition as an example, where $D_{(s)1}^{*}$ denotes the ${}^3D_1$ state of the charmed/charmed-strange meson. The corresponding matrix element for $B_{(s)}\rightarrow D_{(s)1}^{*}$ can be written as
\begin{eqnarray}
B_{\mu}^{B_{(s)}(D^{*}_{(s)1})}\equiv\left\langle D_{(s)1}^{*}(P^{\prime\prime},\epsilon^{\prime\prime*})\left|V_{\mu}-A_{\mu}\right|B_{(s)}(P^{\prime})\right\rangle,
\label{3D1cov}
\end{eqnarray}
following the calculation in Ref. \cite{Cheng:2003sm}, we first obtain the $B_{(s)}\rightarrow D^{*}_{(s)1}$ transition form factors, and then continue to calculate the processes involving other $D$-wave charmed/charmed-strange states, details of other matrix elements are given in Appendix \ref{traceform}. Here, one needs to introduce the vertex wave functions to describe the $B_{(s)}$ and $D_{(s)1}^{*}$ mesons. The expression of a vertex function for an initial $B_{(s)}$ meson has been obtained in Ref. \cite{Cheng:2003sm}. In the following, we will give detailed discussion for the vertex function of the final state $D_{(s)1}^{*}$ meson.

The $D$-wave vertex function has been studied in Ref. \cite{Ke:2011mu}. Here, we listed all $D$-wave vertex functions in Appendix \ref{vertex}, one may refer to Ref. \cite{Ke:2011mu} for more detail. Firstly, we will use ${}^3D_1$ vertex functions to the calculation of $B_{(s)}\rightarrow D_{(s)1}^{*}$ transition.

In the conventional LFQM, the $p'_1$ and $p_2$ are on their mass shell, while in the covariant \cite{Jaus:1999zv} light-front approach, the quark and antiquark are off-shell, but the total momentum $P'=p'_1+p_2$ is still the on-shell momentum of a meson, i.e., $P^{\prime 2}=M^{\prime 2}$ with $M'$ being the mass of an incoming meson. One needs to relate the vertex function deduced in the conventional LFQM to the vertex in the covariant light-front approach. A practical method for this process has been proposed in a covariant light-front approach in Ref. \cite{Jaus:1999zv}.
We obtain the corresponding covariant vertex function as
\begin{eqnarray}
iH_{{}^3D_1}\left[\gamma_{\mu}-\frac{1}{W_{{}^3D_1}}\left(p^\prime_1-p_2\right)_{\mu}\right]\epsilon^{\mu},
\end{eqnarray}
where $H_{{}^{3}D_1}$ and $W_{{}^{3}D_1}$ denote the corresponding scalar functions for ${}^3D_1$ state.
\begin{center}
  \includegraphics[scale=0.3]{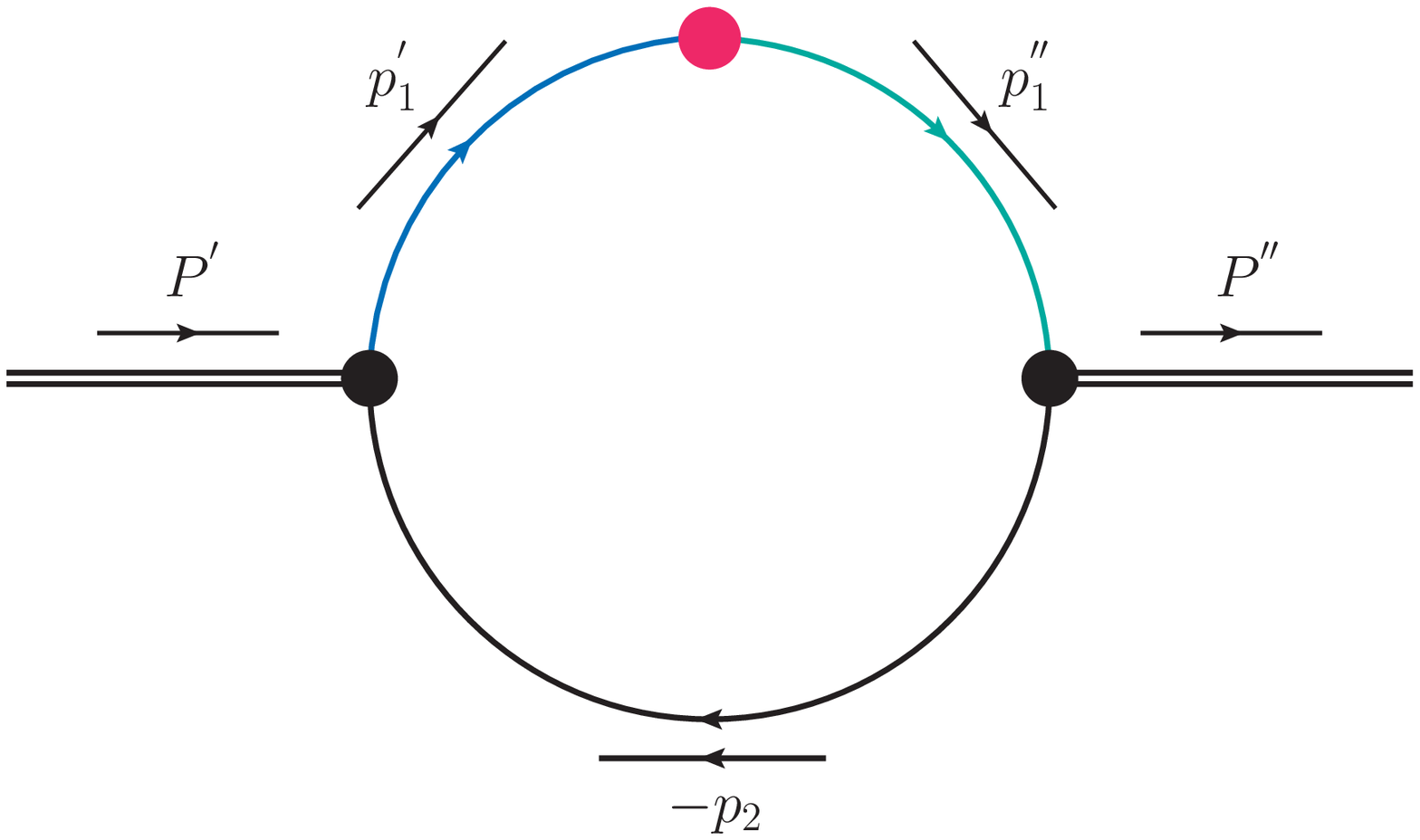}
\figcaption{\label{fig2} (color online).A hadronic one-loop Feynman diagram contribution to the process given Fig. \ref{diagram}. Here the $V-A$ current is attached to a blob in the upper middle of circle.}
\end{center}
%
The explicit expression of the matrix element $B_{\mu}^{B_{(s)}D^{*}_{(s)1}}$, which corresponds to the hadronic one-loop Feynman diagram of Fig.~\ref{fig2}, reads
\begin{eqnarray}
B_{\mu}^{B_{(s)}D^{*}_{(s)1}}=-i^3\frac{N_c}{(2\pi)^4}\int d^4p^{\prime}_1\frac{H^{\prime}_{P}\left(iH^{\prime\prime}_{{}^3D_1}\right)}{N_1^{\prime}N_1^{\prime\prime}N_2}S_{\mu\nu}^{{}^3D_1}\epsilon^{*\prime\prime\nu},
\label{BmatrixElement}
\end{eqnarray}
where $N_c$ is the number of colors, $N_1^{\prime(\prime\prime)}=p_1^{\prime(\prime\prime)2}-m_1^{\prime(\prime\prime)2}+i\epsilon$, $N_2=p_2^{2}-m_2^{2}+i\epsilon$. $H^{\prime}_{P}\gamma_5$ is the vertex function of a pseudoscalar meson, and
\begin{eqnarray}
S^{{}^3D_1}_{\mu\nu}&=&{{\rm Tr}\Bigg\{\left[\gamma_{\nu}-\frac{1}{W^{\prime\prime}_{{}^3D_1}}\left(p^{\prime\prime}_1-p_2\right)_{\nu}\right]}
\left(\slashed{p}^{\prime\prime}_{1}+m^{\prime\prime}_1\right)\gamma_{\mu}\left(1-\gamma_5\right)\nonumber\\&&
\left(\slashed{p}^{\prime}_1+m^{\prime}_1\right)\gamma_5\left(-\slashed{p}_2+m_2\right)\Bigg\}.
\label{T3D1}
\end{eqnarray}

One can integrate over $p_1^{\prime-}$ via a contour integration with $d^4p^{\prime}_1=P^{~\prime +}dp^{\prime -}_1 dx_2d^2p^\prime_\bot/2$ and the integration picks up a residue $p_2=\hat{p}_2$, where the antiquark is set to be on-shell, $\hat{p}_2^2=m_2^2$. The momentum of the quark is given by the momentum conservation, $\hat{p}_1^{\prime}=P^{\prime}-\hat{p}_2$. Consequently, after performing the $p_1^{\prime-}$ integration, we make the replacements:
\begin{eqnarray*}
N_1^{\prime(\prime\prime)}&\rightarrow& \hat{N}_1^{\prime(\prime\prime)}=x_1\left(M^{\prime(\prime\prime)2}-M_0^{\prime(\prime\prime)2}\right),\nonumber\\
H^{\prime}_P&\rightarrow& h_{P}^{\prime},\nonumber\\
H^{\prime\prime}_{{}^3D_1}&\rightarrow& h^{\prime\prime}_{{}^3D_1},\nonumber\\
W^{\prime\prime}_{{}^3D_1}&\rightarrow& \omega^{\prime\prime}_{{}^3D_1},\nonumber\\
\int \frac{d^4p_1^{\prime}}{N_1^{\prime}N_1^{\prime\prime}N_2}H^{\prime}_P
H^{\prime\prime}_{{}^3D_1}S^{{}^3D_1}_{\mu\nu}&\rightarrow& -i\pi\int \frac{dx_2d^2p_{\bot}^{\prime}}{x_2\hat{N}_1^{\prime}\hat{N}_1^{\prime\prime}}
h^{\prime}_Ph^{\prime\prime}_{{}^3D_1}\hat{S}^{{}^3D_1}_{\mu\nu},
\end{eqnarray*}
where the explicit trace expansion of $\hat{S}_{\mu\nu}^{{}^3D_1}$ after integrating Eq.~(\ref{BmatrixElement}) over $p_1^{\prime-}$ is presented in Appendix \ref{traceform}. Additionally, $h^{\prime}_P$ has been given in Ref. \cite{Cheng:2003sm} as,
\begin{eqnarray}
h_{P}^{\prime}=\left(M^{\prime2}-M_{0}^{\prime2}\right)\sqrt{\frac{x_1x_2}{N_c}}
\frac{1}{\sqrt{2}\tilde{M}_0^{\prime}}\varphi,
\end{eqnarray}
where $\varphi$ is the solid harmonic oscillator for $S$-wave and describes the momentum distribution of an initial $B_{(s)}$ meson.

As noted in Ref. \cite{Ke:2011mu}, after carrying out the contour integral over $p_1^{\prime-}$,  
the quantities, $H^{\prime\prime}_{{}^3D_1}$, $W^{\prime\prime}_{{}^3D_1}$, and $\epsilon^{*\prime\prime}$
are replaced by the corresponding
$h^{\prime\prime}_{{}^3D_1}$, $\omega^{\prime\prime}_{{}^3D_1}$ and $\hat{\epsilon}^{*\prime\prime}$,
 respectively. Here, $h_{{}^3D_1}^{\prime\prime}$ is related to $h_{{}^3D_1}^{\prime}$ as,
\begin{eqnarray}
h^{\prime\prime}_{{}^3D_1}=\left(M^{\prime\prime2}-M_0^{\prime\prime2}\right)\sqrt{x_1x_2}h^{\prime}_{{}^3D_1},
\end{eqnarray}
which is derived in Appendix \ref{vertex} and
\begin{eqnarray}
M_0^{\prime\prime2}=\frac{p^{\prime2}_{\bot}+m_1^{\prime\prime2}}{x_1}+
\frac{p^{\prime\prime2}_{\bot}+m_2^2}{x_2},
\end{eqnarray}
with $p^{\prime\prime}_{\bot}=p^{\prime}_{\bot}-x_2q_{\bot}$.

As pointed out in Refs. \cite{Cheng:2003sm,Jaus:1999zv}, $\hat{p}_1^{\prime}$ can be expressed in terms of three external vectors, $P^{\prime}$ and $\tilde{\omega}$:
\begin{eqnarray*}
  \hat p^{\prime\mu}_1 &=& \left(P^\prime - \hat{p}_2 \right)^\mu = x_1P^{\prime\mu} + (0,0,{p}^{~\prime}_\bot)^\mu \nonumber\\&&+
  \frac{1}{2}\left(x_2P^{\prime -} - \frac{{p}^{~2}_{2\bot}+m^2_2}{x_2P^{\prime +}} \right)\tilde\omega^\mu,
\end{eqnarray*}
where $\tilde{\omega}=(\tilde\omega^+,\tilde\omega^-,\tilde\omega_\bot)=(2,0,0_{\bot})$ \cite{Cheng:2003sm,Jaus:1999zv} is a light-like four vector in the light-front coordinate system. Since the constant vector $\tilde{\omega}$ is not Lorentz covariant, if there remain $\tilde\omega$ terms, the corresponding matrix elements are not Lorentz invariant. This $\tilde{\omega}$ dependence also appears in the products of a couple of $\hat{p}_1^{\prime}$'s. This spurious contribution is related to the so-call zero mode effect and should be canceled when calculating physical quantities.

Initiated from a toy model proposed in Ref. \cite{Jaus:1999zv}, Jaus developed a method which allows calculating the zero mode contributions associated with the corresponding matrix element. The $\hat{p}_1^{\prime}$ as well as the products of a couple of $\hat{p}_1^{\prime}$'s can be decomposed into products of vectors $P$, $q$, $\tilde\omega$, and $g_{\mu\nu}$ as shown in Appendix \ref{decomposition} with functions $A^{(m)}_n$, $B^{(m)}_n$, and $C^{(m)}_n$, among which $B^{(m)}_n$ and $C^{(m)}_n$ are related to $\tilde{\omega}$-dependent terms. Based on a toy model, the vertex function of a ground state pseudoscalar meson is described by a multipole ansatz,
\begin{eqnarray}
  H_0(p_1^2,p_2^2)=\frac{g}{N^n_{\Lambda}},
\label{multipole}
\end{eqnarray}
which is different from our conventional vertex functions.
He has proven that at toy model level, the spurious loop integrals of $B^{(2)}_{1}$, $B_{1,2}^{(3)}$ and $C^{(2)}_{1}$, $C_{1,2}^{(3)}$ vanish for the following integrals,
\begin{eqnarray}
\frac{i}{(2\pi)^4}\int d^4p^\prime_1\frac{M^{(m)}_n}{N^\prime_\Lambda N_1^{\prime}N_2N_1^{\prime\prime}N^{\prime\prime}_\Lambda},
\label{toy model}
\end{eqnarray}
where $M^{(m)}_n\equiv B^{(m)}_{n}$ or $C^{(m)}_n$.
This fact matches with a natural consequence of Lorentz invariance of the theory. The $\tilde{\omega}$-dependent terms have been systematically eliminated in the toy model since the $B^{(2)}_{1}$ and $C^{(2)}_{1, 2}$ give trivial contributions to the calculated form factors \cite{Jaus:1999zv}.

However, this method has narrow scope of applications. Note that Jaus proposed this method in a very simple multipole ansatz for the vertex function. One may get totally different contributions from zero mode effects once the form of a vertex function for a meson is changed. For instance, as indicated in Ref. \cite{Choi:2010be}, for the weak transition form factors between pseudoscalar and vector meson, the zero mode effect contributions depend on the form of the vector meson vertex,
\begin{eqnarray}
\Gamma^{\mu}=\gamma^{\mu}-(2k-P_V)^{\mu}/D,
\end{eqnarray}
where the denominator $D$ contains different type of terms. (Readers can also refer to Refs. \cite{Choi:2010zza,Choi:2009ai,DeWitt:2003rs,Bakker:2002mt,Bakker:2000pk,Choi:2005fj} for more details.)

Beyond the toy model, the method of including the zero-mode contributions in Ref. \cite{Jaus:1999zv} was further applied to study the decay constants and form-factors for $S$-wave and $P$-wave mesons \cite{Cheng:2003sm}. In Ref. \cite{Cheng:2003sm}, Cheng et al. used the vertex functions for $S$- and $P$-wave mesons deduced from the conventional light front quark model, which are different from the multipole ansatz proposed by Jaus. They \cite{Cheng:2003sm} applied the method of the toy model to cancel the $C^{(m)}_{n}$ functions. As for $B^{(m)}_n$ functions, they have numerically checked that these $B^{(m)}_n$ give very small contributions to the corresponding form factors. That is, when the multipole vertices are replaced by conventional light front vertex functions, and by setting $B_n^{(m)}$ and $C^{(m)}_{n}$ functions equal to 0,
one can still obtain very good numerical results for decay constants. Indeed, as indicated in Ref. \cite{Jaus:1999zv}, the numerical results obtained by applying conventional vertices are even better than those of introducing the vertices from a multipole ansatz.

It is natural to expect that this method can also be applied to our calculation of the form factors for the transition processes of $D$-wave mesons. In order to calculate the corresponding form factors, one also needs to eliminate the $B^{(m)}_n$ and $C^{(m)}_n$ functions introduced in $D$-wave transition matrix elements. In the following, we will introduce our analysis of including zero-mode contributions.

Following the discussion in Refs. \cite{Cheng:2003sm,Jaus:1999zv}, to avoid the $\tilde{\omega}$ dependence, for $\hat{p}_{1}^{\prime}$ as well as the product of a couple of $\hat{p}^{\prime}_{1}$'s, one needs to do the following replacements:
\begin{eqnarray}
\hat{p}_{1\mu}^{\prime}&\doteq& P_{\mu}A_1^{(1)}+q_{\mu}A_2^{(1)},\label{productOFp1}
\\
\hat{p}_{1\mu}^{\prime}\hat{p}_{1\nu}^{\prime}&\doteq& g_{\mu\nu}A_1^{(2)}+P_{\mu}P_{\nu}A_2^{(2)}+\left(P_{\mu}q_{\nu}+q_{\mu}P_{\nu}\right)
A_3^{(2)}\nonumber\\&&+q_{\mu}q_{\nu}A_4^{(2)},\label{productOFp2}
\\
\hat{p}^{\prime}_{1\mu}\hat{p}^{\prime}_{1\nu}\hat{p}^{\prime}_{1\alpha}&\doteq&\left(g_{\mu\nu}P_{\alpha}+g_{\mu\alpha}P_{\nu}+g_{\nu\alpha}P_{\mu}\right)A_1^{(3)}\nonumber\\&&
+\left(g_{\mu\nu}q_{\alpha}+g_{\mu\alpha}q_{\nu}+g_{\nu\alpha}q_{\mu}\right)
A_2^{(3)} \nonumber\\&&+P_{\mu}P_{\nu}P_{\alpha}A_3^{(3)}+\left(P_{\mu}P_{\nu}q_{\alpha}+P_{\mu}q_{\nu}P_{\alpha}
\right.\nonumber\\&&\left.+q_{\mu}P_{\nu}P_{\alpha}\right)A_4^{(3)}
+\left(q_{\mu}q_{\nu}P_{\alpha}+q_{\mu}P_{\nu}q_{\alpha}\right.\nonumber\\&&\left.+P_{\mu}q_{\nu}q_{\alpha}\right)
 A_5^{(3)}+q_{\mu}q_{\nu}q_{\alpha}A_6^{(3)},\label{productOFp3}
\end{eqnarray}
where $B_n^{(m)}$ and $C^{(m)}_{n}$ functions are disregarded at toy model level and their loop integrals vanish manifestly if conventional vertices are introduced, we will give more detail about our treatment in the following discussion.

For the terms of products that are associated with $\hat{N}_2$, the zero mode contributions will be introduced and the following replacements should be done to eliminate $\tilde{\omega}$-dependent terms
\begin{eqnarray}
\hat{N}_2\rightarrow Z_2 &=&\hat N'_1+m_1^{\prime 2}-m_2^2+(1-2x)M^{\prime 2}
\nonumber \\
&&+\left[q^2+(qP)\right]\frac{p_\perp^\prime q_\perp}{q^2},
\label{productOFp4}\\
\hat{p}_{1\mu}^{\prime}\hat{N}_2&\rightarrow& P_{\mu}(A_1^{(1)}Z_2-A_1^{(2)})+q_{\mu}\left[A_2^{(1)}Z_2+\frac{q\cdot P}{q^2}A_1^{(2)}\right],\nonumber\\\label{productOFp5}
\\
\hat{p}_{1\mu}^{\prime}\hat{p}_{1\nu}^{\prime}\hat{N}_2&\rightarrow & g_{\mu\nu}A_1^{(2)}Z_2+P_{\mu}P_{\nu}(A_2^{(2)}Z_2-2A_1^{(3)})\nonumber\\&&+(P_{\mu}q_{\nu}+q_{\mu}P_{\nu})(A_3^{(2)}Z_2+A_1^{(3)}\frac{q\cdot P}{q^2}-A_2^{(3)})\nonumber\\&&+q_{\mu}q_{\nu}\Big[A_4^{(2)}Z_2+2\frac{q\cdot P}{q^2}A_2^{(1)}A_1^{(2)}\Big],
\label{productOFp6}
\end{eqnarray}
where $P=P^\prime+P^{\prime\prime}$ and $A_j^{(i)}$ and $Z_2$ are functions of $x_1$, $p^{\prime 2}_{\bot}$, $p^{\prime}_{\bot}\cdot q_{\bot}$, and $q^2$. These functions have been obtained in Ref. \cite{Jaus:1999zv}. Again, in the above replacements, $B_n^{(m)}$ and $C^{(m)}_{n}$ can be naturally disregarded at toy model level and their loop integrals vanish manifestly with a standard meson vertex.

Let us take the second rank tensor decomposition $\hat{p}^{\prime}_{1\mu}\hat{p}^{\prime}_{1\nu}$ as an example to explain how to effectively set $B_n^{(m)}$ and $C^{(m)}_{n}$ functions to 0 and eliminate the corresponding $\tilde{\omega}$-dependent terms. There are two $\tilde{\omega}$-dependent functions in the leading order of $\tilde{\omega}$ decomposition, i.e., $B_1^{(2)}$ and $C_1^{(2)}$.

From Ref. \cite{Jaus:1999zv} we have
\begin{eqnarray}
B_{1}^{(2)}&=&A_1^{(1)}C_1^{(1)}-A_1^{(2)},
\end{eqnarray}
by introducing the explicit expression of $C_1^{(1)}$ in Ref. \cite{Jaus:1999zv}, and one can easily obtain
\begin{eqnarray}
B_1^{(2)}&=&-A_1^{(1)}N_2+A_1^{(1)}Z_2-A_1^{(2)}
\end{eqnarray}
in the toy model. The loop integral of $B_1^{(2)}$ naturally vanishes. On the other hand, beyond the toy model, this term should also be eliminated manifestly, i.e., we have the replacement
\begin{eqnarray}
A_1^{(1)}\hat{N}_2\rightarrow A_1^{(1)}Z_2-A_1^{(2)}.
\end{eqnarray}
The same procedure can be applied to $C_{1}^{(2)}$ function, which gives us,
\begin{eqnarray}
A_2^{(1)}\hat{N}_2\rightarrow A_2^{(1)}Z_2+\frac{q\cdot P}{q^2}A_1^{(2)}.
\end{eqnarray}

Here, we still need to emphasize that beyond the toy model, when the conventional light front vertex functions are introduced, elimination of $B_n^{(m)}$ is also necessary when calculating semi-leptonic form factors with $S$-wave and $P$-wave mesons as final states. However, one would obtain very small corrections since these $B_n^{(m)}$ functions give small contributions as described earlier.

Expanding $\hat{S}_{\mu\nu}^{{}^3D_1}$, and replacing the $\hat{p}^{\prime}_{1\mu}\hat{p}^{\prime}_{1\nu}\hat{N}_2$, $\hat{p}^{\prime}_{1\mu}\hat{p}^{\prime}_{1\nu}$, $\hat{p}^{\prime}_{1\mu}\hat{N}_2$, $\hat{p}^{\prime}_{1\mu}$, and $\hat{N}_2$ terms with the above replacements in Eqs.~(\ref{productOFp1})-(\ref{productOFp2}) and Eqs. (\ref{productOFp4})-(\ref{productOFp6}), we can obtain the form factors of a ${}^3D_{1}$ state by comparing the general definition of a matrix element given by Eq. (\ref{GD31}). One needs to note that since in the expansion of $\hat{S}_{\mu\nu}^{{}^3D_1}$, there is no product term of three $\hat{p}_1^{\prime}$'s, Eq. (\ref{productOFp3}) is not used here. This equation just helps us find the tensor decomposition of $\hat{p}^{\prime}_{1\mu}\hat{p}^{\prime}_{1\nu}\hat{N}_2$. This procedure is identical to what we obtain the tensor decomposition of $\hat{p}^{\prime}_{1\mu}\hat{N}_2$ by analyzing the product of $\hat{p}^{\prime}_{1\mu}\hat{p}^{\prime}_{1\nu}$.

After including the zero-mode effect introduced from $B_n^{(m)}$ and $C^{(m)}_{n}$ functions, we present the explicit form factors, $g_D(q^2)$, $f_{D}(q^2)$, $a_{D+}(q^2)$ and $a_{D-}(q^2)$ as
\allowdisplaybreaks
\end{multicols}

\begin{eqnarray}
g_{D}(q^2)&=&-\frac{N_c}{16\pi^3}\int dx_2 d^2 p^\prime_\bot
\frac{2 h^\prime_P h^{\prime\prime}_{{}^3D_1}}{\left(1-x\right)\hat{N}^\prime_1 \hat{N}^{\prime\prime}_1}\left\{\left[A_1^{(1)}\left(2m_2-m_1^{\prime\prime}-m_1^{\prime}\right)+A_2^{(1)}
\left(m_1^{\prime\prime}-m_1^\prime\right)+m_1^\prime\right]-\frac{2}{\omega^{\prime\prime}_{{}^3D_1}}A_1^{(2)}\right\},
\\
f_{D}(q^2)&=&\frac{N_{c}}{16\pi^3}\int dx_2 d^2 p^\prime_\bot
\frac{2 h^\prime_{P} h^{\prime\prime}_{{}^3D_1}}{\left(1-x\right)\hat{N}^\prime_1\hat{N}^{\prime\prime}_1}
\Bigg\{2\bigg[4A_1^{(2)}\left(m_2-m^{\prime}_i\right)+m_2^2\left(m^{\prime\prime}_1+m^{\prime}_1\right)-
m_2\Big[\left(m_1^{\prime\prime}+m_1^{\prime}\right)^2+x\left(M^{\prime\prime2}-M_0^{\prime\prime2}\right)\nonumber\\&&+
x\left(M^{\prime2}-M_0^{\prime2}\right)-q^2\Big]+m_1^{\prime\prime}\left[m_1^{\prime}m_1^{\prime\prime}
+m_1^{\prime2}-M^{\prime2}+x\left(M^{\prime2}-M_0^{\prime2}\right)+Z_2\right]+m_1^{\prime}
\left[-M^{\prime\prime2}+x\left(M^{\prime\prime2}-M_0^{\prime\prime2}\right)+Z_2\right]\bigg]
\nonumber\\&&+4\frac{A_1^{(2)}}{\omega^{\prime\prime}_{{}^3D_1}}\left[2m_2\left(-m_2-m_1^{\prime\prime}+m_1^{\prime}\right)
+2m_1^{\prime\prime}m_1^{\prime}+M^{\prime\prime2}+
M^{\prime2}-q^2-2Z_2\right] \Bigg\},
\\
a_{D+}(q^2)&=&\frac{N_c}{16\pi^3}\int dx_2 d^2 p^{\prime}_{\bot}\frac{h^{\prime}_{P}h^{\prime\prime}_{{}^3D_1}}{\left(1-x\right)\hat{N}^\prime_1\hat{N}^{\prime\prime}_1}\Bigg\{
-2\left[A_1^{(1)}\left(2m_2+m_1^{\prime\prime}-5m_1^{\prime}\right)+m_1^{\prime}\right]-2A_2^{(1)}\left(-m_1^{\prime\prime}-m_1^{\prime}\right)-
2\left(A_2^{(2)}+A_3^{(2)}\right)\left(4m_1^{\prime}-4m_2\right)\nonumber\\&&+
\frac{2}{\omega^{\prime\prime}_{{}^3D_1}}\bigg[\left(A_1^{(1)}-A_2^{(2)}-A_3^{(2)}\right)\left(4m_2^2+4m_2m^{\prime\prime}_1
-4m_2m_1^{\prime}-4m_1^{\prime\prime}m_1^{\prime}-2M^{\prime\prime2}-2M^{\prime2}+2q^2\right)+
\left(-A_1^{(1)}-A_2^{(1)}+1\right)\nonumber\\&&\times\left[m_1^{\prime\prime2}+2m_1^{\prime\prime}m_1^{\prime}+m_1^{\prime2}
+x\left(M^{\prime\prime2}-M_0^{\prime\prime2}\right)+x\left(M^{\prime2}-M_0^{\prime2}\right)-q^2\right]+4\Bigg(A_1^{(1)}Z_2-A_1^{(2)}-
\left(A_2^{(2)}Z_2-2A_1^{(1)}A_1^{(2)}\right)\nonumber\\&&
-\left(A_1^{(1)}A_2^{(1)}Z_2+A_1^{(1)}A_1^{(2)}\frac{m_B^2-m_D^2}{q^2}-A_1^{(2)}A_2^{(1)}\right)\Bigg)\bigg]\Bigg\},\\
a_{D-}(q^2)&=&\frac{N_c}{16\pi^3}\int dx_2 d^2 p^{\prime}_{\bot}\frac{h^{\prime}_{P}h^{\prime\prime}_{{}^3D_1}}{\left(1-x\right)\hat{N}^\prime_1\hat{N}^{\prime\prime}_1}\Bigg\{
-2A_1^{(1)}\left(2m_2-m_1^{\prime\prime}-3m_1^{\prime}\right)-2A_2^{(1)}\left(4m_2+m_1^{\prime\prime}
-7m_1^{\prime}\right)-\left(2A_3^{(2)}+2A_4^{(2)}\right)\left(4m_1^{\prime}-4m_2\right)\nonumber\\&&-6m_1^{\prime}+
\frac{1}{\omega^{\prime\prime}_{{}^3D_1}}\bigg[\left(2A_1^{(1)}+2A_2^{(1)}-2\right)\Big[2m_2^2-4m_2m_1^{\prime}-m_1^{\prime\prime2}
-2m_1^{\prime\prime}m_1^{\prime}+m_1^{\prime2}-2M^{\prime2}-
x\left(M^{\prime\prime2}-M_0^{\prime\prime2}\right)\nonumber\\&&+x\left(M^{\prime2}-M_0^{\prime2}\right)+q^2\Big]+
\left(2A_3^{(2)}+2A_4^{(2)}-2A_2^{(1)}\right)\left(-4m_2^2-4m_2m_1^{\prime\prime}+4m_2m_1^{\prime}
+4m_1^{\prime\prime}m_1^{\prime}+2M^{\prime\prime2}+2M^{\prime2}-2q^2\right)\nonumber\\&&+12\left(A_2^{(1)}Z_2+
\frac{M^{\prime2}-M^{\prime\prime2}}{q^2}A_1^{(2)}\right)-
8\left(A_4^{(2)}Z_2+2\frac{M^{\prime2}-M^{\prime\prime2}}{q^2}A_2^{(1)}A_1^{(2)}\right)-4Z_2+4\Bigg(A_1^{(1)}Z_2
-A_1^{(2)}-2\Big(A_1^{(1)}A_2^{(1)}Z_2\nonumber\\&&+A_1^{(1)}A_1^{(2)}\frac{m_B^2-m_D^2}{q^2}-A_1^{(2)}A_2^{(1)}\Big)\Bigg)
\bigg]\Bigg\}.
\end{eqnarray}
\begin{multicols}{2}
%
The same procedure can also be applied to the transitions relevant to ${}^1D_2$, ${}^3D_2$ as well as ${}^3D_3$ states, whose results are given in Appendix \ref{traceform}. In the following, we continue to discuss these states and focus on the new subjects that should be introduced when dealing with higher spin $D$-wave states.

By analogy to the conventional vertex functions obtained in Appendix \ref{vertex}, we write out the covariant vertex functions for ${}^1D_2$, ${}^3D_2$, and ${}^3D_3$ in one loop Feynman diagrams as,
\begin{eqnarray}
&&iH_{{}^1D_2}\gamma_{5}K_{\mu}K_{\nu}\epsilon^{\mu\nu},
\end{eqnarray}
\begin{eqnarray}
&& iH_{{}^3D_2}\left[\frac{1}{W^a_{{}^3D_2}}\gamma_{\mu}\gamma_{\nu}+
\frac{1}{W^b_{{}^3D_2}}\gamma_{\mu}K_{\nu}+\frac{1}{W^c_{{}^3D_2}}K_{\mu}
K_{\nu}\right]\epsilon^{\mu\nu},
\end{eqnarray}
\begin{eqnarray}
&& iH_{{}^3D_3}\left[K_{\mu}K_{\nu}\left(\gamma_{\alpha}+\frac{2K_{\alpha}}{W_{{}^3D_3}}\right)+K_{\mu}K_{\alpha}
\left(\gamma_{\nu}+\frac{2K_{\nu}}{W_{{}^3D_3}}\right)\right.\nonumber\\&&\left.+K_{\alpha}K_{\nu}\left(\gamma_{\mu}+\frac{2K_{\mu}}{W_{{}^3D_3}}\right)\right]\epsilon^{\mu\nu\alpha},
\end{eqnarray}
respectively, where $H_{{}^{2S+1}D_J}$ and $W_{{}^{2S+1}D_J}$ are the functions of associated states in the
momentum space.

In order to obtain the $B_{(s)}\rightarrow D_{(s)2}^{*(\prime)}$, $D_{(s)3}^{*}$ transition form factors, the matrix elements are denoted as
\begin{eqnarray}
 B_{\mu}^{B_{(s)}(D^{*}_{(s)2})}\equiv\left\langle D^{*}_{(s)2}(P^{\prime\prime},\epsilon^{\prime\prime})\left|V_{\mu}-A_{\mu}\right|B_{(s)}(P^{\prime})\right\rangle,
\\
B_{\mu}^{B_{(s)}(D^{*\prime}_{(s)2})}\equiv \left\langle D^{*\prime}_{(s)2}(P^{\prime\prime},\epsilon^{\prime\prime})\left|V_{\mu}-A_{\mu}\right|B_{(s)}(P^{\prime})\right\rangle,
\\
B_{\mu}^{B_{(s)}(D^{*}_{(s)3})}\equiv \left\langle D^{*}_{(s)3}(P^{\prime\prime},\epsilon^{\prime\prime})\left|V_{\mu}-A_{\mu}\right|B_{(s)}(P^{\prime})\right\rangle.
\end{eqnarray}
It is straightforward to obtain the explicit expressions of the corresponding one loop integrals as
\allowdisplaybreaks
\begin{eqnarray}
B_{\mu}^{B_{(s)}(D^{*}_{(s2)})}&=&-i^3\frac{N_c}{(2\pi)^4}\int
d^4p^{\prime}_1\frac{H^{\prime}_{P}\left(iH^{\prime\prime}_{{}^1D_2}\right)}{N_1^{\prime}N_1^{\prime\prime}N_2}S_{\mu\alpha\beta}^{{}^1D_2}
\epsilon^{*\prime\prime\alpha\beta},\nonumber\\
\\
B_{\mu}^{B_{(s)}(D^{*\prime}_{(s2)})}&=&-i^3\frac{N_c}{(2\pi)^4}\int
d^4p^{\prime}_1\frac{H^{\prime}_{P}\left(iH^{\prime\prime}_{{}^3D_2}\right)}{N_1^{\prime}N_1^{\prime\prime}N_2}S_{\mu\alpha\beta}^{{}^3D_2}
\epsilon^{*\prime\prime\alpha\beta},\nonumber\\
\\
B_{\mu}^{B_{(s)}(D^{*}_{(s3)})}&=&-i^3\frac{N_c}{(2\pi)^4}\int
d^4p^{\prime}_1\frac{H^{\prime}_{P}\left(iH^{\prime\prime}_{{}^3D_3}\right)}{N_1^{\prime}N_1^{\prime\prime}N_2}S_{\mu\alpha\beta\nu}^{{}^3D_3}
\epsilon^{*\prime\prime\alpha\beta\nu}.\nonumber\\
\end{eqnarray}

By integrating over $p_1^{\prime-}$ as discussed in the case of $B_{\mu}^{B_{(s)}(D^{*}_{(s1)})}$, the following replacements should be taken,
\begin{eqnarray*}
N_1^{\prime(\prime\prime)}&\rightarrow&\hat{N}_1^{\prime(\prime\prime)}=x_1\left(M^{\prime(\prime\prime)2}-M_0^{\prime(\prime\prime)2}\right),
\\
H^{\prime}_P&\rightarrow& h_{P}^{\prime},
\\
H^{\prime\prime}_{M}&\rightarrow& h^{\prime\prime}_{M}
=(M^{\prime\prime2}-M_0^{\prime\prime2})\sqrt{x_1x_2}h^{\prime}_M,
\\
W^{\prime\prime}_M&\rightarrow&\omega^{\prime\prime}_M,
\\
\int \frac{d^4p_1^{\prime}}{N_1^{\prime}N_1^{\prime\prime}N_2}H^{\prime}_P
H^{\prime\prime}_{M}S^M &\rightarrow& -i\pi\int \frac{dx_2d^2p_{\bot}^{\prime}}{x_2\hat{N}_1^{\prime}\hat{N}_1^{\prime\prime}}
h^{\prime}_Ph^{\prime\prime}_M\hat{S}^M,
\end{eqnarray*}
where $M$ appearing in the subscripts or superscripts denotes ${}^1D_2$, ${}^3D_2$, and ${}^3D_3$, by which these physical quantities corresponding to different transitions can be easily distinguished. The explict forms of $h^\prime_M$ are given by Eqs.~(\ref{AppAhM}) in Appendix \ref{vertex}. We also present the trace expansions of $\hat{S}^{{}^1D_2}_{\mu\alpha\beta}$, $\hat{S}_{\mu\alpha\beta}^{{}^3D_2}$ and $\hat{S}_{\mu\alpha\beta\nu}^{{}^3D_3}$ in Appendix \ref{traceform}. After performing the contour integral over $p_1^{\prime-}$,  the quantities, $h^{\prime\prime}_{M}$, $\omega^{\prime\prime}_M$, and $\hat{\epsilon}^{\prime\prime}$ replace the corresponding $H^{\prime\prime}_M$, $W^{\prime\prime}_M$, and $\epsilon^{\prime\prime}$, respectively.
The next step is to maintain the $\tilde{\omega}$ independence, the $B_i^{(j)}$ and $C_{i}^{(j)}$ should vanish manifestly by including the zero-mode effect.

Apart from decomposing the tensors like Eqs.~(\ref{productOFp1})-(\ref{productOFp5}) in the $B_{(s)}\rightarrow D_{(s)1}^{*}$ transition discussed above, for $J=2$ states, one also needs to consider the product of four $\hat{p}_1^{\prime}$'s to obtain the reduction of $\hat{p}_{1\mu}^{\prime}\hat{p}_{1\nu}^{\prime}\hat{p}_{1\alpha}^{\prime}\hat{N}_2$, which has been done in Ref. \cite{Cheng:2003sm}\footnote{We should mention that there is a typo in Ref. \cite{Cheng:2003sm} for the $A_{9}^{(4)}$ function, whose correct expression is given by $A_9^{(4)}=A_2^{(1)}A_6^{(3)}-\frac{3}{q^2}A_4^{(4)}$.}, i.e.,
\end{multicols}
\begin{eqnarray}
\hat{p}_{1\mu}^{\prime}\hat{p}_{1\nu}^{\prime}\hat{p}_{1\alpha}^{\prime}\hat{p}_{1\beta}^{\prime}&\doteq& \left(g_{\mu\nu}g_{\alpha\beta}+g_{\mu\alpha}g_{\nu\beta}+g_{\mu\beta}g_{\nu\alpha}\right)A_{1}^{(4)}+
\left(g_{\mu\nu}P_{\alpha}P_{\beta}+g_{\mu\alpha}P_{\nu}P_{\beta}+g_{\mu\beta}P_{\nu}P_{\alpha}
+g_{\nu\alpha}P_{\mu}P_{\beta}+g_{\nu\beta}P_{\mu}P_{\alpha}+g_{\alpha\beta}P_{\mu}P_{\nu}\right)A_2^{(4)}\nonumber\\&&
+\left[g_{\mu\nu}\left(P_{\alpha}q_{\beta}+P_{\beta}q_{\alpha}\right)+g_{\mu\alpha}\left(P_{\nu}q_{\beta}+P_{\beta}q_{\nu}\right)+
g_{\mu\beta}\left(P_{\nu}q_{\alpha}+P_{\alpha}q_{\nu}\right)+g_{\nu\alpha}\left(P_{\mu}q_{\beta}+P_{\beta}q_{\mu}\right)
+g_{\nu\beta}\left(P_{\mu}q_{\alpha}+P_{\alpha}q_{\mu}\right)\nonumber\right.\\&&\left.+g_{\alpha\beta}\left(P_{\mu}q_{\nu}+P_{\nu}q_{\mu}\right)\right]A_3^{(4)}
+\left(g_{\mu\nu}q_{\alpha}q_{\beta}+g_{\mu\alpha}q_{\nu}q_{\beta}+g_{\mu\beta}q_{\nu}q_{\alpha}
+g_{\nu\alpha}q_{\mu}q_{\beta}+g_{\nu\beta}q_{\mu}q_{\alpha}+g_{\alpha\beta}q_{\mu}q_{\nu}\right)A_4^{(4)}\nonumber\\&&
+P_{\mu}P_{\nu}P_{\alpha}P_{\beta}A_5^{(4)}+\left(P_{\mu}P_{\nu}P_{\alpha}q_{\beta}
+P_{\mu}P_{\nu}q_{\alpha}P_{\beta}+P_{\mu}q_{\nu}P_{\alpha}P_{\beta}+q_{\mu}P_{\nu}P_{\alpha}P_{\beta}\right)
A_6^{(4)}\nonumber\\&&+\left(P_{\mu}P_{\nu}q_{\alpha}q_{\beta}+P_{\mu}P_{\alpha}q_{\nu}q_{\beta}
+P_{\mu}P_{\beta}q_{\nu}q_{\alpha}+
P_{\nu}P_{\alpha}q_{\mu}q_{\beta}+P_{\nu}P_{\beta}q_{\mu}q_{\alpha}+P_{\alpha}P_{\beta}q_{\mu}q_{\nu}\right)
A_7^{(4)}\nonumber\\&&+\left(q_{\mu}q_{\nu}q_{\alpha}P_{\beta}+q_{\mu}q_{\nu}P_{\alpha}q_{\beta}
+q_{\mu}P_{\nu}q_{\alpha}q_{\beta}+P_{\mu}q_{\nu}q_{\alpha}q_{\beta}\right)A_8^{(4)}+
q_{\mu}q_{\nu}q_{\alpha}q_{\beta}A_9^{(4)},
\end{eqnarray}

and the corresponding tensor decomposition of $\hat{p}_{1\mu}^{\prime}\hat{p}_{1\nu}^{\prime}\hat{p}_{1\alpha}^{\prime}\hat{N}_2$ is given by

\begin{eqnarray}
\hat{p}_{1\mu}^{\prime}\hat{p}_{1\nu}^{\prime}\hat{p}_{1\alpha}^{\prime}\hat{N}_2\rightarrow&&\left(g_{\mu\nu}P_{\alpha}+g_{\mu\alpha}P_{\nu}+g_{\nu\alpha}P_{\mu}\right)
\left(A_1^{(3)}Z_2-A_1^{(4)}\right)+\left(g_{\mu\nu}q_{\alpha}
+g_{\mu\alpha}q_{\nu}+g_{\nu\alpha}q_{\mu}\right)\Big[A_2^{(3)}Z_2+
\frac{q\cdot P}{3q^2}\left(A_1^{(2)}\right)^2\Big]\nonumber\\&&+P_{\mu}P_{\nu}P_{\alpha}\left(A_3^{(3)}Z_2-2A_2^{(2)}A_1^{(2)}-A_2^{(4)}\right)+
\left(P_{\mu}P_{\nu}q_{\alpha}+P_{\mu}q_{\nu}P_{\alpha}+q_{\mu}P_{\nu}P_{\alpha}\right)\left(A_4^{(3)}Z_2+A_2^{(2)}A_1^{(2)}\frac{m_B^2-m_D^2}{q^2}
-2A_3^{(4)}\right)\nonumber\\&&+\left( q_{\mu}q_{\nu}P_{\alpha}+q_{\mu}P_{\nu}q_{\alpha}+P_{\mu}q_{\nu}q_{\alpha}\right)\left(A_5^{(3)}Z_2+2\frac{m_B^2-m_D^2}{q^2}A_3^{(4)}
-A_4^{(4)}\right)+q_{\mu}q_{\nu}q_{\alpha}\Bigg\{A_6^{(3)}Z_2\nonumber\\&&
+3\frac{q\cdot P}{q^2}\left[A_2^{(1)}A_2^{(3)}-\frac{1}{3q^2}\left(A_1^{(2)}\right)^2\right]\Bigg\}.\label{productOFp7}
\end{eqnarray}
Furthermore, in the $B_{(s)}\rightarrow D_{(s)3}^{*}$ transition, $\hat{p}^{\prime}_{1\mu}\hat{p}^{\prime}_{1\nu}\hat{p}^{\prime}_{1\alpha}\hat{p}^{\prime}_{1\beta}\hat{N}_2$ can be deduced by the product of five $\hat{p}_{1}^{\prime}$'s, where the derivation of the concrete form of $\hat{p}_{1\mu}\hat{p}_{1\nu}^{\prime}\hat{p}_{1\alpha}^{\prime}\hat{p}^\prime_{1\beta}\hat{p}_{1\gamma}^{\prime}$  is given in Appendix \ref{decomposition}. Accordingly, we obtain
\begin{eqnarray}
\hat{p}'_{1\mu}\hat{p}'_{1\nu}\hat{p}'_{1\alpha}\hat{p}'_{1\beta}\hat{N}_2\rightarrow&& I_{1\mu\nu\alpha\beta}A_1^{(4)}Z_2+I_{2\mu\nu\alpha\beta}(A_2^{(4)}Z_2-2A_1^{(1)}A_1^{(4)})+I_{3\mu\nu\alpha\beta}\left(A_3^{(4)}Z_2
+A_1^{(1)}A_1^{(4)}\frac{m_B^2-m_D^2}{q^2}-A_2^{(1)}A_1^{(4)}\right)\nonumber\\&&+
I_{4\mu\nu\alpha\beta}\left(A_4^{(4)}Z_2+2\frac{m_B^2-m_D^2}{q^2}A_2^{(1)}A_1^{(4)}\right)+
I_{5\mu\nu\alpha\beta}\left(A_5^{(4)}Z_2-2A_3^{(3)}A_1^{(2)}-2A_1^{(1)}A_2^{(4)}\right)+
I_{6\mu\nu\alpha\beta}\Big(A_6^{(4)}Z_2\nonumber\\&&+\frac{m_B^2-m_D^2}{q^2}A_3^{(3)}A_1^{(2)}-A_2^{(2)}A_2^{(3)}-2A_1^{(1)}A_3^{(4)}\Big)
+I_{7\mu\nu\alpha\beta}\left(A_7^{(4)}Z_2+2\frac{m_B^2-m_D^2}{q^2}A_2^{(2)}A_2^{(3)}-2A_1^{(1)}A_4^{(4)}\right)\nonumber\\&&
+I_{8\mu\nu\alpha\beta}\left(A_8^{(4)}Z_2+3\frac{m_B^2-m_D^2}{q^2}A_1^{(1)}A_4^{(4)}-A_2^{(1)}A_4^{(4)}+\frac{2A_2^{(1)}A_1^{(4)}}{q^2}\right)\nonumber\\&&
+I_{9\mu\nu\alpha\beta}\left(A_9^{(4)}Z_2+4\frac{m_B^2-m_D^2}{q^2}(A_2^{(1)}A_4^{(4)}-2A_2^{(1)}A_1^{(4)})\right).
\label{productOFp8}
\end{eqnarray}

One can refer to Ref. \cite{Cheng:2003sm, Jaus:1999zv} for the explicit expressions of $A_{i}^{(j)}$ functions.

\begin{multicols}{2}
In fact, after expanding the products of a couple of $\hat{p}_1^\prime$'s and the products of several $\hat{p}_1^{\prime}$'s with $N_2$ to the first order in $\tilde{\omega}$, we find that the zero condition deduced from $B_n^{(m)}$ and $C^{(m)}_{n}$ functions can be independently expressed in terms of lower order of $A^{(j)}_{l(k)}$ functions. To illustrate this point, in Table \ref{replacements}, we give all replacements deduced from $B_n^{(m)}=0$ and $C_n^{(m)}=0$. Strictly speaking these equations hold only when loop integrations of these functions have been done.
The relations presented in Table \ref{replacements} have been applied to Eqs. (\ref{productOFp4})-(\ref{productOFp6}) and Eqs. (\ref{productOFp7})-(\ref{productOFp8}).
\renewcommand\arraystretch{1.5}
\begin{table*}[htbp]
\caption{The corresponding $A^{(j)}_{l(k)}$ replacements related to $B_{m}^{(j+1)}=0$ and $C_{n}^{(j+1)}=0$ conditions.}
\label{replacements}
\centering
\begin{tabular}{|cccc|cccccccc}
\toprule[1pt]
$B_{m}^{(j+1)}(C_{n}^{(j+1)})$ & Related $A_{l(k)}^{(j)}$ functions&$B_{m}^{(j+1)}(C_{n}^{(j+1)})$ & Related $A_{l(k)}^{(j)}$ functions\\
\midrule[1pt]
$C_1^{(1)}$&$\hat{N}_2\rightarrow Z_2$&&\\
\hline
$B_1^{(2)}$&$A_1^{(1)}\hat{N}_2\rightarrow A_1^{(1)}Z_2-A_1^{(2)}$&$C_1^{(2)}$ &$A_2^{(1)}\hat{N}_2\rightarrow A_2^{(1)}Z_2+\frac{q\cdot P}{q^2}A_1^{(2)}$\\
\hline
$B_1^{(3)}$&$A_2^{(2)}\hat{N}_2\rightarrow A_2^{(2)}Z_2-2A_1^{(3)}$&$B_2^{(3)}$&$A_3^{(2)}\hat{N}_2\rightarrow A_3^{(2)}Z_2+A_1^{(3)}\frac{q\cdot P}{q^2}-A_2^{(3)}$\\
$C_1^{(3)}$&$A_1^{(2)}\hat{N}_2\rightarrow A_1^{(2)}Z_2$&$C_2^{(3)}$&$A_4^{(2)}\hat{N}_2\rightarrow A_4^{(2)}Z_2+2\frac{q\cdot P}{q^2}A_2^{(1)}A_1^{(2)}$\\
\hline
$B_1^{(4)}$&$A_1^{(3)}\hat{N}_2\rightarrow A_1^{(3)}Z_2-A_1^{(4)}$&$B_2^{(4)}$&$A_3^{(3)}\hat{N}_2\rightarrow A_3^{(3)}Z_2-2A_2^{(2)}A_1^{(2)}-A_2^{(4)}$\\
$B_3^{(4)}$&$A_4^{(3)}\hat{N}_2\rightarrow A_4^{(3)}Z_2+A_2^{(2)}A_1^{(2)}\frac{q\cdot P}{q^2}-2A_3^{(4)}$&$B_4^{(4)}$&$A_5^{(3)}\hat{N}_2\rightarrow A_5^{(3)}Z_2+2\frac{q\cdot P}{q^2}A_3^{(4)}-A_4^{(4)}$\\
$C_1^{(4)}$&$A_2^{(3)}\hat{N}_2\rightarrow A_2^{(3)}Z_2+\frac{q\cdot P}{3q^2}(A_1^{(2)})^2$&$C_2^{(4)}$&$A_6^{(3)}\hat{N}_2\rightarrow A_6^{(3)}Z_2+3\frac{q\cdot P}{q^2}[A_2^{(1)}A_2^{(3)}-\frac{1}{3q^2}(A_1^{(2)})^2]$\\
\hline
$B_1^{(5)}$&$A_2^{(4)}\hat{N}_2\rightarrow A_2^{(4)}Z_2-2A_1^{(1)}A_1^{(4)}$&$B_2^{(5)}$&$A_3^{(4)}\hat{N}_2\rightarrow A_3^{(4)}Z_2+A_1^{(1)}A_1^{(4)}\frac{q\cdot P}{q^2}-A_2^{(1)}A_1^{(4)}$\\
$B_3^{(5)}$&$A_5^{(4)}\hat{N}_2\rightarrow A_5^{(4)}Z_2-2A_3^{(3)}A_1^{(2)}-2A_1^{(1)}A_2^{(4)}$&$B_4^{(5)}$&$A_6^{(4)}\hat{N}_2\rightarrow A_6^{(4)}Z_2+\frac{q\cdot P}{q^2}A_3^{(3)}A_1^{(2)}-A_2^{(2)}A_2^{(3)}-2A_1^{(1)}A_3^{(4)}$\\
$B_5^{(5)}$&$A_7^{(4)}\hat{N}_2\rightarrow A_7^{(4)}Z_2+2\frac{q\cdot P}{q^2}A_2^{(2)}A_2^{(3)}-2A_1^{(1)}A_4^{(4)}$&$B_6^{(5)}$&$A_9^{(4)}\hat{N}_2\rightarrow A_9^{(4)}Z_2+4\frac{q\cdot P}{q^2}(A_2^{(1)}A_4^{(4)}-2A_2^{(1)}A_1^{(4)})$\\
$C_1^{(5)}$&$A_1^{(4)}\hat{N}_2\rightarrow A_1^{(4)}Z_2$&$C_2^{(5)}$&$A_4^{(4)}\hat{N}_2\rightarrow A_4^{(4)}Z_2+2\frac{q\cdot P}{q^2}A_2^{(1)}A_1^{(4)}$\\
$C_3^{(5)}$&$A_9^{(4)}\hat{N}_2\rightarrow A_9^{(4)}Z_2+4\frac{q\cdot P}{q^2}\left(A_2^{(1)}A_4^{(4)}-2A_2^{(1)}A_1^{(4)}\right)$&&\\
\hline
\bottomrule[1pt]
\end{tabular}
\end{table*}

The replacements presented in Table \ref{replacements} in the integration of Eq. (\ref{toy model}) can be proven in a toy model vertex, whose deductions are given in Appendix \ref{toy model prove}. This indicates that generalization of Jaus's model to higher spin $J$ states is doable. However, when a conventional $D$-wave vertex function is introduced in the loop integration of Eq. (\ref{toy model}), at present, it is difficult to prove these identities. Here, we still need to emphasize that for the conventional light front vertex functions, the replacements listed in Table \ref{replacements} work very well in obtaining the corresponding form factors as well as semi-leptonic decay widths. Besides, the vanishment of $\tilde{\omega}$-dependent terms $B_n^{(m)}$ and $C^{(m)}_{n}$ is not only the conclusion from Jaus's model, but also a physical requirement to obtain physical quantities, i.e., to keep Lorentz invariance intact. So in this work, we still use the replacements listed in Table \ref{replacements} to perform our analysis.

\section{Numerical results}\label{sec3}

\subsection{Numerical results of form factors and decay widths}
In the framework of the light-front quark model \cite{Cheng:2003sm,Chung:1988mu,Ke:2011mu}, one usually adopts a single simple harmonic oscillator (SHO) wave function to approximately describe the corresponding spatial wave function of a meson, where the parameter $\beta$ in the SHO wave function is extracted from the corresponding decay constant.
Due to the limited information on the decay constants of $D$-wave charmed/charmed-strange mesons, we may consult another approach to get it.

In Refs. \cite{Song:2015fha,Song:2015nia}, the mass spectrum of $D/D_s$ mesons have been systematically studied in the framework of the modified Godfrey-Isgur (MGI) model, by which their numerical spatial wave functions can be also obtained. As illustrated in Appendix \ref{vertex}, we may adopt numerical spatial wave functions as an input in our calculation (see Appendix \ref{vertex} for more details). In Table \ref{MGID} and \ref{MGIDS}, we present the numerical masses and eigenvectors of corresponding wave functions of $D$-wave $D^{**}$ and $D_{s}^{**}$ mesons.
Here, the numerical wave functions for the discussed $D$-wave $D/D_s$ mesons can be precisely described by the expansion in the twenty-one SHO bases, where the corresponding expansion coefficients
form the eigenvectors.

Here, we need to emphasis that in the semi-leptonic calculation, no free parameters are included, all parameters have been well fitted in potential model calculations. We also checked the input wave functions obtained from GI model in Ref .\cite{Song:2015fha,Song:2015nia} and obtained very close results of semi-leptonic decay form factors as well as branching ratios. Indicating that once the $\chi^2$ when fitting the mass spectrum is well controlled, we well get consistent results by using the input wave functions from potential model.

Although the observed $D^{*}(2760)$, $D(2750)$ \cite{delAmoSanchez:2010vq, Aaij:2013sza}, $D_{s1}^{*}(2860)$ and $D_{s3}^{*}(2860)$ \cite{Aaij:2014xza,Aaij:2014baa} can be good candidates of $1D$ states in charmed and charmed-strange meson families \cite{DiPierro:2001dwf,Sun:2010pg,Zhong:2010vq,Li:2010vx,Colangelo:2012xi,
Chen:2011rr,Song:2014mha,Godfrey:2014fga,Zhou:2014ytp,Wang:2014jua}, in this work we still take the theoretical masses of $D$-wave $D/D_s$ mesons as an input when studying these semileptonic decays .
\renewcommand\arraystretch{1.5}
\begin{table*}[htbp]
\caption{The predicted masses and eigenvectors corresponding to numerical wave functions
of $D^{**}$ charmed mesons from the modified GI model \cite{Song:2015fha,Song:2015nia}.}
\label{MGID}
\centering
\begin{tabular}{cccccccccccc}
\toprule[1pt]
$n^{2S+1}L_J$&Mass (MeV)&Eigenvector\\
\midrule[1pt]
$1{}^3D_1$&2762&$\left[\begin{array}{cccccc}\{0.74,-0.46,0.35,-0.24,0.17,-0.12,0.09,-0.06,	0.05,\\-0.03,0.02,-0.02,0.01,-0.01,0.01,-0.01,0,0,0,0,0\}\end{array}\right]$\\
$2{}^3D_1$&3131&$\left[\begin{array}{cccccc}\{-0.55,-0.13,0.28,-0.38,0.36,-0.33,0.28,-0.23,0.19,	-0.15,0.12,\\-0.10,0.08,-0.06,0.05,-0.04,0.03,-0.02,0.02,-0.01,0.01\}\end{array}\right]$\\
$1{}^1D_2$&2773& $\left[\begin{array}{cccccc}\{-0.93,0.27,-0.21,0.08,-0.06,0.03,-0.02,0.01,\\-0.01,0,-0.01,0,0,0,0,0,0,0,0,0,0\}\end{array}\right]$\\
$2{}^1D_2$&3128&$\left[\begin{array}{cccccc}\{-0.35,-0.70,0.40,-0.37,0.21,-0.17,0.09,-0.07,0.04,	\\-0.04,0.02,-0.02,0.01,-0.01,0,0,0,0,0,0,0\}\end{array}\right]$\\
$1{}^3D_2$&2779&$\left[\begin{array}{cccccc}\{0.94,-0.26,0.20,-0.07,0.06,-0.02,0.02,-0.01,\\0.01,0,0,0,0,0,0,0,0,0,0,0,0\}\end{array}\right]$\\
$2{}^3D_2$&3135&$\left[\begin{array}{cccccc}\{-0.33,-0.72,0.40,-0.37,0.20,-0.16,0.09,-0.07,\\0.04,-0.03,0.02,-0.02,0.01,-0.01,0,0,0,0,0,0,0\}\end{array}\right]$\\
$1{}^3D_3$&2779&$\left[\begin{array}{cccccc}\{0.90,-0.33,0.22,-0.11,0.07,-0.04,0.03,-0.02,\\0.01,-0.01,0.01,0,0,0,0,0,0,0,0,0,0\}\end{array}\right]$\\
$2{}^3D_3$&3130&$\left[\begin{array}{cccccc}\{0.41,0.62,-0.42,0.38,-0.24,0.18,-0.12,0.09,-0.06,\\0.04,-0.03,0.02,-0.02,0.01,-0.01,0.01,0,0,0,0,0\}\end{array}\right]$\\
\bottomrule[1pt]
\end{tabular}
\end{table*}

\begin{table*}[htbp]
\caption{The predicted masses and eigenvectors corresponding to numerical wave functions
of $D_s^{**}$ charmed-strange mesons from the modified GI model \cite{Song:2015fha,Song:2015nia}.}
\label{MGIDS}
\centering
\begin{tabular}{cccccccccccc}
\toprule[1pt]
$n^{2S+1}L_J$&Mass(MeV)&Eigenvector\\
\midrule[1pt]
$1{}^3D_1$&2865&$\left[\begin{array}{cccccc}\{0.78,-0.44,0.33,	-0.21,0.15,-0.09,0.07,-0.04,0.03,\\-0.02,0.02,-0.01,0.01,-0.01,0,0,0,	0,0,0,0\}\end{array}\right]$\\
$2{}^3D_1$&3244&$\left[\begin{array}{cccccc}\{0.53,0.22,-0.34,0.41,	-0.36,0.31,-0.25,0.20,-0.15,0.12,-0.09,\\0.07,-0.05,0.04,-0.03,0.02,	-0.02,0.01,-0.01,0.01,0\}\end{array}\right]$\\
$1{}^1D_2$&2877& $\left[\begin{array}{cccccc}\{-0.96,0.20,	-0.17,0.05,-0.05,0.01,-0.02,\\0,0,0,0,0,0,0,0,0,0,0,0,0,0
\}\end{array}\right]$\\
$2{}^1D_2$&3247&$\left[\begin{array}{cccccc}\{0.26,0.81,-0.36,0.33,	-0.15,0.12,-0.05,0.05,-0.02,\\0.02,-0.01,0.01,0,0,0,0,0,0,0,0,0
\}\end{array}\right]$\\
$1{}^3D_2$&2882&$\left[\begin{array}{cccccc}\{-0.96,0.19,-0.17,0.04,-0.04,0.01,-0.01,\\0,0,0,0,0,0,0,0,0,0,0,0,0,0
\}\end{array}\right]$\\
$2{}^3D_2$&3252&$\left[\begin{array}{cccccc}\{0.25,0.82,-0.36,0.33,-0.14,0.12,-0.05,0.04,-0.02,\\0.02,-0.01,0.01,0,0,0,0,0,0,0,0,0
\}\end{array}\right]$\\
$1{}^3D_3$&2883&$\left[\begin{array}{cccccc}\{-0.94,0.26,-0.18,0.07,-0.05,0.02,-0.02,\\0.01,-0.01,0,0,0,0,0,0,0,0,0,0,0,0
\}\end{array}\right]$\\
$2{}^3D_3$&3251&$\left[\begin{array}{cccccc}\{-0.32,-0.75,0.40,	-0.33,0.18,-0.13,0.07,-0.05,\\0.03,-0.02,0.01,-0.01,0,0,0,0,0,0,0,0,0
\}\end{array}\right]$\\
\bottomrule[1pt]
\end{tabular}
\end{table*}

In our calculation, other input parameters include the constituent quark masses, $m_{u,d}=220$ MeV, $m_s=419$ MeV, $m_c=1628$ MeV and $m_b=4977$ MeV, which are consistent with those given in the modified GI model \cite{Song:2015fha,Song:2015nia}.
In order to determine the shape parameter $\beta$ for initial pseudoscalar bottom and bottom-strange mesons, we use the direct results of the lattice QCD \cite{Gamiz:2009ku}, where $f_B=190$ MeV and $f_{B_s}=231$ MeV. Then, the parameter $\beta$ can be extracted from these two decay constants by \cite{Cheng:2003sm}
\begin{eqnarray}
\mathcal{}f_p&=&2\frac{\sqrt{2N_c}}{16\pi^3}\int dx_2d^2p_{\bot}^{\prime}\frac{1}{\sqrt{x_2(1-x_2)}\tilde{M}_0^{\prime}}\nonumber\\&&\times\left[m_1^{\prime}x_2+m_2(1-x_2)\right]
\varphi^{\prime}\left(x_2,p_{\bot}^{\prime}\right),
\end{eqnarray}
where $m_1^{\prime}$ and $m_2$ denote the constituent quark masses of $b$ and light quark, respectively.
Finally, we have $\beta_B=0.567$ GeV and $\beta_{B_s}=0.6263$ GeV for bottom and bottom-strange mesons, respectively.

In Appendix \ref{traceform}, we list the detailed expressions for the form factors relevant to the production of ${}^3D_1$, ${}^1D_2$, ${}^3D_2$, and ${}^3D_3$ states.
Following the calculation in Ref. \cite{Cheng:2003sm,Jaus:1999zv}, we choose the $q^+=0$ frame.  Due to the equality $q^2=q^+q^--q_{\bot}^2$, all the results obtained for the form factors are only effective in the $q^2\leq 0$ region. 
This means that we need to extrapolate our results of the form factors to the time-like region.

In this work, we introduce the so-called $z$-series parametrization used in Refs. \cite{Wang:2017jow,Bourrely:2008za,Khodjamirian:2009ys} to obtain our form factors in the time-like region. This parametrization is suggested by including the general and
analytical properties of form factors \cite{Bourrely:2008za}. The explicit expression can be written as \cite{Wang:2017jow}
\begin{eqnarray}
F(q^2)&=&\frac{F(0)}{(1-q^2/m^2_{B_{(s)}})}\left\{1+b_1\left(z(q^2)
-z(0)\right.\right.\nonumber\\&&\left.\left.-\frac{1}{3}\left[z(q^2)^3-z(0)^3\right]\right)
+b_2\left(z(q^2)^2-z(0)^2\right.\right.\nonumber\\&&\left.\left.+\frac{2}{3}\left[z(q^2)^3-z(0)^3\right]
\right)\right\},
\end{eqnarray}
where the conformal transformation is introduced,
\begin{eqnarray}
z(q^2)=\frac{\sqrt{(m_{B}+m_D)^2-q^2}-\sqrt{(m_B+m_D)^2-(m_B-m_D)^2}}
{\sqrt{(m_B+m_D)^2-q^2}+\sqrt{(m_B+m_D)^2-(m_B-m_D)^2}}.\nonumber\\
\end{eqnarray}
Here, in order to give an accurate matching for the transition form factors of $B_{(s)}$ to $D$-wave charmed/charmed-strange mesons, two
parameters $b_1$ and $b_2$ are introduced.
In Table \ref{MGIFF}, the fitting parameters and form factors for $\left\langle D_{(s)}^{**}\left|V-A\right|B_{(s)}\right\rangle$ transitions are collected.

We present the form factors obtained for the $B_{(s)}\rightarrow 1D_{(s)}^{*} (2D_{(s)}^{*}), 1D_{(s)2}^{(\prime)} (2D_{(s)2}^{(\prime)}), 1D^*_{(s)3} (2D^*_{(s)3})$ transitions in Table \ref{MGIFF}. In addition, we also show the $q^2$ dependence of the form factors in Figs. (\ref{1D})-(\ref{2Ds}). From Figs. (\ref{1D})-(\ref{2Ds}), the $z_{3/2-}$ and $o_-$ have positive sign, and $z_{3/2+}$ and $o_+$ have negative sign, which is the requirement from the HQS. We will discuss the relations between our form factors and the expectation from HQS in Sec. \ref{sec4}.

\renewcommand\arraystretch{1.5}
\begin{table*}[htbp]
\caption{The form factors for the semileptonic decays of $B_{(s)}$ to 1$D$- and 2$D$-wave $D_{(s)}$ mesons.}
\label{MGIFF}
\centering
\begin{tabular}{cccccccccccc}
\toprule[1pt]
&$F(q^2=0)$&$F(q^2_{max})$&$b_1$&$b_2$&&$F(q^2=0)$&$F(q^2_{max})$&$b_1$&$b_2$\\
\midrule[1pt]
$g_D^{B\rightarrow D_1^{*}}$&0.0006&-0.0024&156.1&215.9&$g_D^{B_{s}\rightarrow D_{s1}^{*}}$&-0.0061&-0.0021&31.9&-41.3\\
$f_D^{B\rightarrow D_1^{*}}$&0.041&-0.104&92.36&925.3&$f_D^{B_{s}\rightarrow D_{s1}^{*}}$&-0.259&0.011&40.0&148.0\\
$a_{D+}^{B\rightarrow D_1^{*}}$&-0.023&-0.024&7.4&-21.4&$a_{D+}^{B_{s}\rightarrow D_{s1}^{*}}$&0.054&0.064&2.9&-17.8\\
$a_{D-}^{B\rightarrow D_1^{*}}$&0.035&0.039&6.2&-22.0&$a_{D-}^{B_{s}\rightarrow D_{s1}^{*}}$&-0.093&-0.116&1.2&-15.1\\
\hline
$n_{3/2}^{B\rightarrow D_2^{\prime}}$&-0.00174&-0.00169&10.8&-42.8&$n_{3/2}^{B_{s}\rightarrow D^{\prime}_{s2}}$&0.0043&0.0049&4.8&-35.2\\
$m_{3/2}^{B\rightarrow D_2^{\prime}}$&0.013&-0.008&54.0&149.0&$m_{3/2}^{B_{s}\rightarrow D^{\prime}_{s2}}$&-0.027&0.021&69.8&-22.0\\
$z_{3/2+}^{B\rightarrow D_2^{\prime}}$&-0.0015&-0.0012&17.6&-70.6&$z_{3/2+}^{B_{s}\rightarrow D^{\prime}_{s2}}$&0.0044&0.0049&6.2&-41.8\\
$z_{3/2-}^{B\rightarrow D_2^{\prime}}$&0.0018&0.0016&14.2&-58.1&$z_{3/2-}^{B_{s}\rightarrow D^{\prime}_{s2}}$&-0.0052&-0.0059&5.2&-39.2\\
\hline
$n_{5/2}^{B\rightarrow D_2}$&-0.015&-0.020&-2.2&-8.5&$n_{5/2}^{B_{s}\rightarrow D_{s2}}$&0.015&0.020&-2.9&-8.6\\
$m_{5/2}^{B\rightarrow D_2}$&0.304&0.286&11.4&-27.7&$m_{5/2}^{B_{s}\rightarrow D_{s2}}$&-0.327&-0.312&11.3&-36.2\\
$z_{5/2+}^{B\rightarrow D_2}$&0.0039&0.0078&-24.3&68.8&$z_{5/2+}^{B_{s}\rightarrow D_{s2}}$&-0.0043&-0.0082&-23.8&74.5\\
$z_{5/2-}^{B\rightarrow D_2}$&0.0089&0.0101&6.5&-62.7&$z_{5/2-}^{B_{s}\rightarrow D_{s2}}$&-0.010&-0.013&-0.56&-17.54\\
\hline
$y^{B\rightarrow D^{*}_3}$&0.002&0.003&-6.85&7.00&$y^{B_{s}\rightarrow D^{*}_{s3}}$&-0.0021&-0.0031&-7.6&9.5\\
$w^{B\rightarrow D^{*}_3}$&0.077&0.095&2.1&-21.7&$w^{B_{s}\rightarrow D^{*}_{s3}}$&-0.101&-0.127&1.1&-22.5\\
$o_{+}^{B\rightarrow D^{*}_3}$&-0.0015&-0.0024&-8.1&20.0&$o_{+}^{B_{s}\rightarrow D^{*}_{s3}}$&0.0016&0.0023&-6.34&2.73\\
$o_{-}^{B\rightarrow D^{*}_3}$&0.0018&0.0028&-6.0&1.7&$o_{-}^{B_{s}\rightarrow D^{*}_{s3}}$&-0.0020&-0.0029&-6.9&4.5\\
\hline
$g_D^{B\rightarrow 2D_1^{*}}$&-0.0095&-0.0087&15.5&-97.4&$g_D^{B_{s}\rightarrow 2D_{s1}^{*}}$&0.0079&0.0076&14.2&-91.4\\
$f_D^{B\rightarrow 2D_1^{*}}$&-0.631&-0.563&16.0&-49.0&$f_D^{B_{s}\rightarrow 2D_{s1}^{*}}$&0.491&0.438&16.5&-45.5\\
$a_{D+}^{B\rightarrow 2D_1^{*}}$&0.066&0.084&-3.4&-7.7&$a_{D+}^{B_{s}\rightarrow 2D_{s1}^{*}}$&-0.064&-0.082&-5.2&2.5\\
$a_{D-}^{B\rightarrow 2D_1^{*}}$&-0.102&-0.133&-4.8&-1.1&$a_{D-}^{B_{s}\rightarrow 2D_{s1}^{*}}$&0.101&0.132&-6.1&8.6\\
\hline
$n_{3/2}^{B\rightarrow 2D_2^{\prime}}$&0.0076&0.0103&-8.3&17.9&$n_{3/2}^{B_{s}\rightarrow 2D_{s2}^{\prime}}$&-0.0074&-0.0099&-9.5&31.2\\
$m_{3/2}^{B\rightarrow 2D_2^{\prime}}$&-0.058&-0.036&30.6&-128.4&$m_{3/2}^{B_{s}\rightarrow 2D_{s2}^{\prime}}$&0.0296&0.0030&63.1&-198.2\\
$z_{3/2+}^{B\rightarrow 2D_2^{\prime}}$&0.0087&0.0118&-8.3&17.6&$z_{3/2+}^{B_{s}\rightarrow 2D_{s2}^{\prime}}$&-0.0087&-0.0117&-9.6&30.9\\
$z_{3/2-}^{B\rightarrow 2D_2^{\prime}}$&-0.010&-0.013&-7.7&12.9&$z_{3/2-}^{B_{s}\rightarrow 2D_{s2}^{\prime}}$&0.011&0.014&-8.7&22.6\\
\hline
$n_{5/2}^{B\rightarrow 2D_2}$&0.0116&0.0154&-7.0&14.5&$n_{5/2}^{B_{s}\rightarrow 2D_{s2}}$&-0.0117&-0.0152&-7.1&18.9\\
$m_{5/2}^{B\rightarrow 2D_2}$&-0.196&-0.191&11.7&-43.4&$m_{5/2}^{B_{s}\rightarrow 2D_{s2}}$&0.203&0.197&12.2&-35.6\\
$z_{5/2+}^{B\rightarrow 2D_2}$&-0.0061&-0.0091&-15.5&53.0&$z_{5/2+}^{B_{s}\rightarrow 2D_{s2}}$&0.0064&0.0092&-15.9&55.2\\
$z_{5/2-}^{B\rightarrow 2D_2}$&-0.0076&-0.0099&-6.0&15.6&$z_{5/2-}^{B_{s}\rightarrow 2D_{s2}}$&0.0077&0.0098&-5.1&10.7\\
\hline
$y^{B\rightarrow 2D^{*}_3}$&0.0021&0.0030&-11.2&35.0&$y^{B_{s}\rightarrow 2D^{*}_{s3}}$&-0.0021&-0.0029&-11.7&41.2\\
$w^{B\rightarrow 2D^{*}_3}$&0.214&0.273&-3.73&-5.3&$w^{B_{s}\rightarrow 2D^{*}_{s3}}$&-0.189&-0.241&-5.1&2.1\\
$o_{+}^{B\rightarrow 2D^{*}_3}$&-0.0012&-0.0016&-9.5&25.2&$o_{+}^{B_{s}\rightarrow 2D^{*}_{s3}}$&0.0011&0.0015&-9.5&28.2\\
$o_{-}^{B\rightarrow 2D^{*}_3}$&0.0019&0.0027&-10.6&30.0&$o_{-}^{B_{s}\rightarrow 2D^{*}_{s3}}$&-0.0019&-0.0026&-10.9&34.6\\
\bottomrule[1pt]
\end{tabular}
\end{table*}
\end{multicols}
\begin{center}
  \includegraphics[width=18cm]{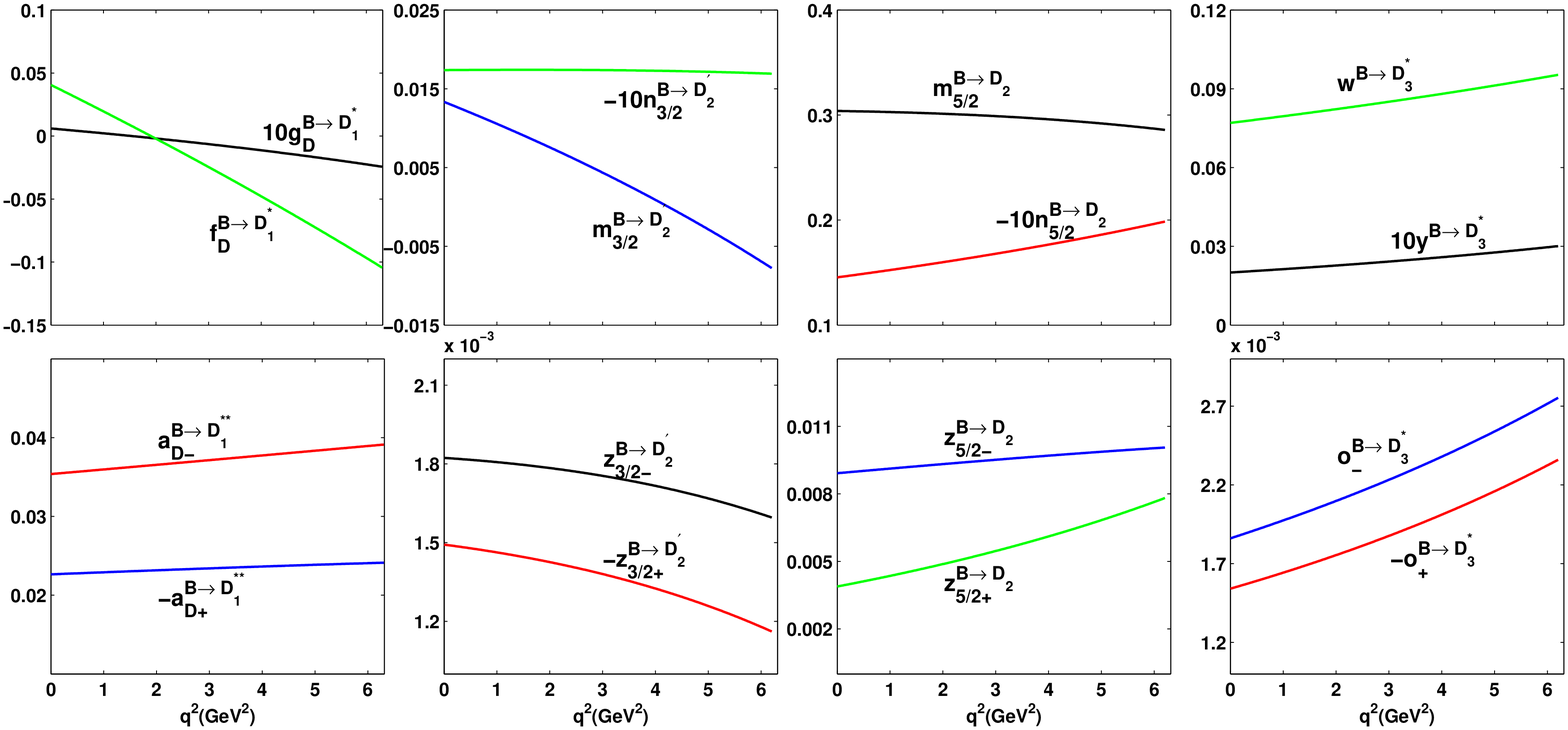}
\figcaption{\label{1D} (color online). The $q^2$ dependence of form factors for $B\rightarrow$ $D^{*}_1$, $D_2$, $D_2^{\prime}$, and $D^{*}_3$ transitions.}
\end{center}
\begin{center}
  \includegraphics[width=18cm]{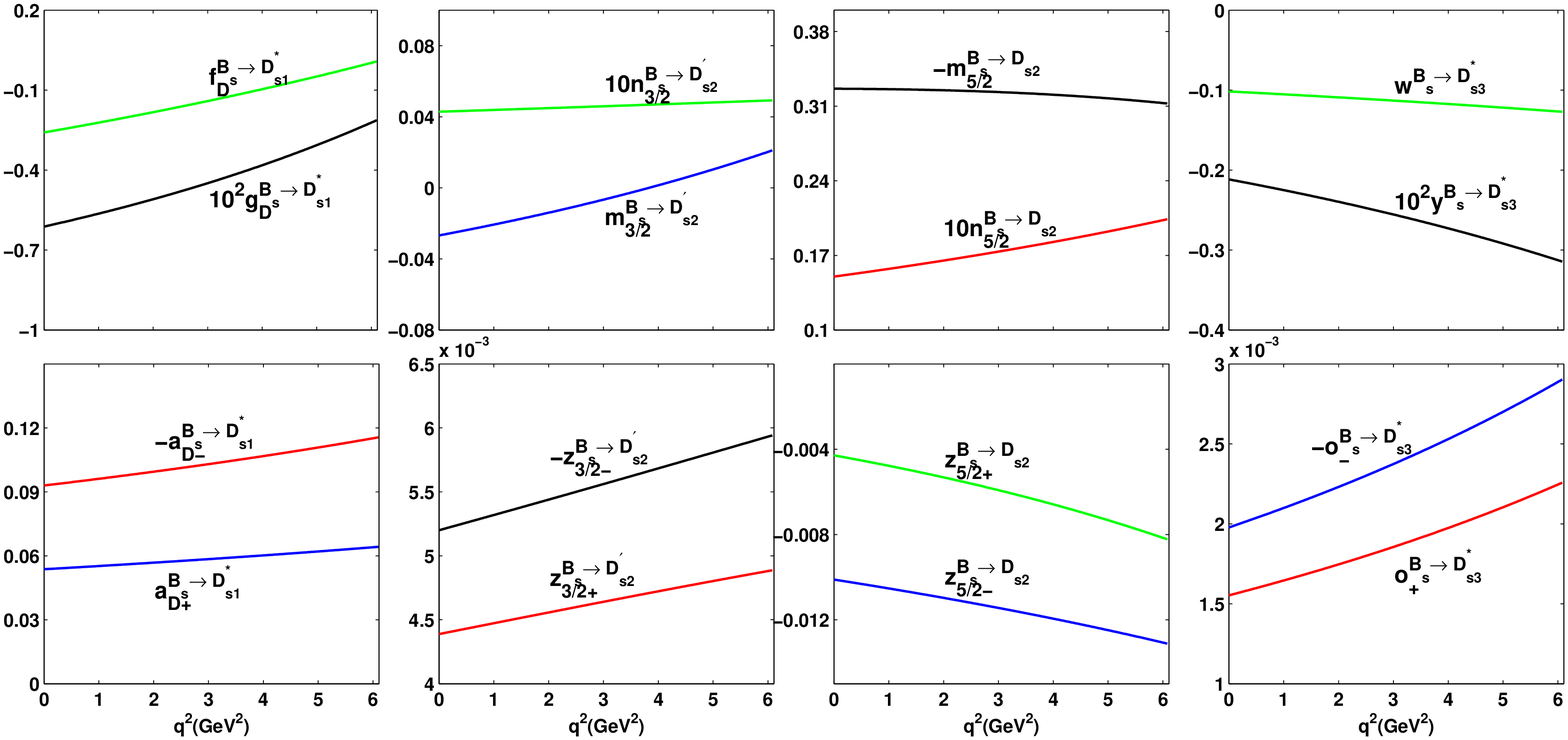}
\figcaption{\label{1Ds} (color online). The $q^2$ dependence of form factors for $B_{s}\rightarrow$ $D^{*}_{s1}$, $D_{s2}$, and $D_{s2}^{\prime}$, $D^{*}_{s3}$ transitions.}
\end{center}
\begin{center}
  \includegraphics[width=18cm]{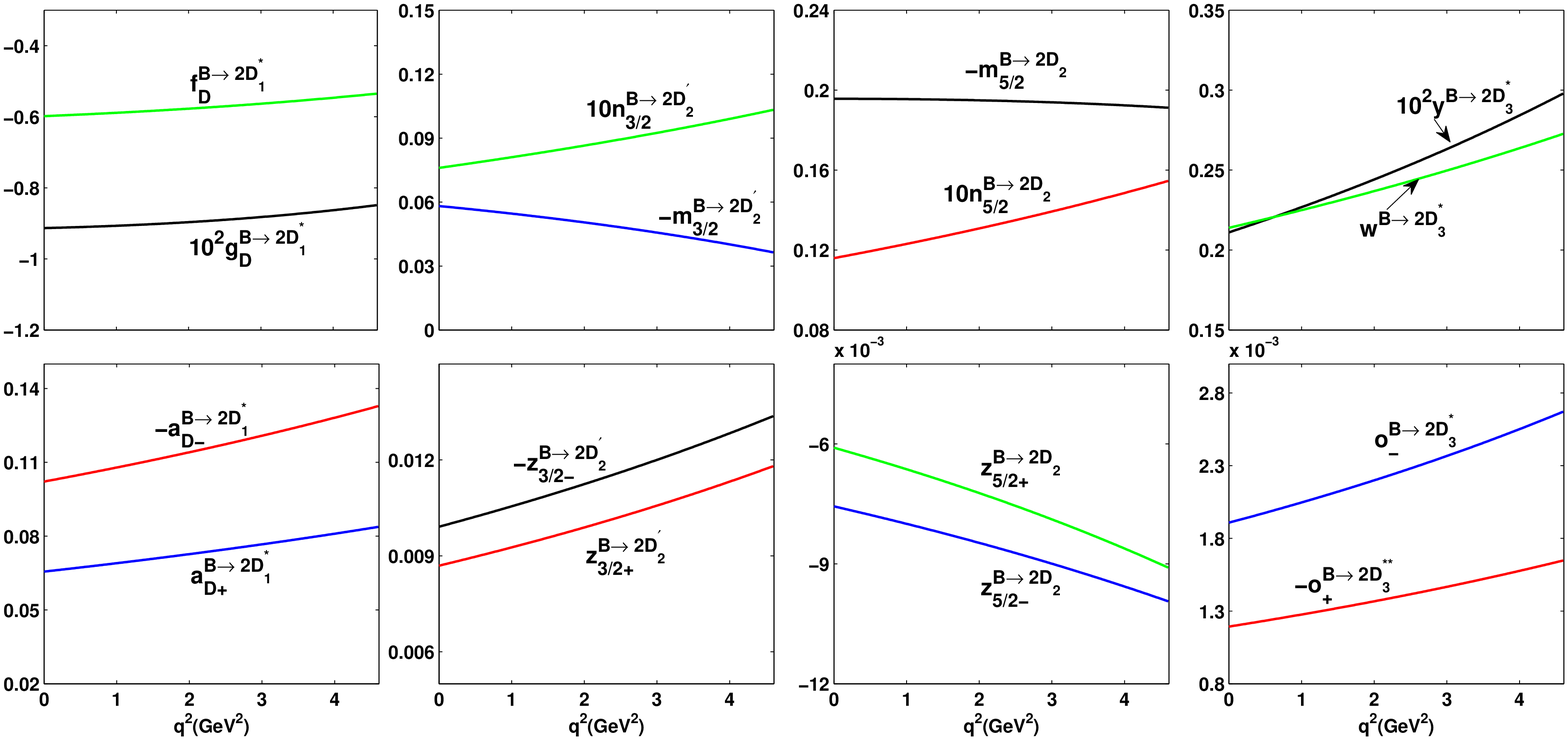}
\figcaption{\label{2D} (color online). The $q^2$ dependence of form factors for  $B\rightarrow$ $2D^{*}_1$, $2D_2$, $2D_2^{\prime}$, and $2D^{*}_3$ transitions.}
\end{center}
\begin{center}
  \includegraphics[width=18cm]{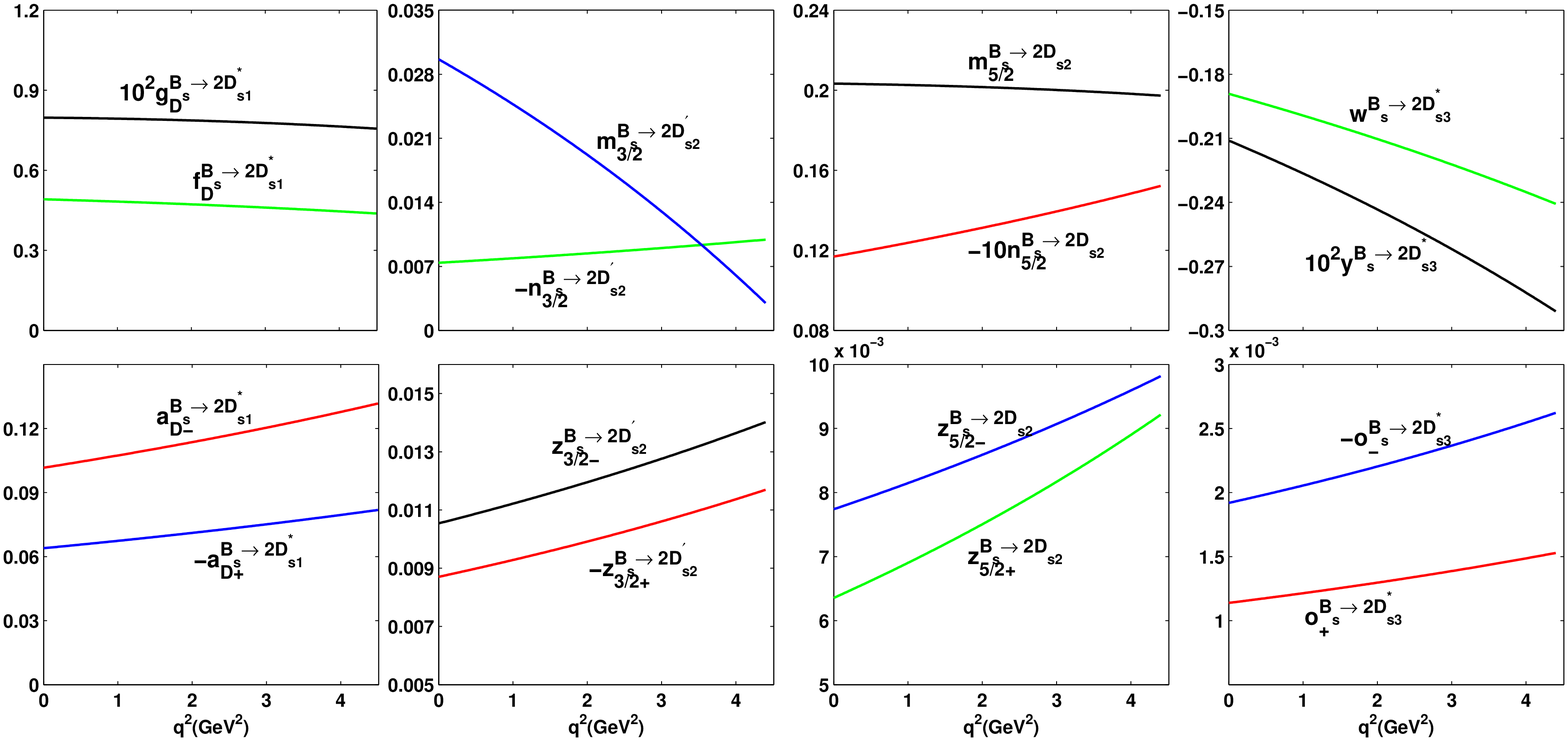}
\figcaption{\label{2Ds} (color online). The $q^2$ dependence of form factors for $B_{s}\rightarrow$ $2D^{*}_{s1}$, $2D_{s2}$, $2D_{s2}^{\prime}$, and $2D^{*}_{s3}$ transitions.}
\end{center}
\begin{multicols}{2}
With the above preparation, we perform the numerical calculation of branching ratios for the $B_{(s)}$ semileptonic decays to $D$-wave $D/D_s$ mesons, which are listed in Table \ref{MGIRESULTS}. The magnitudes of branching ratios presented in Table \ref{MGIRESULTS} are expected to be the typical values for $B_{(s)}$ decay to $D$-wave $D_{(s)}$ via  semi-leptonic processes. 

We also noticed that the production of $D$-wave charmed/charmed-strange via $B_{(s)}$ semi-leptonic decay processes have also been studied by the QCD sum rule \cite{Gan:2009zb,Colangelo:2000jq} and the instantaneous Bethe-Salpeter method \cite{Li:2016efw}, we present the results of these theoretical calculations in Table \ref{QCDSR}. Due to different set of parameters as well as different approaches were introduced, there exist discrepancy between different model calculations, thus the experimental search for the semi-leptonic decays relevant to the production of $D$-wave $D/D_s$ mesons will be an intriguing issue for future experiment.

\begin{table*}[htbp]
\caption{Obtained branching ratios for the $B_{(s)}$ semileptonic decay to the $1D$ and $2D$ states of charmed/charmed-strange mesons. }
\centering
\label{MGIRESULTS}
\begin{tabular}{cccc|cccccccc}
\toprule[1pt]
Decay mode&$\ell=e$&$\ell=\mu$&$\ell=\tau$&Decay mode&$\ell=e$&$\ell=\mu$&$\ell=\tau$\\
\midrule[1pt]
$B\rightarrow D_1^{*}\ell\bar{\nu}_{\ell}$&$6.62\times 10^{-5}$&$6.53\times 10^{-5}$&$1.35\times 10^{-6}$&
$B_{s}\rightarrow D_{s1}^{*}\ell\bar{\nu}_{\ell}$&$1.97\times10^{-4}$&$1.94\times10^{-4}$&$2.37\times 10^{-6}$\\
$B\rightarrow D^{\prime}_2\ell\bar{\nu}_{\ell}$&$1.16\times10^{-6}$&$1.14\times10^{-6}$&$1.21\times10^{-8}$&
$B_{s}\rightarrow D^{\prime}_{s2}\ell\bar{\nu}_{\ell}$&$1.07\times10^{-5}$&$1.05\times10^{-5}$&$1.04\times10^{-7}$\\
$B\rightarrow D_2\ell\bar{\nu}_{\ell}$&$5.39\times10^{-4}$&$5.31\times10^{-4}$&$7.40\times10^{-6}$&
$B_{s}\rightarrow D_{s2}\ell\bar{\nu}_{\ell}$&$5.05\times10^{-4}$&$4.97\times10^{-4}$&$6.62\times10^{-6}$\\
$B\rightarrow D^{*}_3\ell\bar{\nu}_{\ell}$&$7.20\times10^{-5}$&$7.08\times10^{-5}$&$6.50\times10^{-7}$&
$B_{s}\rightarrow D^{*}_{s3}\ell\bar{\nu}_{\ell}$&$1.13\times10^{-4}$&$1.11\times10^{-4}$&$8.78\times10^{-7}$\\
\hline
$B\rightarrow 2D_1^{*}\ell\bar{\nu}_{\ell}$&$7.86\times 10^{-5}$&$7.73\times10^{-5}$&$8.07\times10^{-7}$&
$B_{s}\rightarrow 2D_{s1}^{*}\ell\bar{\nu}_{\ell}$&$6.37\times10^{-5}$&$6.25\times10^{-5}$&$3.76\times10^{-7}$\\
$B\rightarrow 2D^{\prime}_2\ell\bar{\nu}_{\ell}$&$8.77\times10^{-6}$&$8.58\times10^{-6}$&$7.55\times10^{-9}$&
$B_{s}\rightarrow 2D^{\prime}_{s2}\ell\bar{\nu}_{\ell}$&$9.88\times10^{-6}$&$9.65\times10^{-6}$&$4.36\times10^{-9}$\\
$B\rightarrow 2D_2\ell\bar{\nu}_{\ell}$&$8.57\times10^{-5}$&$8.40\times10^{-5}$&$1.04\times10^{-7}$&
$B_{s}\rightarrow 2D_{s2}\ell\bar{\nu}_{\ell}$&$7.02\times10^{-5}$&$6.88\times10^{-5}$&$6.16\times10^{-8}$\\
$B\rightarrow 2D^{*}_3\ell\bar{\nu}_{\ell}$&$1.63\times10^{-4}$&$1.59\times10^{-4}$&$8.64\times10^{-8}$&
$B_{s}\rightarrow 2D^{*}_{s3}\ell\bar{\nu}_{\ell}$&$8.96\times10^{-5}$&$8.74\times10^{-5}$&$3.24\times10^{-8}$\\
\bottomrule[1pt]
\end{tabular}
\end{table*}

\begin{table*}[htbp]
\caption{The branching ratios of semi-leptonic decay of 1$D$ charmed (charmed-strange) meson produced through $B$ meson obtained from theoretical prediction.}
\centering
\label{QCDSR}
\begin{tabular}{ccccccccc}
\toprule[1pt]
{Decay mode}& Ref .\cite{Li:2016efw}& Ref. \cite{Gan:2009zb}&{Decay mode}&Ref .\cite{Li:2016efw}&Ref .\cite{Gan:2014jxa}\\
\midrule[1pt]
$B\rightarrow D_1^*e\bar{\nu}_e$&-&$6.0\times10^{-6}$&$B_s\rightarrow D_{s1}^*e\bar{\nu}_e$&-&$2.85\times10^{-7}$\\
$B\rightarrow D_1^*\mu\bar{\nu}_{\mu}$&-&$6.0\times10^{-6}$&$B_s\rightarrow D_{s1}^*\mu\bar{\nu}_{\mu}$&-&$2.85\times10^{-7}$\\
$B\rightarrow D_1^*\tau\bar{\nu}_{\tau}$&-&-&$B_s\rightarrow D_{s1}^*\tau\bar{\nu}_{\tau}$&-\\
$B\rightarrow D_2^{\prime}e\bar{\nu}_e$&$4.1\times10^{-4}$&$6.0\times10^{-6}$&$B_s\rightarrow D_{s2}^{\prime}e\bar{\nu}_e$&$5.2\times10^{-4}$&$3.4\times10^{-7}$\\
$B\rightarrow D_2^{\prime}\mu\bar{\nu}_{\mu}$&$4.1\times10^{-4}$&$6.0\times10^{-6}$&$B_s\rightarrow D_{s2}^{\prime}\mu\bar{\nu}_{\mu}$&$5.1\times10^{-4}$&$3.4\times10^{-7}$\\
$B\rightarrow D_2^{\prime}\tau\bar{\nu}_{\tau}$&$2.7\times10^{-6}$&-&$B_s\rightarrow D_{s2}^{\prime}\tau\bar{\nu}_{\tau}$&$3.4\times10^{-6}$\\
$B\rightarrow D_2e\bar{\nu}_e$&$1.1\times10^{-3}$&$1.5\times10^{-4}$&$B_s\rightarrow D_{s2}e\bar{\nu}_e$&$1.7\times10^{-3}$&$1.02\times10^{-4}$\\
$B\rightarrow D_2\mu\bar{\nu}_{\mu}$&$1.1\times10^{-3}$&$1.5\times10^{-4}$&$B_s\rightarrow D_{s2}\mu\bar{\nu}_{\mu}$&$1.7\times10^{-3}$&$1.02\times10^{-4}$\\
$B\rightarrow D_2\tau\bar{\nu}_{\tau}$&$8.0\times10^{-6}$&-&$B_s\rightarrow D_{s2}\tau\bar{\nu}_{\tau}$&$1.4\times10^{-5}$&\\
$B\rightarrow D_3^*e\bar{\nu}_e$&$1.0\times10^{-3}$&$2.1\times10^{-4}$&$B_s\rightarrow D_{s3}^*e\bar{\nu}_e$&$1.5\times10^{-3}$&$3.46\times10^{-4}$\\
$B\rightarrow D_3^*\mu\bar{\nu}_{\mu}$&$1.0\times10^{-3}$&$2.1\times10^{-4}$&$B_s\rightarrow D_{s3}^*\mu\bar{\nu}_{\mu}$&$1.4\times10^{-3}$&$3.46\times10^{-4}$\\
$B\rightarrow D_3^*\tau\bar{\nu}_{\tau}$&$5.4\times10^{-6}$&-&$B_s\rightarrow D_{s3}^*\tau\bar{\nu}_{\tau}$&$9.5\times10^{-6}$\\
\midrule[1pt]

\bottomrule[1pt]
\end{tabular}
\end{table*}

\section{Relations in the heavy quark limit}\label{sec4}

For the processes discussed in this work, the corresponding transition amplitudes can be expressed by our deduced light-front form factors. Indeed, in the heavy quark limit, these light front form factors can be related by Isgur-Wise (IW) functions $\xi(\omega)$ and $\zeta(\omega)$. The heavy-quark limit will provide rigorous conditions to our calculations.

The $B_{(s)}\rightarrow D^{*}_{(s)1}$ and $B_{(s)}\rightarrow D^{\prime}_{(s)2}$ transitions in the heavy quark limit are related to the IW function $\xi(\omega)$ defined in Ref. \cite{Gan:2014jxa} by the following equation:
\begin{eqnarray}
\xi(\omega)&=&-\frac{6\sqrt{2}}{\sqrt{3}}\sqrt{m_Bm_D}\frac{1}{\omega-1}g_D(q^2)=\frac{-\sqrt{6}}{(\omega^2-1)\sqrt{m_Bm_D}}f_D(q^2)\nonumber\\
&=&-\frac{\sqrt{6m_B^3}}{3\sqrt{m_D}}(a_{D+}(q^2)+a_{D-}(q^2))
=\frac{\sqrt{6m_Bm_D}}{\omega+2}(a_{D+}(q^2)-a_{D-}(q^2))\nonumber\\
&=&2\sqrt{m_B^3m_D}n_{\frac{3}{2}}(q^2)=-\frac{1}{\omega-1}\sqrt{\frac{m_B}{m_D}}m_{\frac{3}{2}}(q^2)\nonumber\\
&=&\sqrt{m_B^3m_D}(z_{\frac{3}{2}+}(q^2)-z_{\frac{3}{2}-}(q^2)),
\label{xi}
\end{eqnarray}
and obey the additional HQS relation
\begin{eqnarray}
z_{\frac{3}{2}+}(q^2)+z_{\frac{3}{2}-}(q^2)=0,
\label{zHQET}
\end{eqnarray}
where
$\omega=(m_B^2+m_D^2-q^2)/(2m_Bm_D)$.

The $B_{(s)}\rightarrow D_{(s)2}$ and $B_{(s)}\rightarrow D^{*}_{(s)3}$ transition form factors are related to the IW function $\zeta(\omega)$ by the following relations
\end{multicols}
\begin{eqnarray}
\zeta(\omega)&=&-\frac{5\sqrt{5}}{\sqrt{3}}\frac{\sqrt{m_B^3 m_D}}{\omega+1}n_{5\over2}(q^2)=-\frac{5}{2}\sqrt{\frac{5}{3}}\sqrt{\frac{m_B}{m_D}}\frac{1}{1-\omega^2}m_{\frac{5}{2}}(q^2)\nonumber\\
&=&\frac{m_B^5}{m_D}\sqrt{\frac{5}{3}}(z_{\frac{5}{2}+}(q^2)+z_{\frac{5}{2}-}(q^2))=5\sqrt{\frac{5}{3}}\sqrt{m_B^3m_D}\frac{1}{3-2\omega}
(z_{\frac{5}{2}+}(q^2)-z_{\frac{5}{2}-}(q^2))\nonumber\\
&=&2y(q^2)\sqrt{m_B^5m_D}=\sqrt{\frac{m_B^3}{m_D}}\frac{w(q^2)}{\omega+1}=-\sqrt{\frac{m_B^7}{m_D}}(o_{+}(q^2)-o_{-}(q^2)),
\label{zeta}
\end{eqnarray}
\begin{multicols}{2}
and
\begin{eqnarray}
o_{+}(q^2)+o_{-}(q^2)=0.
\label{oHQET}
\end{eqnarray}
These relations are model independent, which are the consequences of heavy quark symmetry. In the following, we will check whether our results obtained from light front quark model can satisfy these relations.

To present our numerical results, we rewrite Eq. (\ref{xi}) and Eq. (\ref{zeta}) as
\end{multicols}
\begin{eqnarray}
\xi=\xi_{(g_D)}=\xi_{(f_D)}=\xi_{(a_{D+}+a_{D-})}=\xi_{(a_{D+}-a_{D-})}
=\xi_{(n_{3/2})}=\xi_{(m_{3/2})}=\xi_{(z_{3/2+}-z_{3/2-})},
\label{xis}
\end{eqnarray}
and
\begin{eqnarray}
\zeta=\zeta_{(n_{5/2})}=\zeta_{(m_{5/2})}=\zeta_{(z_{5/2+}+z_{5/2-})}=\zeta_{(z_{5/2+}-z_{5/2-})}
=\zeta_{(y)}=\zeta_{(w)}=\zeta_{(o_+-o_-)}.
\label{zetas}
\end{eqnarray}
\begin{multicols}{2}
For example, in Eq. (\ref{xis}),
\begin{eqnarray}
\xi_{(g_D)}&\equiv& -\frac{6\sqrt{2}}{\sqrt{3}}\sqrt{m_Bm_D}\frac{1}{\omega-1}g_D(q^2),\\
\xi_{(f_D)}&\equiv& \frac{-\sqrt{6}}{(\omega^2-1)\sqrt{m_Bm_D}}f_D(q^2).
\end{eqnarray}
Other $\xi_{(F)}$ terms are the corresponding expressions given in Eq. (\ref{xi}). The same notation are also applied between Eq. (\ref{zeta}) and Eq. (\ref{zetas}).
By using the above notations, we present the numerical results of IW functions deduced from our $B_{(s)}\rightarrow 1D^*_{(s)1} (2D^*_{(s)1}), 1D^{(\prime)}_{(s)2} (2D^{(\prime)}_{(s)2}), 1D^*_{(s)3} (2D^*_{(s)3})$ light front form factors in Table \ref{HQET}.

In Table \ref{HQET}, we present the calculated IW function values at $q^2=0$ and $q^2=q^2_{max}$. Here, we discuss the results of obtained IW functions in $B\rightarrow 1D^*_{1}, 1D^{(\prime)}_{2}, 1D^*_{3}$ processes, the results for $B\rightarrow 2D^*_{1}, 2D^{(\prime)}_{2}, 2D^*_{3}$ and $B_s\rightarrow 1D^*_{s1}(2D^*_{s1}), 1D^{(\prime)}_{s2}(2D^{(\prime)}_{s2}), 1D^*_{s3}(2D^*_{s3})$ can be analysed in a similar way. From Table \ref{HQET}, we find the $\xi_{(g_D)}$, $\xi_{(f_D)}$, $\xi_{(a_{D+}+a_{D-})}$, $\xi_{n_{3/2}}$, $\xi_{m_{3/2}}$ and $\xi_{(z_{3/2+}-z_{3/2-})}$ are similar to one another, which can approximately meet the requirement in Eq. (\ref{xi}). In addition, the values of $z_{3/2+}$ have an opposite sign to that of $z_{3/2-}$, which is approximately satisfied in Table \ref{HQET}. However, the value of $\xi_{(a_{D+}-a_{D-})}$ is about twice as big as other $\xi$ form factors, which implies a violation of Eq. (\ref{xi}). The transition form factors in $B\rightarrow D_2$ and $B\rightarrow D^*_3$ processes can be related by IW function $\zeta$. We find the numerical results of $\zeta_{n_{5/2}}$, $\zeta_{(y)}$, $\zeta_{(w)}$, and $\zeta_{(o_+-o_-)}$ are close to each other, and the relation in Eq. (\ref{oHQET}) for the form factors $o_+(q^2)$ and $o_{-}(q^2)$ still holds. We also find a discrepancy in the obtained $\zeta_{(m_{5/2})}$ function, which is about 2-3 times bigger than other $\zeta$ functions.

Indeed, from Table \ref{HQET} we noticed that the discrepancies between the results obtained in light front quark model and the expectations from heavy quark limit also exist in $B\rightarrow 2D^*_{1}, 2D^{(\prime)}_{2}, 2D^*_{3}$ and $B_s\rightarrow 1D^*_{s1}(2D^*_{s1}), 1D^{(\prime)}_{s2}(2D^{(\prime)}_{s2}), 1D^*_{s3}(2D^*_{s3})$ transition processes. Here, we need to specify that the relations in Eqs. (\ref{xi})-(\ref{oHQET}) are derived in the heavy quark limit, while we introduced definite masses for $c$ and $b$ quarks in our calculation, this may indicate that the $1/m_Q$ correction would play an essential role for some of form factors.

\begin{table*}[htbp]
\caption{The calculated IW functions deduced from light front quark model. Here, we use the values at $q^2=0$ and $q^2=q^2_{max}$ to present our results.}\label{HQET}
\begin{tabular}{c|cc|c|cc|c|cc|c|cc}
\toprule[1pt]
\midrule[1pt]
&$q^2=0$&$q^2=q_{max}^2$&&$q^2=0$&$q^2=q_{max}^2$&&$q^2=0$&$q^2=q_{max}^2$&&$q^2=0$&$q^2=q_{max}^2$\\
\midrule[1pt]
$\xi_{(g_D)}^{B\rightarrow D^*_1}$&0.051&...&$\zeta_{(n_{5/2})}^{B\rightarrow D^{\prime}_2}$&0.85&1.27&
$\xi_{(g_D)}^{B\rightarrow 2D^*_1}$&1.27&...&$\zeta_{(n_{5/2})}^{B\rightarrow 2D^{\prime}_2}$&0.75&1.05\\
$\xi_{(f_D)}^{B\rightarrow D^*_1}$&0.053&...&$\zeta_{(m_{5/2})}^{B\rightarrow D^{\prime}_2}$&2.81&...&
$\xi_{(f_D)}^{B\rightarrow 2D^*_1}$&1.18&...&$\zeta_{(m_{5/2})}^{B\rightarrow 2D^{\prime}_2}$&2.68&...\\
$\xi_{(a_{D+}+a_{D-})}^{B\rightarrow D^*_1}$&0.076&0.088&$\zeta_{(z_{{5/2+} }+z_{{5/2-}})}^{B\rightarrow D^{\prime}_2}$&0.66&0.98&
$\xi_{(a_{D+}+a_{D-})}^{B\rightarrow 2D^*_1}$&0.21&0.27&$\zeta_{(z_{{5/2+} }+z_{{5/2-}})}^{B\rightarrow 2D^{\prime}_2}$&0.63&0.86\\
$\xi_{(a_{D+}-a_{D-})}^{B\rightarrow D^*_1}$&0.168&0.198&&&&
$\xi_{(a_{D+}-a_{D-})}^{B\rightarrow 2D^*_1}$&0.53&0.72&&&\\
\midrule[1pt]
$\xi_{(n_{3/2})}^{B\rightarrow D^{\prime}_2}$&0.070&0.068&$\zeta_{(y)}^{B\rightarrow D^*_3}$&0.43&0.64&
$\xi_{(n_{3/2})}^{B\rightarrow 2D^{\prime}_2}$&0.33&0.44&$\zeta_{(y)}^{B\rightarrow 2D^*_3}$&0.48&0.67\\
$\xi_{(m_{3/2})}^{B\rightarrow D^{\prime}_2}$&0.086&...&$\zeta_{(w)}^{B\rightarrow D^*_3}$&0.25&0.35&
$\xi_{(m_{3/2})}^{B\rightarrow 2D^{\prime}_2}$&0.53&...&$\zeta_{(w)}^{B\rightarrow 2D^*_3}$&0.69&0.93\\
$\xi_{(z_{{3/2+}}-z_{{3/2-}})}^{B\rightarrow D^{\prime}_2}$&0.067&0.056&$\zeta_{(o_{+}-o_{-})}^{B\rightarrow D^*_3}$&0.68&1.00&
$\xi_{(z_{{3/2+}}-z_{{3/2-}})}^{B\rightarrow 2D^{\prime}_2}$&0.40&0.54&$\zeta_{(o_{+}-o_{-})}^{B\rightarrow 2D^*_3}$&0.59&0.83\\
$z_{{3/2+}}$&0.0015&0.0012&$o_{+}$&-0.0015&-0.0024&$z_{{3/2+}}$&-0.0086&-0.0117&$o_{+}$&-0.0012&-0.0016\\
$z_{{3/2-}}$&-0.0018&-0.0016&$o_{-}$&0.0019&0.0028&
$z_{{3/2-}}$&0.0099&0.0134&$o_{-}$&0.0019&0.0027\\
\midrule[1pt]
$\xi_{(g_D)}^{B_s\rightarrow D^*_{s1}}$&0.58&...&$\zeta_{(n_{5/2})}^{B_s\rightarrow D^{\prime}_{s2}}$&0.92&1.37&
$\xi_{(g_D)}^{B_s\rightarrow 2D^*_{s1}}$&1.23&...&$\zeta_{(n_{5/2})}^{B_s\rightarrow 2D^{\prime}_{s2}}$&0.79&1.08\\
$\xi_{(f_D)}^{B_s\rightarrow D^*_{s1}}$&0.36&...&$\zeta_{(m_{5/2})}^{B_s\rightarrow D^{\prime}_{s2}}$&3.21&...&
$\xi_{(f_D)}^{B_s\rightarrow 2D^*_{s1}}$&1.02&...&$\zeta_{(m_{5/2})}^{B_s\rightarrow 2D^{\prime}_{s2}}$&3.01&...\\
$\xi_{(a_{D+}+a_{D-})}^{B_s\rightarrow D^*_{s1}}$&0.23&0.31&$\zeta_{(z_{{5/2+}}+z_{{5/2-}})}^{B_s\rightarrow D^{\prime}_{s2}}$&0.73&1.07&
$\xi_{(a_{D+}+a_{D-})}^{B_s\rightarrow 2D^*_{s1}}$&0.21&0.28&$\zeta_{(z_{{5/2+} }+z_{{5/2-}})}^{B_s\rightarrow 2D^{\prime}_{s2}}$&0.67&0.90\\
$\xi_{(a_{D+}-a_{D-})}^{B_s\rightarrow D^*_{s1}}$&0.44&0.58&&&&
$\xi_{(a_{D+}-a_{D-})}^{B_s\rightarrow 2D^*_{s1}}$&0.54&0.72&&&\\
\midrule[1pt]
$\xi_{(n_{3/2})}^{B_s\rightarrow D^{\prime}_{s2}}$&0.18&0.21&$\zeta_{(y)}^{B_s\rightarrow D^*_{s3}}$&0.48&0.71&
$\xi_{(n_{3/2})}^{B_s\rightarrow 2D^{\prime}_{s2}}$&0.33&0.44&$\zeta_{(y)}^{B_s\rightarrow 2D^*_{s3}}$&0.51&0.70\\
$\xi_{(m_{3/2})}^{B_s\rightarrow D^{\prime}_{s2}}$&0.18&...&$\zeta_{(w)}^{B_s\rightarrow D^*_{s3}}$&0.34&0.47&
$\xi_{(m_{3/2})}^{B_s\rightarrow 2D^{\prime}_{s2}}$&0.29&...&$\zeta_{(w)}^{B_s\rightarrow 2D^*_{s3}}$&0.61&0.83\\
$\xi_{(z_{{3/2+}}-z_{{3/2-}})}^{B_s\rightarrow D^{\prime}_{s2}}$&0.20&0.23&$\zeta_{(o_{+}-o_{-})}^{B_s\rightarrow D^*_{s3}}$&0.74&1.09&
$\xi_{(z_{{3/2+}}-z_{{3/2-}})}^{B_s\rightarrow 2D^{\prime}_{s2}}$&0.43&0.57&$\zeta_{(o_{+}-o_{-})}^{B_s\rightarrow 2D^*_{s3}}$&0.61&0.82\\
$z_{{3/2+}}$&0.0044&0.0049&$o_{+}$&0.0016&0.0023&$z_{{3/2+}}$&-0.0086&-0.0115&$o_{+}$&0.0011&0.0015\\
$z_{{3/2-}}$&-0.0052&-0.0059&$o_{-}$&-0.0020&-0.0029&
$z_{{3/2-}}$&0.0105&0.0140&$o_{-}$&-0.0019&-0.0026\\
\bottomrule[1pt]\bottomrule[1pt]
\end{tabular}
\end{table*}

\section{Summary}

In the past years, a great progress on observing $D$-wave $D/D_s$ mesons has been made in experiment  \cite{delAmoSanchez:2010vq, Aaij:2013sza,Aaij:2014xza,Aaij:2014baa}. These observations enrich the $D/D_s$ meson families. Although all the possible candidates of $D$-wave $D/D_s$ are produced through the nonleptonic weak decays of $B/B_s$ mesons, in this work we study the possibility of production of the $1D$ and $2D$ $D/D_s$ meson families via the semileptonic decays, among which, though, $2D$ states of $D/D_s$ meson families are still absent in experiment.

For getting the numerical results of the discussed semileptonic decays, we adopt the light-front quark model, which has been extensively applied to
study decay processes including semileptonic decays \cite{Cheng:2003sm,Xu:2014mqa,Li:2010bb,Cheng:2004yj,Ke:2009ed,Ke:2009mn,Lu:2007sg,Wang:2007sxa,Wang:2008ci,Shen:2008zzb,Wang:2008xt,Wang:2009mi,Chen:2009qk,Wang:2010npa}. Our study under the framework of the LFQM shows that the whole deduction relevant to the $D$-wave $D/D_s$ mesons produced via $B/B_s$ mesons is much more complicated than that
relevant to production of the $S$-wave and $P$-wave $D/D_s$ mesons \cite{Cheng:2003sm,Xu:2014mqa,Li:2010bb}. In this paper, we have given detailed derivation of many formulas necessary for obtaining the final semileptonic decay widths. The numerical results obtained for the discussed semileptonic decays have shown that the semileptonic decays of $B/B_s$ mesons are suitable for finding the $D$-wave charmed and charmed-strange mesons. Further, we have presented that our light front form factors can approximately satisfy the requirement from HQS expectations.

Before the present work, there were some theoretical studies on the $B_{(s)}$ semi-leptonic decays to $D$-wave charmed mesons by the QCD sum rule \cite{Gan:2009zb,Colangelo:2000jq} and the instantaneous Bethe-Salpeter method \cite{Li:2016efw}. We have noticed that different theoretical groups have given different results for the $B_{(s)}$ semi-leptonic decays to $D$-wave charmed mesons. Thus, experimental search for our predicted semileptonic decays will provide a crucial test for the theoretical frameworks applied to study on the $B_{(s)}$ semi-leptonic decays.

As indicated by our numerical results, the semileptonic decays of pseudoscalar $B/B_s$ mesons can be an ideal platform to carry out the investigation on $D$-wave charmed and charmed-strange mesons. With the running of LHCb at 13 TeV and forthcoming BelleII, we expect the experimental progress on this issue furthermore.

\acknowledgments{Kan Chen would like to thank Qi Huang and Hao Xu for helpful discussion. We also would like to thank Yu-Ming Wang for the suggestion of form factors adopted in this work.}

\end{multicols}

\vspace{10mm}

\begin{multicols}{2}

\end{multicols}

\begin{small}
\appendix
\renewcommand{\theequation}{\Alph{section}.\arabic{equation}}

\section{Trace expansion and form factors}\label{traceform}

In this Appendix, we present the detailed expansions of $\hat{S}_{\mu\nu}^{{}^3D_1}$, $\hat{S}_{\mu\alpha\beta}^{{}^1D_2}$, $\hat{S}_{\mu\alpha\beta}^{{}^3D_2}$ and $\hat{S}^{{}^3D_3}_{\mu\alpha\beta\nu}$. Form factors associated with these expressions are also collected here.

When integrating over $p_1^{\prime-}$, we need to do the following integrations
\begin{eqnarray}
\label{HT3D1}
\hat{B}^{B_{(s)}(D^{*}_{(s)1})}_{\mu}&=&\frac{N_c}{16\pi^3}\int_0^{1}dx\int d^2p^{\prime}_{\bot}\frac{h^{\prime}_0h^{\prime\prime}_{{}^3D_1}}{\left(1-x\right)N^{\prime}_1N^{\prime\prime}_1}\hat{S}^{{}^3D_1}_{\mu\nu}\epsilon^{*\prime\prime\nu},
\nonumber\\
\\
\hat{B}^{B_{(s)}(D^{*}_{(s)2})}_{\mu}&=&\frac{N_c}{16\pi^3}\int_0^{1}dx\int d^2p^{\prime}_{\bot}\frac{h^{\prime}_0h^{\prime\prime}_{{}^1D_2}}{\left(1-x\right)N^{\prime}_1N^{\prime\prime}_1}
\hat{S}^{{}^1D_2}_{\mu\alpha\beta}\epsilon^{*\prime\prime\alpha\beta}.
\nonumber\\
\\
\hat{B}^{B_{(s)}(D_{(s)2}^{*\prime})}_{\mu}&=&\frac{N_c}{16\pi^3}\int_0^{1}dx\int d^2p^{\prime}_{\bot}\frac{h^{\prime}_0h^{\prime\prime}_{{}^3D_2}}{\left(1-x\right)N^{\prime}_1N^{\prime\prime}_1}
\hat{S}^{{}^3D_2}_{\mu\alpha\beta}\epsilon^{*\prime\prime\alpha\beta},
\nonumber\\
\\
\hat{B}^{B_{(s)}(D^{*}_{(s)3})}_{\mu}&=&\frac{N_c}{16\pi^3}\int_0^{1}dx\int d^2p'_{\bot}\frac{h'_0h'_{{}^3D_3}}{(1-x)N'_1N''_1}\hat{S}^{{}^3D_3}_{\mu\alpha\beta\nu}\epsilon''^{*\alpha\beta\nu},
\nonumber\\
\end{eqnarray}
where the trace expansions of  $\hat{S}_{\mu\nu}^{{}^3D_1}$, $\hat{S}_{\mu\alpha\beta}^{{}^1D_2}$, $\hat{S}_{\mu\alpha\beta}^{{}^3D_2}$ and $\hat{S}^{{}^3D_3}_{\mu\alpha\beta\nu}$ are
\allowdisplaybreaks

\begin{eqnarray}
\hat{S}_{\mu\nu}^{{}^3D_1}&=&{\rm Tr}\left\{\left[\gamma_{\nu}-\frac{1}{\omega^{\prime\prime}_{{}^3D_1}}\left(p^{\prime\prime}_1-p_2\right)_{\nu}\right]
\left(\slashed{p}^{\prime\prime}_{1}+m^{\prime\prime}_1\right)\gamma_{\mu}\left(1-\gamma_5\right)
\left(\slashed{p}^{\prime}_1+m^{\prime}_1\right)\gamma_5\left(-\slashed{p}_2+m_2\right)\right\}\nonumber\\
&=&-2i\epsilon_{\mu\nu\alpha\beta}\left[p_1^{\prime\alpha}P^{\beta}
\left(m_1^{\prime\prime}-m_1^{\prime}\right)+p_1^{\prime\alpha}q^{\beta}
\left(m_1^{\prime\prime}+m_1^{\prime}-2m_2\right)+q^{\alpha}P^{\beta}m_1^{\prime}\right]
+\frac{1}{\omega^{\prime\prime}_{{}^3D_1}}\left(4p_{1\nu}^{\prime}-3q_{\nu}-P_{\nu}\right)\nonumber\\&&\times
i\epsilon_{\mu\alpha\beta\rho}p_1^{\prime\alpha}q^{\beta}P^{\rho}+2g_{\mu\nu}
\left[m_2\left(q^2-N_1^{\prime}-N_1^{\prime\prime}-m_1^{\prime2}-m_1^{\prime\prime2}\right)-
m_1^{\prime}\left(M^{\prime\prime2}-N_1^{\prime\prime}-N_2-m_1^{\prime\prime2}
-m_2^{2}\right)\nonumber\right.\\&&\left.-m_1^{\prime\prime}\left(M^{\prime2}-N_1^{\prime}-N_2-m_1^{\prime2}
-m_2^{2}\right)-2m_1^{\prime}m_1^{\prime\prime}m_2\right]+8p_{1\mu}^{\prime}p_{1\nu}^{\prime}
\left(m_2-m_1^{\prime}\right)-2\left(P_{\mu}q_{\nu}+q_{\mu}P_{\nu}+2q_{\mu}q_{\nu}\right)m_1^{\prime}\nonumber\\&&
+2p_{1\mu}^{\prime}P_{\nu}\left(m_1^{\prime}-m_1^{\prime\prime}\right)+2p_{1\mu}^{\prime}
q_{\nu}\left(3m_1^{\prime}-m_1^{\prime\prime}-2m_2\right)+2P_{\mu}p_{1\nu}^{\prime}
\left(m_1^{\prime}+m_1^{\prime\prime}\right)+2q_{\mu}p_{1\nu}^{\prime}\left(3m_1^{\prime}
+m_1^{\prime\prime}-2m_2\right)\nonumber\\&&+\frac{1}{2\omega^{\prime\prime}_{{}^3D_1}}\left(4p_{1\nu}^{\prime}-3q_{\nu}-P_{\nu}\right)
\left\{2p_{1\mu}^{\prime}\left[M^{\prime2}+M^{\prime\prime2}-q^2-2N_2+
2\left(m_1^{\prime}-m_2\right)\left(m_1^{\prime\prime}+m_2\right)\right]\nonumber\right.\\&&\left.+q_{\mu}\left[q^2-2M^{\prime2}
+N_1^{\prime}-N_1^{\prime\prime}+2N_2-\left(m_1^{\prime}+m_1^{\prime\prime}\right)^2+
2\left(m_1^{\prime}-m_2\right)^2\right]+P_{\mu}\left[q^2-N_1^{\prime}-
N_1^{\prime\prime}-\left(m_1^{\prime}+m_1^{\prime\prime}\right)^2\right]\right\},
\\
\hat{S}_{\mu\alpha\beta}^{{}^1D_2}&=&{\rm Tr}\left\{\left[\gamma_5 \frac{\left(p_2-p_1^{\prime\prime}\right)_{\alpha}}{2}\frac{\left(p_2-p_1^{\prime\prime}\right)_{\beta}}{2}\right]\left(\slashed{p}^{\prime\prime}_{1}
+m^{\prime\prime}_1\right)\gamma_{\mu}\left(1-\gamma_5\right)
\left(\slashed{p}^{\prime}_1+m^{\prime}_1\right)\gamma_5\left(-\slashed{p}_2+m_2\right)\right\}\nonumber\\
&=&-\frac{1}{8}i\epsilon_{\mu\nu\sigma\delta}\left(P_{\alpha}-4p^{\prime}_{1\alpha}+3q_{\alpha}\right)
\left(P_{\beta}-4p^{\prime}_{1\beta}+3q_{\beta}\right)P^{\nu}p^{\prime\sigma}_1q^{\delta}-
\left\{\frac{1}{16}\left(P_{\alpha}-4p^{\prime}_{1\alpha}+3q_{\alpha}\right)\nonumber\right.\\&&\left.\times\left(P_{\beta}-4p^{\prime}_{1\beta}+3q_{\beta}\right)
\left[-q_{\mu}\left(2m_2^2-4m_2m^{\prime}_1-m^{\prime\prime2}_1+2m^{\prime}_1m^{\prime\prime}_1+m^{\prime2}_1-2M^{\prime2}+2N_2-N^{\prime\prime}_1
+N^{\prime}_1+q^2\right)\nonumber\right.\right.\\&&\left.\left.+2p^{\prime}_{1\mu}\left[2\left(m_2-m^{\prime\prime}_1\right)\left(m_2-m^{\prime}_1\right)-M^{\prime\prime2}
-M^{\prime2}+2N_2+q^2\right]+P_{\mu}
\left[\left(m^{\prime\prime}_1-m^{\prime}_1\right)^2+N^{\prime\prime}_1+N^{\prime}_1-q^2\right]\right]\right\},
\\
\hat{S}_{\mu\alpha\beta}^{{}^3D_2}&=&{\rm Tr}\left\{\gamma_5
\left[\frac{1}{\omega_{{}^3D_2}^{a\prime\prime}}\gamma_{\alpha}\gamma_{\beta}+\frac{1}{\omega_{{}^3D_2}^{b\prime\prime}}
\gamma_{\alpha}\frac{\left(p_2-p^{\prime\prime}_1\right)_{\beta}}{2}+\frac{1}{\omega_{{}^3D_2}^{c\prime\prime}}\frac{\left(p_2-p^{\prime\prime}_1\right)_{\alpha}}{2}
\frac{\left(p_2-p^{\prime\prime}_1\right)_{\beta}}{2}\right]\nonumber\right.\\&&\left.
\times\left(\slashed{p}^{\prime\prime}_{1}+m^{\prime\prime}_1\right)\gamma_{\mu}\left(1-\gamma_5\right)
\left(\slashed{p}^{\prime}_1+m^{\prime}_1\right)\gamma_5\left(-\slashed{p}_2+m_2\right)\right\}\nonumber\\
&=&-i\frac{1}{2\omega^{b\prime\prime}_{{}^3D_2}}\epsilon_{\alpha\mu\sigma\delta}
\left(P_{\beta}-4p^{\prime}_{1\beta}+3q_{\beta}\right)\left\{\left(m^{\prime}_1+m^{\prime\prime}_1\right)P^{\sigma}p^{\prime\delta}_1-
q^{\delta}\left[p^{\prime\sigma}_1(2m_2+m^{\prime\prime}_1-m^{\prime}_1)+m^{\prime}_1P^{\sigma}\right]\right\}\nonumber\\&&
-\frac{i}{8\omega^{c\prime\prime}_{{}^3D_2}}\epsilon_{\mu\lambda\sigma\delta}
P^{\lambda}p^{\prime\sigma}_1q^{\delta}\left(P_{\alpha}-4p^{\prime}_{1\alpha}+3q_{\alpha}\right)
\left(P_{\beta}-4p^{\prime}_{1\beta}+3q_{\beta}\right)-\frac{1}{2\omega^{b\prime\prime}_{{}^3D_2}}\left(P_{\beta}-4p^{\prime}_{1\beta}+3q_{\beta}\right)\nonumber\\&&\times
\left\{g_{\alpha\mu}\left[m_2^2\left(m^{\prime\prime}_1-m^{\prime}_1\right)+m_2\left(\left(m^{\prime\prime}_1-m^{\prime}_1\right)^2+N^{\prime\prime}_1
+N^{\prime}_1-q^2\right)-m^{\prime\prime2}_1m^{\prime}_1
+m^{\prime\prime}_1\left(m^{\prime2}_1-M^{\prime2}+Z_2+N^{\prime}_1\right)\nonumber\right.\right.\\&&\left.\left.
+m^{\prime}_1\left(M^{\prime\prime2}-Z_2-N^{\prime\prime}_1\right)\right]+p^{\prime}_{1\alpha}\left[q_{\mu}\left(2m_2+m^{\prime\prime}_1-3m^{\prime}_1\right)
+4\left(m^{\prime}_1-m_2\right)p^{\prime}_{1\mu}
+\left(m^{\prime\prime}_1-m^{\prime}_1\right)P_{\mu}\right]\nonumber\right.\\&&\left.+2m_2p^{\prime}_{1\mu}q_{\alpha}-m^{\prime\prime}_1
\left(P_{\alpha}p^{\prime}_{1\mu}
+p^{\prime}_{1\mu}q_{\alpha}\right)+m^{\prime}_1\left[-P_{\alpha}p^{\prime}_{1\mu}+P_{\mu}q_{\alpha}
+q_{\mu}(P_{\alpha}+2q_{\alpha})-3p^{\prime}_{1\mu}q_{\alpha}\right]\right\}\nonumber\\&&
-\frac{1}{16\omega^{c\prime\prime}_{{}^3D_2}}\left(P_{\alpha}-4p^{\prime}_{1\alpha}+3q_{\alpha}\right)
\left(P_{\beta}-4p^{\prime}_{1\beta}+3q_{\beta}\right)\left\{-q_{\mu}\left(2m_2^2-4m_2m^{\prime}_1-m^{\prime\prime2}_1+
2m^{\prime\prime}_1m^{\prime}_1+m^{\prime2}_1\nonumber\right.\right.\\&&\left.-2M^{\prime2}+2Z_2-N^{\prime\prime}_1+
N^{\prime}_1+q^2\right)+2p^{\prime}_{1\mu}
\left[2(m_2-m^{\prime\prime}_1)(m_2-m^{\prime}_1)
-M^{\prime\prime2}-M^{\prime2}+2Z_2+q^2\right]\nonumber\\&&\left.+P_{\mu}\left[\left(m^{\prime\prime}_1-m^{\prime}_1\right)^2
+N^{\prime\prime}_1+N^{\prime}_1-q^2\right]\right\},
\\
\hat{S}^{{}^3D_3}_{\mu\alpha\beta\nu}&=&
{\rm Tr}\left\{\left[\frac{\left(p_2-p^{\prime\prime}_1\right)_{\alpha}}{2}\frac{\left(p_2-p^{\prime\prime}_1\right)_{\beta}}{2}
\left(\gamma_\nu+\frac{\left(p_2-p^{\prime\prime}_1\right)_{\nu}}{\omega^{\prime\prime}_{{}^3D_3}}\right)
+\frac{\left(p_2-p^{\prime\prime}_1\right)_{\alpha}}{2}\frac{\left(p_2-p^{\prime\prime}_1\right)_{\nu}}{2}(\gamma_\beta+
\frac{\left(p_2-p^{\prime\prime}_1\right)_{\beta}}{\omega^{\prime\prime}_{{}^3D_3}})\right.\right.\nonumber\\&&\left.\left.
+\frac{\left(p_2-p^{\prime\prime}_1\right)_{\nu}}{2}\frac{\left(p_2-p^{\prime\prime}_1\right)_{\beta}}{2}\left(\gamma_\alpha
+\frac{\left(p_2-p^{\prime\prime}_1\right)_{\alpha}}{\omega^{\prime\prime}_{{}^3D_3}}\right)\right]
\left(\slashed{p}^{\prime\prime}_{1}+m^{\prime\prime}_1\right)\gamma_{\mu}\left(1-\gamma_5\right)
\left(\slashed{p}^{\prime}_1+m^{\prime}_1\right)\gamma_5\left(-\slashed{p}_2+m_2\right)\right\}\nonumber\\
&=&-\frac{3i}{8}\epsilon
_{\mu\nu\sigma\delta}\left(P_{\alpha}-4p^{\prime}_{1\alpha}+3q_{\alpha}\right)\left(P_{\beta}-4p^{\prime}_{1\beta}+3q_{\beta}\right)
\left\{q^{\delta}\left[m^{\prime}_1P^{\sigma}-p^{\prime\sigma}_{1}(-2m_2+m^{\prime\prime}_1+m^{\prime}_1)\right]+\left(m^{\prime\prime}_1-m^{\prime}_1\right)
P^{\sigma}p^{\prime\delta}
_{1}\right\}\nonumber\\&&+\frac{3i}{16\omega^{\prime\prime}_{{}^3D_3}}\epsilon_{\mu\sigma\delta\lambda}P^{\sigma}p^{\prime\delta}_1
q^{\lambda}\left(P_{\alpha}-4p^{\prime}_{1\alpha}+3q_{\alpha}\right)\left(P_{\beta}-4p^{\prime}_{1\beta}+3q_{\beta}\right)\left(P_{\nu}-4p^{\prime}_{1\nu}+3q_{\nu}\right)
+\frac{3}{8}\left(P_\alpha-4p^{\prime}_{1\alpha}+3q_{\alpha}\right)\nonumber\\&&\times\left(P_{\beta}-4p^{\prime}_{1\beta}+3q_{\beta}\right)
\left\{g_{\mu\nu}\left[m_2^2\left(m^{\prime\prime}_1+m^{\prime}_1\right)-m_2\left[\left(m^{\prime\prime}_1+m^{\prime}_1\right)^2+N^{\prime\prime}_1+N^{\prime}_1-q^2\right]
+m^{\prime\prime2}_1m^{\prime}_1+m^{\prime\prime}_1\left(m^{\prime2}_1-M^{\prime2}+Z_2+N^{\prime}_1\right)\right.\right.\nonumber\\&&\left.\left.+
m^{\prime}_1\left(-M^{\prime\prime2}+Z_2+N^{\prime\prime}_1\right)\right]+p^{\prime}_{1\mu}\left[-q_{\nu}(2m_2+m^{\prime\prime}_1-3m^{\prime}_1)+
4(m_2-m^{\prime}_1)p^{\prime}_{1\nu}+
(m^{\prime}_1-m^{\prime\prime}_1)P_{\nu}\right]-2m_2p^{\prime}_{1\nu}q_{\mu}+m^{\prime\prime}_1\left(P_{\mu}p^{\prime}_{1\nu}\nonumber\right.\right.\\&&
\left.\left.
+p^{\prime}_{1\nu}q_{\mu}\right)+m^{\prime}_1\left(P_{\mu}p^{\prime}_{1\nu}-P_{\nu}q_{\mu}-q_{\nu}P_{\mu}-2q_{\nu}q_{\mu}+3p^{\prime}_{1\nu}q_{\mu}\right)\right\}+
\frac{3}{32\omega^{\prime\prime}_{{}^3D_3}}\left(P_{\alpha}-4p^{\prime}_{1\alpha}+3q_{\alpha}\right)\left(P_{\beta}-4p^{\prime}_{1\beta}+3q_{\beta}\right)\nonumber\\&&\times
\left(P_{\nu}-4p^{\prime}_{1\nu}+3q_{\nu}\right)\left\{-q_{\mu}\left[2m_2^2-2m^{\prime}_1\left(2m_2+m^{\prime\prime}_1\right)-m^{\prime\prime2}_1+m^{\prime2}_1-2M^{\prime2}+2Z_2
-N^{\prime\prime}_1+N^{\prime}_1+q^2\right]\nonumber\right.\\&&\left.+
2p^{\prime}_{1\mu}\left[2(m_2+m^{\prime\prime}_1)\left(m_2-m^{\prime}_1\right)-M^{\prime\prime2}-M^{\prime2}+2Z_2+q^2\right]+P_{\mu}\left[\left(m^{\prime\prime}_1+m^{\prime}_1\right)^2
+N^{\prime\prime}_1+N^{\prime}_1-q^2\right]\right\}.
\end{eqnarray}

The corresponding form factors for a ${}^1D_2$ state are
\allowdisplaybreaks
\begin{eqnarray}
n(q^2)&=&\frac{N_c}{16\pi^3}\int dx_2 d^2
p^{\prime}_{\bot}\frac{4h^{\prime}_Ph^{\prime\prime}_{{}^1D_2}}{(1-x)\hat{N}_1^{\prime}\hat{N}_1^{\prime}}\left\{
A_1^{(2)}-A_1^{(3)}-A_2^{(3)}\right\},
\\
m(q^2)&=&\frac{N_c}{16\pi^3}\int dx_2 d^2
p^{\prime}_{\bot}\frac{4h^{\prime}_P h^{\prime\prime}_{{}^1D_2}}{(1-x)\hat{N}_1^{\prime}\hat{N}_1^{\prime\prime}}\Bigg\{\left(A_1^{(2)}-A_1^{(3)}-A_{2}^{(3)}\right)
\left[2\left(m_2-m^{\prime\prime}_1\right)\left(m_2-m_1^{\prime}\right)-M^{\prime\prime2}-M^{\prime2}+q^2\right]+
2A_1^{(2)}Z_2\nonumber\\&&-2\left(A_2^{(3)}Z_2+\frac{M^{\prime2}-M^{\prime\prime2}}{3q^2}(A_1^{(2)})^2\right)
-2\left(A_1^{(3)}Z_2-A_1^{(4)}\right)\Bigg\},
\\
z_{+}(q^2)&=&\frac{N_c}{16\pi^3}\int dx_2 d^2 p^{\prime}_{\bot}\frac{h^{\prime}_{P}h^{\prime\prime}_{{}^1D_2}}{(1-x)\hat{N}_1^{\prime}\hat{N}_1^{\prime\prime}}\bigg\{
\left(-4A_1^{(1)}-2A_2^{(1)}+5A_2^{(2)}+6A_3^{(2)}-2A_3^{(3)}+A_4^{(2)}-4A_4^{(3)}-2A_5^{(3)}+1\right)
\left(2m_1^{\prime}m_1^{\prime\prime}+q^2\right)\nonumber\\&&+\left(2A_1^{(1)}+2A_2^{(1)}-A_2^{(2)}-2A_3^{(2)}
-A_4^{(2)}-1\right)\left[m_1^{\prime\prime2}+m_1^{\prime2}+x(M^{\prime2}-M_0^{\prime2})+
x\left(M^{\prime\prime2}-M_0^{\prime\prime2}\right)\right]+\Big[-4m_2^2+2M^{\prime\prime2}
+2M^{\prime2}\nonumber\\&&+4m_2\left(m_1^{\prime\prime}+m_1^{\prime}\right)\Big]\left(A_1^{(1)}-2A_2^{(2)}-2A_3^{2}
+A_3^{(3)}+2A_4^{(3)}+A_5^{(3)}\right)-4\Bigg[\big(A_1^{(1)}Z_2-A_1^{(2)}\big)-2\left(A_2^{(2)}Z_2-2A_1^{(3)}\right)\nonumber\\&&-2\left(A_3^{(2)}Z_2
+A_1^{(3)}\frac{q\cdot P}{q^2}-A_2^{(3)}\right)+\left(A_3^{(3)}Z_2-2A_2^{(2)}A_1^{(2)}-A_2^{(4)}\right)+2\left(A_4^{(3)}Z_2
+A_2^{(2)}A_1^{(2)}\frac{q\cdot P}{q^2}-A_1^{(1)}A_2^{(3)}-A_3^{(4)}\right)\nonumber\\&&+\left(A_5^{(3)}Z_2+2\frac{q\cdot P}{q^2}A_1^{(1)}A_2^{(3)}-A_4^{(4)}\right)\Bigg]\bigg\},
\\
z_{-}(q^2)&=&\frac{N_c}{16\pi^3}\int dx_2 d^2
p^{\prime}_{\bot}\frac{h^{\prime}_{P}h^{\prime\prime}_{{}^1D_2}}{(1-x)\hat{N}_1^{\prime}\hat{N}_1^{\prime\prime}}\Bigg\{
\left(-2A_1^{(1)}-4A_2^{(1)}+A_2^{(2)}+6A_3^{(2)}+5A_4^{(2)}-2A_4^{(3)}-4A_5^{(3)}
-2A_6^{(3)}+1\right)\left(2m_2^2+2m_1^{\prime\prime}m_1^{\prime}+q^2\right)\nonumber\\&&+\left(2A_1^{(1)}+
3A_2^{(1)}-A_2^{(2)}-4A_3^{(2)}-3A_4^{(2)}+A_4^{(3)}+2A_5^{(3)}+A_6^{(3)}-1\right)
\left(4m_2m_1^{\prime}+2M^{\prime2}\right)+\Big(2A_1^{(1)}+2A_2^{(1)}-A_2^{(2)}-2A_3^{(2)}
\nonumber\\&&-A_4^{(2)}-1\Big)\left[m_1^{\prime\prime2}-m_1^{\prime2}+x\left(M^{\prime\prime2}-M_0^{\prime\prime2}\right)-
x\left(M^{\prime2}-M_0^{\prime2}\right)\right]+\left(A_2^{(1)}-2A_3^{(2)}-2A_4^{(2)}+A_4^{(3)}+2A_5^{(3)}+A_6^{(3)}\right)
\big(2M^{\prime\prime2}\nonumber\\&&+4m_2m^{\prime\prime}\big)-8\left(A_2^{(1)}Z_2+
\frac{M^{\prime2}-M^{\prime\prime2}}{q^2}A_1^{(2)}\right)+
10\left(A_4^{(2)}Z_2+2\frac{M^{\prime2}-M^{\prime\prime2}}{q^2}A_2^{(1)}A_1^{(2)}\right)+2Z_2
-4\Bigg[A_6^{(3)}Z_2+3\frac{M^{\prime2}-M^{\prime\prime2}}{q^2}\nonumber\\&&\times\bigg[A_2^{(1)}A_2^{(3)}-
\frac{1}{3q^2}(A_1^{(2)})^2\bigg]\Bigg]+2\Bigg[-2\left(A_1^{(1)}Z_2-A_1^{(2)}\right)+\left(A_2^{(2)}Z_2-2A_1^{(3)}\right)+6\left(A_3^{(2)}Z_2+A_1^{(3)}\frac{q\cdot P}{q^2}-A_2^{(3)}\right)-2\big(A_4^{(3)}Z_2\nonumber\\&&+A_2^{(2)}A_1^{(2)}\frac{q\cdot P}{q^2}-A_1^{(1)}A_2^{(3)}-A_3^{(4)}\big)
-4\left(A_5^{(3)}Z_2+2\frac{q\cdot P}{q^2}A_1^{(1)}A_2^{(3)}-A_4^{(4)}\right)\Bigg]\Bigg\}.
\end{eqnarray}

The corresponding form factors for a ${}^3D_2$ state are
\begin{eqnarray}
n^{\prime}(q^2)&=&\frac{N_c}{16\pi^3}\int dx_2 d^2p^{\prime}_{\bot}
\frac{2h^{\prime}_{P}h^{\prime\prime}_{{}^3D_2}}{(1-x)\hat{N}_1^{\prime}\hat{N}_1^{\prime\prime}}\Bigg\{
\frac{1}{\omega^{b\prime\prime}_{{}^3D_2}}\bigg[A_1^{(1)}\left(2m_2+m_1^{\prime\prime}-2m_1^{\prime}\right)
+m_1^{\prime}\left(-2A_2^{(1)}+A_2^{(2)}+2A_3^{(2)}+A_4^{(2)}+1\right)\nonumber\\&&-m_1^{\prime\prime}
\left(A_2^{(1)}+A_2^{(2)}-A_4^{(2)}\right)-m_2\left(2A_2^{(2)}+2A_3^{(2)}\right)\bigg]+
\frac{2}{\omega^{c\prime\prime}_{{}^3D_2}}\left(A_1^{(2)}-A_1^{(3)}-A_2^{(3)}\right)\Bigg\},\\
m^{\prime}(q^2)&=&\frac{N_c}{16\pi^3}\int dx_2 d^2p^{\prime}_{\bot}
\frac{2h^{\prime}_{P}h^{\prime\prime}_{{}^3D_2}}{(1-x)\hat{N}_1^{\prime}\hat{N}_1^{\prime\prime}}\Bigg\{
\frac{1}{\omega^{b\prime\prime}_{{}^3D_2}}\bigg[\left[\left(m_2+m_1^{\prime\prime}\right)\left(m_2-m_1^{\prime}\right)
\left(m_1^{\prime}-m_1^{\prime\prime}\right)+m_2 q^2\right]\left(1-A_2^{(1)}-A_1^{(1)}\right)+A_1^{(2)}\Big[6m_2\nonumber\\&&
-2\left(m_1^{\prime\prime}+4m_1^{\prime}\right)\Big]+
8\left(A_1^{(3)}+A_2^{(3)}\right)\left(m_1^{\prime}-m_2\right)+
\left(Z_2-A_2^{(1)}Z_2-\frac{M^{\prime2}-M^{\prime\prime2}}{q^2}A_1^{(2)}\right)
\left(m_1^{\prime}-m_1^{\prime\prime}\right)+\Big(M^{\prime\prime2}m_1^{\prime}-M^{\prime2}
m_1^{\prime\prime}\nonumber\\&&+x\left(M^{\prime2}-M_0^{\prime2}\right)\left(m_2+m_1^{\prime\prime}\right)
+x\left(M^{\prime\prime2}-M_0^{\prime\prime2}\right)\left(m_2-m_1^{\prime}\right)\Big)
\left(A_1^{(1)}+A_2^{(1)}-1\right)+(m_2-m^{\prime}_1)(A_1^{(1)}Z_2-A_1^{(2)})\bigg]\nonumber\\&&+\frac{2}{\omega^{c\prime\prime}_{{}^3D_2}}\Bigg[\left(A_1^{(2)}-A_1^{(3)}
-A_2^{(3)}\right)\times\left[2(m_2-m_1^{\prime\prime})(m_2-m_1^{\prime})-M^{\prime\prime2}
-M^{\prime2}+q^2\right]+2A_1^{(2)}Z_2-2\left(A_2^{(3)}Z_2+
\frac{M^{\prime2}-M^{\prime\prime2}}{3q^2}\left(A_1^{(2)}\right)^2\right)
\nonumber\\&&-2\left(A_1^{(3)}Z_2-A_1^{(4)}\right)\Bigg]\Bigg\},\\
z_{+}^{\prime}(q^2)&=&\frac{N_c}{16\pi^3}\int dx_2 d^2p^{\prime}_{\bot}\frac{h^{\prime}_{P}h^{\prime\prime}_{{}^3D_2}}{(1-x)\hat{N}_1^{\prime}\hat{N}_1^{\prime\prime}}
\Bigg\{\frac{2}{\omega^{b\prime\prime}_{{}^3D_2}}\bigg[m_1^{\prime}\left(6A_1^{(1)}+2A_2^{(1)}
-9A_2^{(2)}-10A_3^{(2)}+4A_3^{(3)}-A_4^{(2)}+8A_4^{(3)}+4A_5^{(3)}-1\right)\nonumber\\&&+
m_1^{\prime\prime}\left(A_1^{(1)}-A_2^{(1)}-A_2^{(2)}+A_4^{(2)}\right)+m_2\left(-2A_1^{(1)}+
6A_2^{(2)}+6A_3^{(2)}-4A_3^{(3)}-8A_4^{(3)}-4A_5^{(3)}\right)\bigg]+\frac{1}
{\omega^{c\prime\prime}_{{}^3D_2}}\Bigg[\Big[m_2^{\prime2}\nonumber\\&&-m_2\left(m_1^{\prime\prime}+m_1^{\prime}\right)+
m_1^{\prime\prime}m_1^{\prime}\Big]\left(-4A_1^{(1)}+8A_2^{(2)}+8A_3^{(2)}-4A_3^{(3)}
-8A_4^{(3)}-4A_5^{(3)}\right)+\left(m_1^{\prime}-m_1^{\prime\prime}\right)^2\Big(2A_1^{(1)}
+2A_2^{(1)}-A_2^{(2)}\nonumber\\&&-2A_3^{(2)}-A_4^{(2)}-1\Big)+\left(M^{\prime2}+M^{\prime\prime2}-q^2\right)
\left(2A_1^{(1)}-4A_2^{(2)}-4A_3^{(2)}+2A_3^{(3)}+4A_4^{(3)}+2A_5^{(3)}\right)\nonumber\\&&+
\left[x(M^{\prime2}-M_0^{\prime2})+x(M^{\prime\prime2}-M_0^{\prime\prime2})-q^2\right]
\left(2A_1^{(1)}+2A_2^{(1)}-A_2^{(2)}-2A_3^{(2)}-A_4^{(2)}-1\right)-\bigg(4 \Big(\left(A_1^{(1)}Z_2-A_1^{(2)}\right)\nonumber\\&&-2\left(A_2^{(2)}Z_2-2A_1^{(3)}\right)-2\left(A_3^{(2)}Z_2+A_1^{(3)}
\frac{q\cdot P}{q^2}-A_2^{(3)}\right)+\left(A_3^{(3)}Z_2-2A_2^{(2)}A_1^{(2)}-A_2^{(4)}\right)+
2\big(A_4^{(3)}Z_2+A_2^{(2)}A_1^{(2)}\frac{q\cdot P}{q^2}\nonumber\\&&-A_1^{(1)}A_2^{(3)}-A_3^{(4)}\big)+\left(A_5^{(3)}Z_2+2\frac{q\cdot P}{q^2}A_1^{(1)}A_2^{(3)}-A_4^{(4)}\right)\Big)\bigg]\Bigg]\Bigg\},
\\
%
z_{-}^{\prime}(q^2)&=&\frac{N_c}{16\pi^3}\int dx_2
d^2p^{\prime}_{\bot}\frac{h^{\prime}_{P}h^{\prime\prime}_{{}^3D_2}}{(1-x)\hat{N}_1^{\prime}\hat{N}_1^{\prime\prime}}
\Bigg\{\frac{2}{\omega^{b\prime\prime}_{{}^3D_2}}\bigg[m_1^{\prime}\left(6A_1^{(1)}+10A_2^{(1)}-3A_2^{(2)}
-14A_3^{(2)}-11A_4^{(2)}+4A_4^{(3)}+8A_5^{(3)}+4A_{6}^{(3)}-3\right)\nonumber\\&&+m_1^{\prime\prime}
\left(-A_1^{(1)}+A_2^{(1)}+A_2^{(2)}-A_4^{(2)}\right)+m_2\left(-2A_1^{(1)}-4A_2^{(1)}+2A_2^{(2)}
+10A_3^{(2)}+8A_4^{(2)}-4A_4^{(3)}-8A_5^{(3)}-4A_6^{(3)}\right)\bigg]\nonumber\\&&+
\frac{1}{\omega^{c\prime\prime}_{{}^3D_2}}\Bigg[\left[2\left(m_2-m_1^{\prime}\right)^2
-\left(m_1^{\prime}-m_1^{\prime\prime}\right)^2\right]\left(-2A_1^{(1)}-2A_2^{(1)}+A_2^{(2)}+
2A_3^{(2)}+A_4^{(2)}+1\right)+\Big[m_2^{(2)}-m_2\left(m_1^{\prime}+m_1^{\prime\prime}\right)\nonumber\\&&+
m_1^{\prime}m_1^{\prime\prime}\Big]\left(-4A_2^{(1)}+8A_3^{(2)}+8A_4^{(2)}-4A_4^{(3)}-8A_5^{(3)}-4A_6^{(3)}\right)+
\left(x\left(M^{\prime\prime2}-M_0^{\prime\prime2}\right)-x\left(M^{\prime2}-M_0^{\prime2}\right)-q^2
+2M^{\prime2}\right)\nonumber\\&&\times\left(2A_1^{(1)}+2A_2^{(1)}-A_2^{(2)}-2A_3^{(2)}-A_4^{(2)}-1\right)+
\left(M^{\prime\prime2}+M^{\prime2}-q^2\right)\left(2A_2^{(1)}-4A_3^{(2)}-4A_4^{(2)}+2A_4^{(3)}+4A_5^{(3)}+2A_6^{(3)}\right)
\nonumber\\&&+2Z_2
-8\left(A_2^{(1)}Z_2+\frac{M^{\prime2}-M^{\prime\prime2}}{q^2}A_1^{(2)}\right)+
10\left(A_4^{(2)}Z_2+2\frac{M^{\prime2}-M^{\prime\prime2}}{q^2}A_2^{(1)}A_1^{(2)}\right)-4\bigg[A_6^{(3)}Z_2+
3\frac{M^{\prime2}-M^{\prime\prime2}}{q^2}\nonumber\\&&
\times\Big[A_2^{(1)}A_2^{(3)}-\frac{1}{3q^2}(A_1^{(2)})^2\Big]\bigg]+
\bigg[2\Bigg(-2\left(A_1^{(1)}Z_2-A_1^{(2)}\right)+\left(A_2^{(2)}Z_2-2A_1^{(3)}\right)+6\left(A_3^{(2)}Z_2+A_1^{(3)}\frac{q\cdot P}{q^2}-A_2^{(3)}\right)-2\big(A_4^{(3)}Z_2\nonumber\\&&+A_2^{(2)}A_1^{(2)}\frac{q\cdot P}{q^2}-A_1^{(1)}A_2^{(3)}-A_3^{(4)}\big)-4\left(A_5^{(3)}Z_2+2\frac{q\cdot P}{q^2}A_1^{(1)}A_2^{(3)}-A_3^{(4)}\right)\Bigg)\bigg]\Bigg]\Bigg\},
\end{eqnarray}

The corresponding form factors for a ${}^3D_3$ state are
\begin{eqnarray}
y(q^2)&=&\frac{N_c}{16\pi^3}\int dx_2d^2p^{\prime}_{\bot}\frac{3h^{\prime}_{P}h^{\prime\prime}_{{}^3D_3}}{(1-x)\hat{N}_1^{\prime}\hat{N}_1^{\prime\prime}}
\Bigg\{2m_1^{\prime}\left(3A_1^{(1)}+3A_2^{(1)}-3A_2^{(2)}-6A_3^{(2)}+A_3^{(3)}
-3A_4^{(2)}+3A_4^{(3)}+3A_5^{(3)}+A_6^{(3)}-1\right)\nonumber\\&&+2m_1^{\prime\prime}
(A_1^{(1)}-A_2^{(1)}-2A_2^{(2)}+A_3^{(3)}+2A_4^{(2)}+A_4^{(3)}-A_5^{(3)}-A_6^{(3)})+
2m_2\bigg(-2A_1^{(1)}+4A_2^{(2)}+4A_3^{(2)}-2A_3^{(3)}-4A_4^{(3)}\nonumber\\&&-2A_5^{(3)}\bigg)+
\frac{12}{\omega^{\prime\prime}_{{}^3D_3}}\left(A_1^{(2)}-2A_1^{(3)}-2A_2^{(3)}+A_2^{(4)}+2A_3^{(4)}+A_4^{(4)}\right)\Bigg\},
\\
w(q^2)&=&\frac{N_c}{16\pi^3}\int
dx_2d^2p^{\prime}_{\bot}\frac{3h^{\prime}_{P}h^{\prime\prime}_{{}^3D_3}}{(1-x)\hat{N}_1^{\prime}\hat{N}_1^{\prime\prime}}
\Bigg\{2\Bigg[\left(2A_1^{(1)}+2A_2^{(1)}-A_2^{(2)}-2A_3^{(2)}-A_4^{(2)}-1\right)
\bigg[m_2^{2}\left(-m_1^{\prime\prime}-m_1^{\prime}\right)+m_2\Big[\left(m_1^{\prime\prime}+m_1^{\prime}\right)^2
\nonumber\\&&+
x\left(M^{\prime\prime2}-M_0^{\prime\prime2}\right)+x\left(M^{\prime2}-M_0^{\prime2}\right)-q^2\Big]
-m_1^{\prime\prime2}m_1^{\prime}-m_1^{\prime\prime}\left[m_1^{\prime2}-M^{\prime2}+
x\left(M^{\prime2}-M_0^{\prime2}\right)\right]+m_1^{\prime}\Big[M^{\prime\prime2}\nonumber\\&&
-x\left(M^{\prime\prime2}-M_0^{\prime\prime2}
\right)\Big]\bigg]
+4m_2\left[2A_1^{(2)}-5A_1^{(3)}-5A_2^{(3)}+3\left(A_2^{(4)}+2A_3^{(4)}+A_4^{(4)}\right)\right]+
m_1^{\prime}\bigg(-12A_1^{(2)}+24A_1^{(3)}\nonumber\\&&+24A_2^{(3)}-12A_2^{(4)}-24A_3^{(4)}
-12A_4^{(4)}-2\left(A_2^{(1)}Z_2+\frac{M^{\prime2}-M^{\prime\prime2}}{q^2}A_1^{(2)}\right)+
A_4^{(2)}Z_2+2\frac{M^{\prime2}-M^{\prime\prime2}}{q^2}A_2^{(1)}A_1^{(2)}+Z_2\bigg)
\nonumber\\&&+m_1^{\prime\prime}\left(4A_1^{(2)}-4A_1^{(3)}-4A_2^{(3)}-2\left(A_2^{(1)}Z_2+
\frac{M^{\prime2}-M^{\prime\prime2}}{q^2}A_1^{(2)}\right)+A_4^{(2)}Z_2+
2\frac{M^{\prime2}-M^{\prime\prime2}}{q^2}A_2^{(1)}A_1^{(2)}+Z_2\right)\nonumber\\&&-(m^{\prime\prime}_1 + m^{\prime}_1)(2(A_1^{(1)}Z_2-A_1^{(2)})-(A_2^{(2)}Z_2-2A_1^{(3)})-
2(A_3^{(2)}Z_2+A_1^{(3)}\frac{q\cdot P}{q^2}-A_2^{(3)}))\Bigg]
-\frac{12}{\omega^{\prime\prime}_{{}^3D_3}}\Bigg[(A_1^{(2)}-2A_1^{(3)}-2A_2^{(3)}+A_2^{(4)}\nonumber\\&&+2A_3^{(4)}+A_4^{(4)} )\left[2\left(m_2+m_1^{\prime\prime}\right)\left(m_2-m_1^{\prime}\right)-M^{\prime2}-M^{\prime\prime2}+q^2\right]
+2A_1^{(2)}Z_2
+2\left(Z_2A_4^{(4)}+\frac{2}{q^2}\left(M^{\prime2}-M^{\prime\prime2}\right)A_2^{(1)}A_1^{(4)}\right)
\nonumber\\&&-4\left(A_2^{(3)}Z_2+\frac{M^{\prime2}-M^{\prime\prime2}}{3q^2}\left(A_1^{(2)}\right)^2\right)-2\bigg(\left(A_2^{(4)}Z_2-2A_1^{(1)}A_1^{(4)}\right)
+2\left(A_3^{(4)}Z_2+A_1^{(1)}A_1^{(4)}\frac{q\cdot P}{q^2}-A_2^{(1)}A_1^{(4)}\right)\nonumber\\&&-2\left(A_1^{(3)}Z_2-A_1^{(4)}\right)\bigg)\Bigg]\Bigg\},\\
o_{+}(q^2)&=&\frac{N_c}{16\pi^3}\int
dx_2d^2p^{\prime}_{\bot}\frac{3h^{\prime}_{P}h^{\prime\prime}_{{}^3D_3}}{(1-x)\hat{N}_1^{\prime}\hat{N}_1^{\prime\prime}}
\Bigg\{2m_1^{\prime}\bigg(7A_1^{(1)}+3A_2^{(1)}-15A_2^{(2)}-18A_3^{(2)}+13A_3^{(3)}
-3A_4^{(2)}+27A_4^{(3)}+15A_5^{(3)}-4A_5^{(4)}\nonumber\\&&+A_6^{(3)}-4\left(3\left(A_6^{(4)}+A_7^{(4)}\right)
+A_8^{(4)}\right)-1\bigg)+2m_1^{\prime\prime}\left[-A_1^{(1)}+A_2^{(1)}+2A_2^{(2)}-A_3^{(3)}
-2A_4^{(2)}-A_4^{(3)}+A_5^{(3)}+A_6^{(3)}\right]\nonumber\\&&+2m_2\left(-2A_1^{(1)}+8A_2^{(2)}
+8A_3^{(2)}-10A_3^{(3)}-20A_4^{(3)}-10A_5^{(3)}+4A_5^{(4)}+12A_6^{(4)}+12A_7^{(4)}
+4A_8^{(4)}\right)\nonumber\\&&+\frac{2}{\omega^{\prime\prime}_{{}^3D_3}}\Bigg[\left[m_1^{\prime2}+m_1^{\prime\prime2}
+x\left(M^{\prime2}-M_0^{\prime2}\right)+x\left(M^{\prime\prime2}-M_0^{\prime\prime2}\right)-q^2
+2m_1^{\prime}m_1^{\prime\prime}\right]\Big(-3A_1^{(1)}-3A_2^{(1)}+3A_2^{(2)}+6A_3^{(2)}
\nonumber\\&&-A_3^{(3)}+3A_4^{(2)}-3A_4^{(3)}-3A_5^{(3)}-A_6^{(3)}+1\Big)+\left(M^{\prime\prime2}
+M^{\prime2}-2m_2^2-q^2+2m_1^{\prime}m_2-2m_2m_1^{\prime\prime}
+2m_1^{\prime}m_1^{\prime\prime}\right)\nonumber\\&&\times\left(-2A_1^{(1)}+6A_2^{(2)}+6A_3^{(2)}-6A_3^{(3)}
-12A_4^{(3)}-6A_5^{(3)}+2A_5^{(4)}+6A_6^{(4)}+6A_7^{(4)}+2A_8^{(4)}\right)
-4\bigg[\big(A_5^{(4)}Z_2-2A_3^{(3)}A_1^{(2)}\nonumber\\&&-2A_1^{(1)}A_2^{(4)}\big)+3\left(A_6^{(4)}Z_2+\frac{q\cdot P}{q^2}A_3^{(3)}A_1^{(2)}-A_2^{(2)}A_1^{(2)}A_2^{(1)}-2A_1^{(1)}A_3^{(4)}\right)
+3\left(A_7^{(4)}Z_2+2\frac{q\cdot P}{q^2}A_2^{(2)}A_1^{(2)}A_2^{(1)}-2A_1^{(1)}A_4^{(4)}\right)\nonumber\\&&+
\left(A_8^{(4)}Z_2+3\frac{q\cdot P}{q^2}A_1^{(4)}A_4^{(4)}-A_2^{(1)}A_4^{(4)}+\frac{2A_2^{(1)}A_1^{(4)}}{q^2}\right)- \left(A_1^{(1)}Z_2-A_1^{(2)}\right)+3\left(A_2^{(2)}Z_2-2A_1^{(3)}\right)+3\bigg(A_3^{(2)}Z_2+A_1^{(3)}\frac{q\cdot P}{q^2}\nonumber\\&&-A_2^{(3)}\bigg)-3\left(A_3^{(3)}Z_2-2A_2^{(2)}A_1^{(2)}-A_2^{(4)}\right)-6\left(A_4^{(3)}Z_2+A_2^{(2)}A_1^{(2)}\frac{q\cdot P}{q^2}-A_1^{(1)}A_2^{(3)}-A_3^{(4)}\right)-3\big(A_5^{(3)}Z_2\nonumber\\&&+2\frac{q\cdot P}{q^2}A_1^{(1)}A_2^{(3)}-A_4^{(4)}\big)\bigg]\Bigg]\Bigg\},
\\
%
o_{-}(q^2)&=&\frac{N_c}{16\pi^3}\int
dx_2d^2p^{\prime}_{\bot}\frac{3h^{\prime}_{P}h^{\prime\prime}_{{}^3D_3}}{(1-x)\hat{N}_1^{\prime}\hat{N}_1^{\prime\prime}}
\Bigg\{2m_2\bigg(-2A_1^{(1)}-4A_2^{(1)}+4A_2^{(2)}+16A_3^{(2)}-2A_3^{(3)}
+12A_4^{(2)}-16A_4^{(3)}-26A_5^{(3)}\nonumber\\&&-12A_6^{(3)}+4A_6^{(4)}+12A_7^{(4)}
+12A_8^{(4)}+4A_9^{(4)}\bigg)+2m_1^{\prime}\bigg[9A_1^{(1)}+13A_2^{(1)}-9A_2^{(2)}
-30A_3^{(2)}+3A_3^{(3)}-21A_4^{(2)}\nonumber\\&&+21A_4^{(3)}+33A_5^{(3)}+15A_6^{(3)}
-4\left(A_6^{(4)}+3\left(A_7^{(4)}+A_8^{(4)}\right)+A_9^{(4)}\right)-3\bigg]+2m_1^{\prime\prime}
\Big(A_1^{(1)}-A_2^{(1)}-2A_2^{(2)}\nonumber\\&&+A_3^{(3)}+2A_4^{(2)}+A_4^{(3)}-A_5^{(3)}
-A_6^{(3)}\Big)+\frac{2}{\omega^{\prime\prime}_{{}^3D_3}}\Bigg[\left(-2M^{\prime2}-4m_2m_1^{\prime}
+4m_2^{2}-4m_1^{\prime}m_1^{\prime\prime}+2q^2-2M^{\prime\prime2}+4m_2m_1^{\prime\prime}\right)\nonumber\\&&
\times\left(3A_1^{(1)}+4A_2^{(1)}-3A_2^{(2)}-9A_3^{(2)}+A_3^{(3)}-6A_4^{(2)}
+6A_4^{(3)}+9A_5^{(3)}+4A_6^{(3)}-A_6^{(4)}-3A_7^{(4)}-3A_8^{(4)}
-A_9^{(4)}-1\right)\nonumber\\&&+\left[m_1^{\prime2}-m_1^{\prime\prime2}+x\left(M^{\prime2}-M_0^{\prime2}\right)
-x\left(M^{\prime\prime2}-M_0^{\prime\prime2}\right)-q^2-2m_2^{2}+2M^{\prime\prime2}
+2m_1^{\prime}m_1^{\prime\prime}-4m_2m_1^{\prime\prime}\right]
\Big(3A_1^{(1)}+3A_2^{(1)}\nonumber\\&&-3A_2^{(2)}-6A_3^{(2)}+A_3^{(3)}-3A_4^{(2)}
+3A_4^{(3)}+3A_5^{(3)}+A_6^{(3)}-1\Big)-4\bigg[Z_2A_9^{(4)}+\frac{4}{q^2}A_4^{(4)}A_2^{(1)}
\left(M^{\prime2}-M^{\prime\prime2}\right)\nonumber\\&&-\frac{8}{q^4}A_2^{(1)}A_1^{(4)}\left(M^{\prime2}
-M^{\prime\prime2}\right)\bigg]+10\left(A_2^{(1)}Z_2+\frac{M^{\prime2}-M^{\prime\prime2}}{q^2}A_1^{(2)}\right)-
18\left(A_4^{(2)}Z_2+2\frac{M^{\prime2}-M^{\prime\prime2}}{q^2}A_2^{(1)}A_1^{(2)}\right)
\nonumber\\&&-2Z_2+14\Bigg[A_6^{(3)}Z_2+3\frac{M^{\prime2}-M^{\prime\prime2}}{q^2}
\bigg[A_2^{(1)}A_2^{(3)}-\frac{1}{3q^2}\left(A_1^{(2)}\right)^2\bigg]\Bigg]
-2\bigg[2\bigg(A_6^{(4)}Z_2+\frac{q\cdot P}{q^2}A_3^{(3)}A_1^{(2)}-A_2^{(2)}A_1^{(2)}A_2^{(1)}\nonumber\\&&-2A_1^{(1)}A_3^{(4)}\bigg)+6\left(A_7^{(4)}Z_2+2\frac{q\cdot P}{q^2}A_2^{(2)}A_1^{(2)}A_2^{(1)}-2A_1^{(1)}A_4^{(4)}\right)+6\left(A_8^{(4)}Z_2+3\frac{q\cdot P}{q^2}A_1^{(1)}A_4^{(4)}-A_2^{(1)}A_4^{(4)}+\frac{2A_2^{(1)}A_1^{(4)}}{q^2}\right)
\nonumber\\&&-3\left(A_1^{(1)}Z_2-A_1^{(2)}\right)+3\left(A_2^{(2)}Z_2-2A_1^{(3)}\right)+12\left(A_3^{(2)}Z_2+A_1^{(3)}\frac{q\cdot P}{q^2}-A_2^{(3)}\right)-\left(A_3^{(3)}Z_2-2A_2^{(2)}A_1^{(2)}-A_2^{(4)}\right)
-9\big(A_4^{(3)}Z_2\nonumber\\&&+A_2^{(2)}A_1^{(2)}\frac{q\cdot P}{q^2}-A_1^{(1)}A_2^{(3)}-A_3^{(4)}\big)-15\left(A_5^{(3)}Z_2+2\frac{q\cdot P}{q^2}A_1^{(1)}A_2^{(3)}-A_4^{(4)}\right)\bigg]\Bigg]\Bigg\}.
\end{eqnarray}

\section{
Conventional vertex functions for $D$-wave mesons
}\label{vertex}
In the conventional light-front approach, a meson with momentum $P$ and spin $J$ can be defined as
\begin{eqnarray}
\left|M(P,{}^{2S+1}L_J,J_z)\right\rangle=\int \left\{d^3\tilde{p}_1d^3\tilde{p}_2\right\}2\left(2\pi\right)^3\delta^3\left(\tilde{P}-\tilde{p}_1-
\tilde{p}_2\right)\sum_{\lambda_1\lambda_2}\Psi_{LS}^{JJ_z}\left(\tilde{p}_1,\tilde{p}_2,
\lambda_1,\lambda_2\right)\left|q_1(p_1,\lambda_1)\bar{q}_2(p_2,\lambda_2)\right\rangle,
\end{eqnarray}
where $q_1$ and $\bar{q}_2$ denote the quark and antiquark within the meson, respectively, and $p_1$ and $p_2$ are the on-shell light-front momenta of quark and antiquark, respectively.
The symbol `` $\tilde{}$ '' means an operation on a momentum to extract only plus and transverse components, i.e.,
\begin{eqnarray}
\tilde{p}=\left(p^{+},p_{\bot}\right),p_{\bot}=\left(p^1,p^2\right),p^-=\frac{m^2+p^2_{\bot}}{p^+},
\end{eqnarray}
and
\begin{eqnarray}
&&\left\{d^3p\right\}\equiv\frac{dp^+d^2p_{\bot}}{2(2\pi)^3},
\quad\left|q(p_1,\lambda_1)\bar{q}(p_2,\lambda_2)\right\rangle=b_{\lambda_1}^{\dagger}(p_1)
d_{\lambda_2}^{\dagger}(p_2)\left|0\right\rangle,\nonumber\\
&&\left\{b_{\lambda^{\prime}}(p^{\prime}),b_{\lambda}^{\dagger}(p)\right\}
=\left\{d_{\lambda^{\prime}}(p^{\prime}),d^{\dagger}_{\lambda}(p)\right\}=
2\left(2\pi^3\right)\delta^3\left(\tilde{p}^{\prime}-\tilde{p}\right)\delta_{\lambda\lambda^{\prime}}.\nonumber
\end{eqnarray}

In the light-front coordinate, the definition of the light-front relative momentum $(x,p_{\bot})$ reads
\begin{eqnarray}
p_1^{+}&=&x_1P^{+},\quad p_2^{+}=x_2P^+,\quad x_1+x_2=1,\nonumber\\
p_{1\bot}&=&x_1P_{\bot}+p_{\bot}, \quad p_{2\bot}=x_2P_{\bot}-p_{\bot}.
\end{eqnarray}

The meson wave function $\Psi_{LS}^{JJ_z}$ in momentum space is expressed as
\begin{eqnarray}
\Psi_{LS}^{JJ_z}\left(\tilde{p}_1,\tilde{p}_2,\lambda_1,\lambda_2\right)&=&\frac{1}{\sqrt{N_c}}\left\langle
LS;L_zS_z|LS;JJ_z\right\rangle R_{\lambda_1\lambda_2}^{SS_z}\left(x_2,p_{\bot}\right)\varphi_{LL_z}\left(x_2,p_{\bot}\right),
\label{psi}
\end{eqnarray}
where $\varphi_{LL_z}\left(x_2,p_{\bot}\right)$ describes the momentum distribution of the constituent quarks inside a meson with the orbital angular momentum $L$, and $\left\langle LS;L_zS_z|LS;JJ_z\right\rangle$ is the corresponding Clebsch-Gordan (CG) coefficient. In Eq. (\ref{psi}), $R_{\lambda_1\lambda_2}^{SS_z}$ transforms a light-front helicity $(\lambda_1,\lambda_2)$ eigenstate into a state with spin $(S,S_z)$
\begin{eqnarray}
&&R_{\lambda_1\lambda_2}^{SS_z}\left(x_2,p_{\bot}\right)=\frac{1}{\sqrt{2}\tilde{M}_0
\left(M_0+m_1+m_2\right)}\bar{u}\left(p_1,\lambda_1\right)\left(\bar{\slashed{P}}+M_0\right)\Gamma_Sv\left(p_2,\lambda_2\right)\nonumber\\
&&{\text{with}}\quad
\left\{
\begin{array}{lc}
 \Gamma_0=\gamma_5
& \text{for}\qquad S=0\\
\Gamma_1=-\slashed{\epsilon}(S_z)& \text{for}\qquad S=1\\
\end{array}\right. ,
\end{eqnarray}
where $\bar{P}$ is the momentum of meson in the rest frame, $\bar{P}=p_1+p_2$, $M_0^{2}=\frac{m_1^2+p_{\bot}^2}{x_1}+\frac{m_2^2+p_{\bot}^2}{x_2}$ and $\tilde{M}_0\equiv\sqrt{M_0^2-(m_1-m_2)^2}$.
According to the spinor representation of $\bar{u}$ and $v$ in Appendix of Ref. \cite{Brodsky:1997de}, one has calculated the explicit expression of $R_{\lambda_1\lambda_2}^{SS_z}$ in Ref. \cite{Yu:2007hp}.

With the help of the potential model, for a definite meson state with quantum numbers $n_r{}^{2S+1}L_J$, its mass and
the corresponding numerical spatial wave function can
be calculated. The obtained spatial wave function can be produced by the expansion of a list of SHO wave functions (the number of the SHO wave function bases taken is set to be $N$), where the expansion coefficients form
the corresponding eigenvector.

The meson wave function $\Psi_{2S}^{JJ_z}(p_1,p_2,\lambda_1,\lambda_2)$ in momentum space reads
\begin{eqnarray}
\Psi_{2S}^{JJ_z}(p_1,p_2,\lambda_1,\lambda_2)&=&\frac{1}{\sqrt{N}_c}\left\langle 2S;L_zS_z|2S;JJ_z\right\rangle R_{\lambda_1\lambda_2}^{SS_z}\left(x,p_{\bot}\right)\varphi^{\prime}_{2L_z}\left(x,p_{\bot}\right)\nonumber\\
&=&\sum_n^N \frac{\beta^2}{\sqrt{2}} a_n R^{\prime}_{n2}(x,p_{\bot})\pi\sqrt{\frac{30 e_1e_2}{x(1-x)M_0}}
\frac{1}{\sqrt{N}_c}\frac{1}{\sqrt{2}\tilde{M}_0\left(M_0+m_1+m_2\right)}\nonumber\\
&&\times \bar{u}\left(p_1,\lambda_1\right)\left(\bar{\slashed{P}}+M_0\right)\Gamma_{({}^{2S+1}D_J)}v\left(p_2,\lambda_2\right),\nonumber\\
&=&\frac{1}{\sqrt{N_c}}\frac{\varphi_N}{\sqrt{2}\tilde{M}_0\left(M_0+m_1+m_2\right)}
\bar{u}\left(p_1,\lambda_1\right)\left(\bar{\slashed{P}}+M_0\right)\Gamma_{({}^{2S+1}D_J)}v\left(p_2,\lambda_2\right).
\label{psip}
\end{eqnarray}
Here, $\varphi_N=\sum_n^N \frac{\beta^2}{\sqrt{2}} a_n R^{\prime}_{n2}(x,p_{\bot})\pi\sqrt{\frac{30 e_1e_2}{x(1-x)M_0}}$, where
\begin{eqnarray}
R'_{nl}(|p|)=\frac{(-1)^n(-i)^l}{\beta^{3/2}}\sqrt{\frac{2n!}{\Gamma\left(n+l+3/2\right)}}
\left(\frac{1}{\beta}\right)^le^{-\frac{p^2}{2\beta^2}}L_n^{l+\frac{1}{2}}\left(\frac{p^2}{\beta^2}\right), \label{B9}
\end{eqnarray}
$a_n$ is the expansion coefficients of the corresponding eigenvectors, and
$\Gamma_{({}^{2S+1}D_J)}$ denotes the corresponding vertex structure of $D$-wave mesons.

One can further simplify these wave functions by using the Dirac equations $\slashed{p}_1u(p_1)=m_1u(p_1)$ and $\slashed{p}_2v(p_2)=-m_2v(p_1)$. After this simplification, the wave function of $D$-wave mesons can be written as \footnote{We need to mention that a minus sign is needed in front of $\omega^c_{{}^3D_2}$ which is a typo in Ref. \cite{Ke:2011mu}.}
\begin{eqnarray}
\Psi_{2S}^{JJ_z}\left(\tilde{p}_1,\tilde{p}_2,\lambda_1,\lambda_2\right)=\bar{u}\left(p_1,\lambda_1\right)h'_{({}^{2S+1}D_J)}\Gamma'_{({}^{2S+1}D_J)}v\left(p_2,\lambda_2\right)\nonumber\\
\end{eqnarray}
with
\begin{eqnarray}
h'_{{}^3D_1}&=&-\sqrt{\frac{1}{N_c}}\frac{1}{\sqrt{2}\tilde{M_0}}\frac{\sqrt{6}}{12\sqrt{5}M_{0}^2\beta^2}\left[M_0^2-(m_1-m_2)^2\right]\left[M_0^2-(m_1+m_2)^2\right]\varphi_N,\nonumber\\
h'_{{}^1D_2}&=&\sqrt{\frac{1}{N_c}}\frac{1}{\tilde{M}_0\beta^2}\varphi_N,\nonumber\\
h'_{{}^3D_2}&=&\sqrt{\frac{1}{N_c}}\sqrt{\frac{2}{3}}\frac{1}{\tilde{M}_0\beta^2}\varphi_N,\nonumber\\
h'_{{}^3D_3}&=&-\sqrt{\frac{1}{N_c}}\frac{1}{3}\frac{1}{\tilde{M}_0\beta^2}\varphi_N,
\label{AppAhM}
\\
\nonumber \\
\Gamma'_{{}^3D_1}&=&\left[\gamma_{\mu}-\frac{1}{\omega_{{}^3D_1}}(p_1-p_2)_{\mu}\right]\epsilon^{\mu},\nonumber\\
\Gamma'_{{}^1D_2}&=&\gamma_5K_{\mu}K_{\nu}\epsilon^{\mu\nu},\nonumber\\
\Gamma'_{{}^3D_2}&=&\gamma_5\left[\frac{1}{\omega^a_{{}^3D_2}}\gamma_{\mu}\gamma_{\nu}+\frac{1}{\omega^b_{{^3D_2}}}\gamma_\mu K_{\nu}+\frac{1}{\omega^c_{{}^3D_2}}K_{\mu}K_{\nu}\right]\epsilon^{\mu\nu},\nonumber\\
\Gamma'_{{}^3D_3}&=&\left[K_{\mu} K_{\nu}\left(\gamma_{\alpha}+\frac{2K_{\alpha}}{\omega_{{}^3D_3}}\right)+K_{\mu}K_{\alpha}\left(\gamma_{\nu}+\frac{2K_{\nu}}{\omega_{{}^3D_3}}\right)+K_{\alpha}K_{\nu}(\gamma_{\mu}+
\frac{2K_{\mu}}{\omega_{{}^3D_3}})\right]\epsilon^{\mu\nu\alpha},
\end{eqnarray}
and
\begin{eqnarray}
\omega_{{}^3D_1}&=&\frac{\left(m_1+m_2\right)^2-M_0^2}{2M_0+m_1+m_2},\nonumber\\
\omega^{a}_{{}^3D_2}&=&\frac{12M_0^2}{\left[M_0^2-(m_1+m_2)^2\right]\left[M_0^2-(m_1-m_2)^2\right]},\nonumber\\
\omega^{b}_{{}^3D_2}&=&-\frac{2M_0}{M_0^2-\left(m_1-m_2\right)^2},\nonumber\\
\omega^{c}_{{}^3D_2}&=&-\frac{M_0}{m_2-m_1},\nonumber\\
\omega_{{}^3D_3}&=&M_0+m_1+m_2.
\end{eqnarray}

\section{Tensor decomposition}\label{decomposition}


The second-order tensor decomposition of $\hat{p}_{1\mu}^{\prime}\hat{p}^{\prime}_{1\nu}$ and the third-order of $\hat{p}_{1\mu}^{\prime}\hat{p}_{1\nu}^{\prime}\hat{p}_{1\alpha}^{\prime}$ were given
in Ref. \cite{Jaus:1999zv}. The fourth-order tensor decomposition of $\hat{p}_{1\mu}^{\prime}\hat{p}_{1\nu}^{\prime}\hat{p}_{1\alpha}^{\prime}\hat{p}_{1\beta}^{\prime}$ was also obtained in Ref. \cite{Cheng:2003sm}\footnote{When reproducing the coefficients $A_i^{(4)}$, $B_{j}^{(4)}$ and $C_{k}^{(4)}$, we find a typo in Ref. \cite{Cheng:2003sm} whose correct expression is given by $A_9^{(4)}=A_2^{(1)}A_6^{(3)}-\frac{3}{q^2}A_4^{(4)}$.}.
For a ${}^3D_3$ state, we need the fifth-order tensor decomposition of $\hat{p}_{1\mu}^{\prime}\hat{p}_{1\nu}^{\prime}\hat{p}_{1\alpha}^{\prime}\hat{p}_{1\beta}^{\prime}\hat{N}_2$. Here, we just include the leading-order contribution from $\tilde{\omega}$, and get the following expression,
\begin{eqnarray*}
\hat{p}^{\prime}_{1\mu}\hat{p}^{\prime}_{1\nu}\hat{p}^{\prime}_{1\alpha}\hat{p}^{\prime}_{1\beta}\hat{p}^{\prime}_{1\delta}\doteq \sum_{i=1}^{12}L_{i\mu\nu\alpha\beta\delta}A_i^{(5)}+\sum_{j=1}^{6}M_{j\mu\nu\alpha\beta\delta}B_{j}^{(5)}+\sum_{k=1}^{3}N_{k\mu\nu\alpha\beta\delta}C_{k}^{(5)}+O\left(\tilde{\omega}^{2}\right)
\end{eqnarray*}
with
\begin{eqnarray*}
L_{1\mu\nu\alpha\beta\delta}&=&\left(ggP\right)_{\mu\nu\alpha\beta\delta}=g_{\mu\nu}\left(gP\right)_{\alpha\beta\delta}+g_{\mu\alpha}\left(gP\right)_{\nu\beta\delta}+g_{\mu\beta}\left(gP\right)_{\nu\alpha\delta}
+g_{\mu\delta}\left(gP\right)_{\nu\alpha\beta}+\left(g_{\nu\alpha}g_{\beta\delta}+g_{\nu\beta}g_{\alpha\delta}+g_{\nu\delta}g_{\alpha\beta}\right)P_{\mu},\nonumber\\
L_{2\mu\nu\alpha\beta\delta}&=&\left(ggq\right)_{\mu\nu\alpha\beta\delta}=g_{\mu\nu}\left(gq\right)_{\alpha\beta\delta}+g_{\mu\alpha}\left(gq\right)_{\nu\beta\delta}+g_{\mu\beta}\left(gq\right)_{\nu\alpha\delta}
+g_{\mu\delta}\left(gq\right)_{\nu\alpha\beta}+\left(g_{\nu\alpha}g_{\beta\delta}+g_{\nu\beta}g_{\alpha\delta}+g_{\nu\delta}g_{\alpha\beta}\right)q_{\mu},\nonumber\\
L_{3\mu\nu\alpha\beta\delta}&=&\left(gPPP\right)_{\mu\nu\alpha\beta\delta}=g_{\mu\nu}P_{\alpha}P_{\beta}P_{\delta}+permutations,\nonumber\\
L_{4\mu\nu\alpha\beta\delta}&=&\left(gPPq\right)_{\mu\nu\alpha\beta\delta}=g_{\mu\nu}\left(PPq\right)_{\alpha\beta\delta}+permutations,\nonumber\\
L_{5\mu\nu\alpha\beta\delta}&=&\left(gPqq\right)_{\mu\nu\alpha\beta\delta}=g_{\mu\nu}\left(Pqq\right)_{\alpha\beta\delta}+permutations,\nonumber\\
L_{6\mu\nu\alpha\beta\delta}&=&\left(gqqq\right)_{\mu\nu\alpha\beta\delta}=g_{\mu\nu}q_{\alpha}q_{\beta}q_{\delta}+permutations,\nonumber\\
L_{7\mu\nu\alpha\beta\delta}&=&\left(PPPPP\right)_{\mu\nu\alpha\beta\delta}=P_{\mu}P_{\nu}P_{\alpha}P_{\beta}P_{\delta},\nonumber\\
L_{8\mu\nu\alpha\beta\delta}&=&\left(PPPPq\right)_{\mu\nu\alpha\beta\delta}=P_{\mu}P_{\nu}P_{\alpha}P_{\beta}q_{\delta}+P_{\mu}P_{\nu}P_{\alpha}q_{\beta}P_{\delta}+P_{\mu}P_{\nu}q_{\alpha}P_{\beta}P_{\delta}+
P_{\mu}q_{\nu}P_{\alpha}P_{\beta}P_{\delta}+q_{\mu}P_{\nu}P_{\alpha}P_{\beta}P_{\delta},\nonumber\\
L_{9\mu\nu\alpha\beta\delta}&=&\left(PPPqq\right)_{\mu\nu\alpha\beta\delta}=\left(PPP\right)_{\mu\nu\alpha}\left(qq\right)_{\beta\delta}+permutations,\nonumber
\\
L_{10\mu\nu\alpha\beta\delta}&=&\left(PPqqq\right)_{\mu\nu\alpha\beta\delta}=\left(PP\right)_{\mu\nu}\left(qqq\right)_{\alpha\beta\delta}+permutation,\nonumber\\
L_{11\mu\nu\alpha\beta\delta}&=&\left(Pqqqq\right)_{\mu\nu\alpha\beta\delta}=P_{\mu}q_{\nu}q_{\alpha}q_{\beta}q_{\delta}+q_{\mu}P_{\nu}q_{\alpha}q_{\beta}q_{\delta}+q_{\mu}q_{\nu}P_{\alpha}q_{\beta}q_{\delta}+
q_{\mu}q_{\nu}q_{\alpha}P_{\beta}q_{\delta}+q_{\mu}q_{\nu}q_{\alpha}q_{\beta}P_{\delta},\nonumber\\
L_{12\mu\nu\alpha\beta\delta}&=&\left(qqqqq\right)_{\mu\nu\alpha\beta\delta}=q_{\mu}q_{\nu}q_{\alpha}q_{\beta}q_{\delta},\nonumber
\\
\nonumber \\
%
M_{1\mu\nu\alpha\beta\delta}&=&\left(gPP\tilde{\omega}\right)_{\mu\nu\alpha\beta\delta}=\frac{1}{\tilde{\omega} P}\left[g_{\mu\nu}\left(PP\tilde{\omega}\right)_{\alpha\beta\delta}+permutations\right],\nonumber\\
M_{2\mu\nu\alpha\beta\delta}&=&\left(gPq\tilde{\omega}\right)_{\mu\nu\alpha\beta\delta}=\frac{1}{\tilde{\omega} P}\left[g_{\mu\nu}\left(Pq\tilde{\omega}\right)_{\alpha\beta\delta}+permutations\right],\nonumber\\
M_{3\mu\nu\alpha\beta\delta}&=&\left(PPPP\tilde{\omega}\right)_{\mu\nu\alpha\beta\delta}=\frac{1}{\tilde{\omega} P}\left[P_{\mu}P_{\nu}P_{\alpha}P_{\beta}\tilde{\omega}_{\delta}+P_{\mu}P_{\nu}P_{\alpha}\tilde{\omega}_{\beta}P_{\delta}+P_{\mu}P_{\nu}\tilde{\omega}_{\alpha}P_{\beta}P_{\delta}+
P_{\mu}\tilde{\omega}_{\nu}P_{\alpha}P_{\beta}P_{\delta}+\tilde{\omega}_{\mu}P_{\nu}P_{\alpha}P_{\beta}P_{\delta}\right],\nonumber\\
M_{4\mu\nu\alpha\beta\delta}&=&\left(PPPq\tilde{\omega}\right)_{\mu\nu\alpha\beta\delta}=\frac{1}{\tilde{\omega} P}\left[\left(PPP\right)_{\mu\nu\alpha}\left(q\tilde{\omega}\right)_{\beta\delta}+permutations\right],\nonumber\\
M_{5\mu\nu\alpha\beta\delta}&=&\left(PPqq\tilde{\omega}\right)_{\mu\nu\alpha\beta\delta}=\frac{1}{\tilde{\omega} P}\left[\left(PP\right)_{\mu\nu}(qq)_{\alpha\beta}\tilde{\omega}_{\delta}+permutations\right],\nonumber\\
M_{6\mu\nu\alpha\beta\delta}&=&\left(Pqqq\tilde{\omega}\right)_{\mu\nu\alpha\beta\delta}=\frac{1}{\tilde{\omega} P}\left[\left(qqq\right)_{\mu\nu\alpha}\left(P\tilde{\omega}\right)_{\beta\delta}+permutations\right].\nonumber
\\
\\
%
N_{1\mu\nu\alpha\beta\delta}&=&\left(gg\tilde{\omega}\right)_{\mu\nu\alpha\beta\delta}=\frac{1}{\tilde{\omega} P}\left[g_{\mu\nu}\left(g\tilde{\omega}\right)_{\alpha\beta\delta}+g_{\mu\alpha}\left(g\tilde{\omega}\right)_{\nu\beta\delta}+g_{\mu\beta}\left(g\tilde{\omega}\right)_{\nu\alpha\delta}
+g_{\mu\delta}\left(g\tilde{\omega}\right)_{\nu\alpha\beta}+\left(g_{\nu\alpha}g_{\beta\delta}+g_{\nu\beta}g_{\alpha\delta}+g_{\nu\delta}g_{\alpha\beta}\right)\tilde{\omega}_{\mu}\right],\nonumber\\
N_{2\mu\nu\alpha\beta\delta}&=&\left(gqq\tilde{\omega}\right)_{\mu\nu\alpha\beta\delta}=\frac{1}{\tilde{\omega} P}\left[g_{\mu\nu}(qq\tilde{\omega})_{\alpha\beta\delta}+permutations\right],\nonumber\\
N_{3\mu\nu\alpha\beta\delta}&=&\left(qqqq\tilde{\omega}\right)_{\mu\nu\alpha\beta\delta}=\frac{1}{\tilde{\omega} P}\left[q_{\mu}q_{\nu}q_{\alpha}q_{\beta}\tilde{\omega}_{\delta}+
q_{\mu}q_{\nu}q_{\alpha}\tilde{\omega}_{\beta}q_{\delta}+q_{\mu}q_{\nu}\tilde{\omega}_{\alpha}q_{\beta}q_{\delta}+q_{\mu}\tilde{\omega}_{\nu}q_{\alpha}q_{\beta}q_{\alpha}+
\tilde{\omega}_{\mu}q_{\nu}q_{\alpha}q_{\beta}q_{\delta}\right].
\end{eqnarray*}

By contracting $\hat{p}_{1\mu}^{\prime}\hat{p}_{1\nu}^{\prime}\hat{p}_{1\alpha}^{\prime}\hat{p}_{1\beta}^{\prime}
\hat{p}_{1\delta}^{\prime}$ with $\tilde{\omega}^{\delta}$, $q^{\delta}$, and $g^{\beta\delta}$, and making a comparison with the concrete expression of $\hat{p}_{1\mu}^{\mu}\hat{p}_{1\nu}^{\prime}\hat{p}_{1\alpha}^{\prime}\hat{p}_{1\beta}^{\prime}$ and $\hat{p}_{1\mu}^{\prime}\hat{p}_{1\nu}^{\prime}\hat{p}_{1\alpha}^{\prime}$, we obtain all the coefficients, i.e.,
\begin{eqnarray*}
A_1^{(5)}&=&A_1^{(1)}A_{1}^{(4)}, \quad A_2^{(5)}=A_{2}^{(1)}A_1^{(4)}, \quad A_3^{(5)}=A_{1}^{(1)}A_2^{(4)},\nonumber\\
A_4^{(5)}&=&A_1^{(1)}A_3^{(4)},\quad A_5^{(5)}=A_1^{(1)}A_4^{(4)},\quad A_6^{(5)}=
A_{2}^{(1)}A_4^{(4)}-\frac{2A_2^{(1)}A_1^{(4)}}{q^2},\nonumber\\
A_7^{(5)}&=&A_1^{(1)}A_{5}^{(4)}, \quad A_8^{(5)}=A_{1}^{(1)}A_{6}^{4},\quad  A_{9}^{(5)}=A_{1}^{(1)}A_{7}^{(4)},\nonumber\\ A_{10}^{(5)}&=&A_1^{(1)}A_{8}^{(4)},\quad A_{11}^{(5)}=A_{1}^{(1)}A_{9}^{(4)},\quad A_{12}^{(5)}=A_{2}^{(1)}A_{9}^{(4)}
-\frac{4A_{2}^{(1)}A_{4}^{(4)}}{q^2}+\frac{8A_2^{(1)}A_{1}^{(4)}}{\left(q^2\right)^2},\nonumber\\
\\
B_1^{(5)}&=&A_1^{(1)}B_1^{(4)}-A_{1}^{(1)}A_{1}^{(4)},\quad B_2^{(5)}=A_1^{(1)}C_1^{(4)}-A_{2}^{(1)}A_{1}^{(4)},\quad B_3^{(5)}=A_1^{(1)}B_2^{(4)}-A_1^{(1)}A_2^{(4)},\nonumber\\
B_4^{(5)}&=&A_1^{(1)}B_3^{(4)}-A_{1}^{(1)}A_{3}^{(4)},\quad B_5^{(5)}=A_1^{(1)}B_4^{(4)}-A_1^{(1)}A_4^{(4)},\quad B_6^{(5)}=
A_1^{(1)}C_2^{(4)}-A_2^{(1)} A_4^{(4)}+\frac{2A_2^{(1)}A_1^{(4)}}{q^2},\nonumber
\\
C_1^{(5)}&=&A_1^{(4)}C_1^{(1)},\quad C_2^{(5)}=C_1^{(1)}A_4^{(4)}+\frac{2\left(P\cdot q\right)A_2^{(1)}A_1^{(4)}}{q^2},\nonumber\\
C_3^{(5)}&=&C_1^{(1)}A_9^{(4)}+\frac{4A_2^{(1)}A_4^{(4)}}{q^2}\left(P\cdot q\right)-\frac{8A_2^{(1)}A_1^{(4)}}{\left(q^2\right)^2}(P\cdot q).
\end{eqnarray*}

\section{Helicity amplitudes and decay widths}\label{decay width}
In this appendix, we will give the explicit forms of helicity form factors for the corresponding semi-leptonic decays, one can easily obtain the associated decay widths from the expressions of these helicity form factors.

In this work, we study the production of $D$-wave charmed/charmed-strange mesons and their partners through the semileptonic decay of $B_{(s)}$ mesons. The effective weak Hamiltonian involved in the $B^-(0^-)\rightarrow {D}^{*0}\ell^- \bar\nu$ and $\bar B_s^0(0^-)\rightarrow D_s^{*+}\ell^- \bar\nu$ transitions is
\begin{eqnarray}
H_{\text{eff}}=\frac{G_F}{\sqrt{2}}V_{cb}\left[\bar{c}\gamma_{\mu}(1-\gamma_5)b\right]
\left[\bar{\ell}\gamma^{\mu}(1-\gamma_5)\nu\right],
\end{eqnarray}
where $G_F$ is the Fermi coupling constant and $V_{cb}$ denotes the Cabibbo-Kobayashi-Maskawa (CKM) matrix element.

The concrete expression of decay widths for $\left\langle D_{(s)1}^{*}\left|V_{\mu}-A_{\mu}\right|B_{(s)}\right\rangle$ matrix element can be obtained  by using the helicity form factors as
\begin{eqnarray}
H^{\pm}_{\pm}(q^2)&=&if_{D}(q^2)\mp i g_D(q^2)\sqrt{\lambda(m^2_{B_{(s)}},m^2_{D^*_{(s)1}},q^2)},\\
H_{0}^{0}(q^2)&=&-\frac{i}{\sqrt{q^2}}\left\{\frac{m^2_{B_{(s)}}-m^2_{D^*_{(s)1}}-q^2}{2m_{D^*_{(s)1}}}
f_D(q^2)+\frac{\lambda(m^2_{B_{(s)}},m^2_{D^*_{(s)1},q^2},q^2)}{2m_{D^*_{(s)1}}}a_{D+}(q^2)\right\},\\
H_{s}^0(q^2)&=&-\frac{i}{\sqrt{q^2}}\sqrt{\lambda(m^2_{B_{(s)}},m^2_{D^*_{(s)1},q^2},q^2)}\frac{1}{2m_{D^*_{(s)1}}}
\left(f_{D}(q^2)+\left(m^2_{B_{(s)}}-m^2_{D^*_{(s)1}}\right)a_{D+}(q^2)+q^2a_{D-}(q^2)\right).
\end{eqnarray}
Here, $\lambda(a^2,b^2,c^2)=(a^2-b^2-c^2)^2-4b^2c^2$. With above equations, one can write out the semi-leptonic decay width in terms of helicity amplitudes as
\begin{eqnarray}
\frac{d\Gamma(\bar{B}_{(s)}\rightarrow D^*_{(s)1}l\bar{\nu})}{dq^2}&=&\frac{d\Gamma_L(\bar{B}_{(s)}\rightarrow D^*_{(s)1}l\bar{\nu})}{dq^2}+\frac{d\Gamma^+(\bar{B}_{(s)}\rightarrow D^*_{(s)1}l\bar{\nu}))}{dq^2}+\frac{\Gamma^-(\bar{B}_{(s)}\rightarrow D^*_{(s)1}l\bar{\nu}))}{dq^2}\\&=&(\frac{q^2-m_l^2}{q^2})^2\frac{\sqrt{\lambda(m^2_{B_{(s)}},m^2_{D^*_{(s)1}},q^2)}G^2_F V^2_{cb}}{384m^3_{B_{(s)}}\pi^3}\left\{3m^2_l\left|H_{s}^0\right|^2+(m_l^2+2q^2)\left(\left|H_{0}^{0}\right|^2+\left|H^+_{+}\right|^2+\left|H^-_{-}\right|^2\right)\right\}
\end{eqnarray}

When it comes to production of charmed/charmed-strange mesons with $J=2$ via the semileptonic decays of bottom/bottom-strange mesons, one can obtain following total decay width
\begin{eqnarray}
\frac{d\Gamma_{L}\left(\bar B_{(s)}\rightarrow{}D^{(\prime)}_{(s)2}l\bar{\nu}\right)}{dq^2}&=&\frac{2}{3}
\frac{\lambda\left(m^2_{B_{(s)}},m^2_{D^{(\prime)}_{(s)2}},q^2\right)}{4m^2_{D^{(\prime)}_2}}\frac{d\Gamma_L\left(\bar B_{(s)}\rightarrow D^{*}_{(s)1} l\bar{\nu}\right)}{dq^2}\Bigg|_{g_D, f_D, a_{D+}, a_{D-}\rightarrow n_{\frac{5}{2}(\frac{3}{2})}, m_{\frac{5}{2}(\frac{3}{2})}, z_{\frac{5}{2}+(\frac{3}{2}+)}, z_{\frac{5}{2}-(\frac{3}{2}-)}},\\
\frac{d\Gamma^{\pm}\left(\bar B_{(s)}\rightarrow D^{(\prime)}_{(s)2} l\bar{\nu}\right)}{dq^2}&=&\frac{1}{2}\frac{\lambda\left(m^2_{B_{(s)}},m^2_{D^{(\prime)}_{(s)2}},q^2\right)}{4m^2_{D^{(\prime)}_{(s)2}}}
\frac{d\Gamma^{\pm}\left(\bar B_{(s)}\rightarrow D^{*}_{(s)1} l\bar{\nu}\right)}{dq^2}\Bigg|_{g_D, f_D, a_{D+}, a_{D-}\rightarrow n_{\frac{5}{2}(\frac{3}{2})}, m_{\frac{5}{2}(\frac{3}{2})}, z_{\frac{5}{2}+(\frac{3}{2}+)}, z_{\frac{5}{2}-(\frac{3}{2}-)}},
\end{eqnarray}
We should emphasize that in our calculation the mixing between ${}^1D_2$ and ${}^3D_2$ states has been considered. When presenting the form factors $F_{\frac{3}{2}(\frac{5}{2})}$ in $B_{(s)}\rightarrow D^{(\prime)}_{(s)2}(2D^{(\prime)}_{(s)2})$ transition, such mixing is included by

\begin{eqnarray}
F_{\frac{3}{2}}(q^2)&=&-\sqrt{\frac{2}{5}}F(q^2)+\sqrt{\frac{3}{5}}F^{\prime}(q^2),\nonumber\\
F_{\frac{5}{2}}(q^2)&=&\sqrt{\frac{3}{5}}F(q^2)+\sqrt{\frac{2}{5}}F^{\prime}(q^2).
\end{eqnarray}
where $F^{(\prime)}\equiv n^{(\prime)}, m^{(\prime)}, z_{+}^{(\prime)}, z_{-}^{(\prime)}$.

In a similar way, we can also obtain the decay width for the production of ${}^3D_3$ states, i.e.,
\begin{eqnarray}
\frac{d\Gamma_{L}(B_{(s)}\rightarrow D^{*}_{(s)3}l\bar{\nu})}{dq^2}&=&\frac{1}{15}\frac{\lambda^2\left(m^2_{B_{(s)}},m^2_{D^{*}_{(s)3}},q^2\right)}{4m^4_{D^{*}_{(s)3}}}
\frac{d\Gamma_{L}\left(B_{(s)}\rightarrow D^{*}_{(s)1} l\bar{\nu}\right)}{dq^2}\Bigg|_{g_D, f_D, a_{D+}, a_{D-}\rightarrow y, w, o_+, o_-},\\
\frac{d\Gamma^{\pm}(B_{(s)}\rightarrow D^{*}_{(s)3} l\bar{\nu})}{dq^2}&=&\frac{1}{10}\frac{\lambda^2\left(m^2_{B_{(s)}},m^2_{D^{*}_{(s)3}},q^2\right)}{4m^4_{D^{*}_{(s)3}}}
\frac{d\Gamma^{\pm}\left(B_{(s)}\rightarrow D^{*}_{(s)1} l\bar{\nu}\right)}{dq^2}\Bigg|_{g_D, f_D, a_{D+}, a_{D-}\rightarrow y, w, o_+, o_-}.
\end{eqnarray}

\section{
Polarization Tensor algebra
}\label{polarization tensor}

When we consider the polarization vector of a massive vector boson, the four-momentum in any other inertial system can be obtained by performing a Lorentz transformation. Hence, it is enough to consider the four-momentum in the rest frame,
\begin{eqnarray}
p^{\mu}=\left(M,0,0,0\right).
\end{eqnarray}

In the rest frame there exist three possible helicities for a spin 1 particle, i.e., three independent polarization vectors have the form,
\begin{eqnarray}
\epsilon^{\mu}\left(\lambda=+1\right)&=&\left(0,-\frac{1}{\sqrt{2}},-\frac{i}{\sqrt{2}},0\right),\nonumber\\
\epsilon^{\mu}(\lambda=0)&=&\left(0,0,0,1\right),\nonumber\\
\epsilon^{\mu}(\lambda=-1)&=&\left(0,\frac{1}{\sqrt{2}},-\frac{i}{\sqrt{2}},0\right),
\end{eqnarray}
which satisty $p\cdot\epsilon\left(\lambda\right)=0$.

In the following, for convenience of readers, we present the tensor algebra in the rectangular coordinate system.  One can also do this in the light-front frame by adjusting the corresponding metric tensor, where the tensor algebra will still give identical results.
The normalization of polarization vectors is given by
\begin{eqnarray}
\epsilon^{*\mu}\left(\lambda\right)\epsilon_{\mu}\left(\lambda'\right)=
-\delta_{\lambda\lambda'}.
\end{eqnarray}
Due to the Lorentz covariance, the sum over the polarization states is
\begin{eqnarray}
\sum_{\lambda}\epsilon^{*\mu}\left(\lambda\right)\epsilon^{\nu}\left(\lambda\right)=-g^{\mu\nu}+\frac{P^{\mu}P^{\nu}}{M_0^2}=-G^{\mu\nu},
\end{eqnarray}
where $G^{\mu\nu}=g^{\mu\nu}-\frac{P^{\mu}P^{\nu}}{M^2}$, and
\begin{eqnarray}
\sum_{m}\epsilon^{*}_{\alpha\beta}(m)\epsilon_{\alpha'\beta'}(m)=\frac{1}{2}(G_{\alpha\alpha'}G_{\beta\beta'}+G_{\alpha\beta'}G_{\alpha'\beta})-\frac{1}{3}
G_{\alpha\beta}G_{\alpha'\beta'}.
\end{eqnarray}

In general, the higher-rank polarization tensor must satisfy the following conditions
\begin{eqnarray}
\text{Transversility}&:& p^{\mu_{i}}\epsilon_{\mu_1...\mu_i...\mu_n}\left(\lambda\right)=0,\nonumber\\
\text{Symmetric}&:& \epsilon_{\mu_1...\mu_i...\mu_j...\mu_n}(\lambda)=\epsilon_{\mu_1...\mu_j...\mu_i...\mu_n}(\lambda),\nonumber\\
\text{Traceless}&:& g^{\mu_i\mu_j}\epsilon_{\mu_1...\mu_i...\mu_j...\mu_n}(\lambda)=0, \\
\text{Normalization}&:& \epsilon^*_{\mu_1...\mu_{n}}(\lambda)\epsilon^{\mu_1...\mu_{n}}(\lambda')=(-1)^{n}\delta_{\lambda\lambda'},\nonumber\\
\text{Conjugation}&:& \epsilon^*_{\mu_1...\mu_n}(\lambda)=(-1)^{\lambda}\epsilon^{\mu_1...\mu_n}(-\lambda).\nonumber
\label{condition}
\end{eqnarray}

Now, we take the CG coefficient of ${}^3D_1$ state as an example to show how to rewrite a CG coefficient in the tensor contract form. Firstly, we concentrate on a tensor
$\epsilon_{\alpha\beta}\left(L_z\right)\epsilon_{\rho}\left(S_z\right)\left\langle 21;mS_z|1J_z\right\rangle$.
Due to constraint from Lorentz covariance, the decomposition of this tensor must be a linear combination of $g_{\alpha\beta}$, $p_{\alpha}$ or $G_{\alpha\beta}$, $p_{\alpha}$, i.e., \begin{eqnarray}
\epsilon_{\alpha\beta}\left(L_z\right)\epsilon_{\rho}\left(S_z\right)\left\langle 21;L_zS_z|1J_z\right\rangle
=AG_{\alpha\beta}\epsilon_{\rho}\left(J_z\right)+BG_{\alpha\rho}\epsilon_{\beta}\left(J_{z}\right)+CG_{\beta\rho}\epsilon_{\alpha}\left(J_z\right)+Dp_{\alpha}p_{\beta}\epsilon_{\rho}\left(J_z\right)+Ep_{\alpha}p_{\rho}\epsilon_{\beta}\left(J_z\right)+Fp_{\beta}p_{\rho}\epsilon_{\alpha}\left(J_z\right).
\end{eqnarray}
Multiplying $p^{\alpha}$ on both sides of the above equation, introducing the transverse condition in Eq. (\ref{condition}) and having $p^{\alpha}G_{\alpha\beta}=0$, we easily find  $D=E=F=0$. Then, the symmetry property of this CG coefficient results in $B=C$.
Thus, the original tensor is reduced into a simpler form
\begin{eqnarray}
\epsilon_{\alpha\beta}\left(L_z\right)\epsilon_{\rho}\left(S_z\right)\left\langle 21;L_zS_z|1J_z\right\rangle=AG_{\alpha\beta}\epsilon_{\rho}\left(J_z\right)+BG_{\alpha\rho}\epsilon_{\beta}\left(J_z\right)+BG_{\beta\rho}\epsilon_{\alpha}\left(J_z\right).
\label{lzszcg}
\end{eqnarray}
Next, the traceless condition is applied by multiplying $g^{\alpha\beta}$ on both sides of Eq. (\ref{lzszcg}) and we find $B=-3A/2$. Now, we multiply $\epsilon^{*\alpha\beta}\left(L_z\right)\epsilon^{*\rho}\left(S_z\right)$ on both sides of Eq. (\ref{lzszcg}), by which we obtain,
\begin{eqnarray}
\left\langle 21;L_zS_z|1J_z\right\rangle=3A\epsilon^{*\alpha\beta}\left(L_z\right)\epsilon_{\alpha}^{*}\left(S_z\right)\epsilon_{\beta}\left(J_z\right).
\label{A}
\end{eqnarray}
Here, only one undetermined constant is left. We can assign the specific value for $L_z$, $S_z$ and $J_z$ on the left side of Eq. (\ref{A}), and introduce the expression of $\epsilon^{*\alpha\beta}\left(L_z\right)$, $\epsilon_{\alpha}^{*}\left(S_z\right)$ and $\epsilon_{\beta}\left(J_z\right)$ in the right side to solve $A$. Finally, we obtain $A=-\sqrt{1/15}$, and have
\begin{eqnarray}
\left\langle 2 1;L_z S_z|1 J_z\right\rangle=-\sqrt{\frac{3}{5}}\epsilon^{*\mu\nu}\left(L_z\right)\epsilon_{*\mu}\left(S_z\right)\epsilon_{\nu}\left(J_z\right).
\end{eqnarray}
This example provides us the detailed deduction of a CG coefficient how to transform it into a tensor contraction form. In the same way, one can deal with other CG coefficients.

When calculating the semileptonic decay width, we use the second-order and the third-order tensors. The polarization tensor of a higher spin state with angular momentum $j$ and helicity $\lambda$ can be constructed by lower-rank polarization tensors and CG coefficients. A general relation reads
\begin{eqnarray}
\epsilon_{\mu_1...\mu_n}(\lambda)&=&\sum_{\lambda_{n-1},\lambda_n}\langle n-1,\lambda_{n-1};n,\lambda_n\rangle\epsilon_{\mu_1\mu_2...\mu_{n-1}}(\lambda_{n-1})
\epsilon_{\mu_n}(\lambda_n).
\end{eqnarray}
with $\lambda=\lambda_n+\lambda_{n-1}$.
One can generalize this equation to obtain the polarization tensors of higher spin states. Thus, we have
\begin{eqnarray}
\epsilon^{\mu\nu}(\pm 2)&=&\epsilon^{\mu}(\pm1)\epsilon^{\nu}(\pm1),\nonumber\\
\epsilon^{\mu\nu}(\pm 1)&=&\sqrt{\frac{1}{2}}\left[\epsilon^{\mu}(\pm)\epsilon^{\nu}(0)
+\epsilon^{\mu}(0)\epsilon^{\nu}(\pm1)\right],\nonumber\\
\epsilon^{\mu\nu}(0)&=&\sqrt{\frac{1}{6}}\left[\epsilon^{\mu}(+1)\epsilon^{\nu}(-1)
+\epsilon^{\mu}(-1)\epsilon^{\nu}(+1)\right]
+\sqrt{\frac{2}{3}}\epsilon^{\mu}(0)\epsilon^{\nu}(0),
\end{eqnarray}
for $J=2$ state,
and
\begin{eqnarray}
\epsilon^{\alpha\beta\gamma}(\pm3)&=&\epsilon^{\alpha\beta}(\pm2)\epsilon^{\gamma}(\pm1),\nonumber\\
\epsilon^{\alpha\beta\gamma}(\pm2)&=&\frac{1}{\sqrt{3}}
\epsilon^{\alpha\beta}(\pm2)\epsilon^{\gamma}(0)+\sqrt{\frac{2}{3}}\epsilon^{\alpha\beta}(\pm1)
\epsilon^{\gamma}(\pm1),\nonumber\\
\epsilon^{\alpha\beta\gamma}(\pm1)&=&\frac{1}{\sqrt{15}}\epsilon^{\alpha\beta}(\pm2)
\epsilon^{\gamma}(\mp1)+2\sqrt{\frac{2}{15}}\epsilon^{\alpha\beta}(\pm1)\epsilon^{\nu}(0)+
\sqrt{\frac{2}{5}}\epsilon^{\alpha\beta}(0)\epsilon^{\gamma}(\pm1),\nonumber\\
\epsilon^{\alpha\beta\gamma}(0)&=&\frac{1}{\sqrt{5}}\epsilon^{\alpha\beta}(+1)\epsilon^{\gamma}(-1)
+\sqrt{\frac{3}{5}}\epsilon^{\alpha\beta}(0)\epsilon^{\gamma}(0)+
\frac{1}{\sqrt{5}}\epsilon^{\alpha\beta}(-1)\epsilon^{\gamma}(+1),
\end{eqnarray}
for $J=3$ state.

\begin{multicols}{2}
\section{Proof of Lorentz invariance of matrix elements in a multipole ansatz}\label{toy model prove}
In this Appendix, we will show explicitly that loop integrals of $B_2^{(4)}$ and $B_3^{(5)}$ vanish analytically for the toy model, i.e., with multipole ansatz for the vertex functions, which can be extended to other $B^{(m)}_{n}$ and $C^{(m)}_n$.
Let us consider the integral,
\begin{eqnarray}
I[M] \equiv \frac{i}{(2\pi)^4}\int d^4p^\prime_1\frac{M}{N^\prime_\Lambda N_1^{\prime}N_2N_1^{\prime\prime}N^{\prime\prime}_\Lambda},
\label{inte}
\end{eqnarray}
where $M$ is the function related to $A^{(m)}_{n}, B^{(m)}_{n}, C^{(m)}_n$ as well as $N_2$ and $Z_2$.
By inserting the identities in Table~\ref{replacements}, we can prove that the loop integrals of these $B^{(m)}_{n}$ and $C^{(m)}_{n}$ functions vanish in the multipole ansatz.

Starting from Jaus's result \cite{Jaus:1999zv}, the complete momentum integral of $N_2$ is given by
\begin{eqnarray}
I[N_2]&=& \frac{i}{(2\pi)^4}\int d^4p_1^{\prime}\frac{N_2}{N^{\prime}_\Lambda N_1^{\prime}N_2N_1^{\prime\prime}N_\Lambda^{\prime\prime}}\\ \nonumber
&=& \frac{1}{16\pi^2(\Lambda^{\prime2}-m_1^{\prime2})(\Lambda^{\prime\prime2}-
m_1^{\prime\prime2})}\int_0^1dy \text{ln}\frac{C^0_{11}C^{0}_{\Lambda\Lambda}}{C^0_{1\Lambda}C^{0}_{\Lambda 1}},
\label{N2}
\end{eqnarray}
where
\begin{eqnarray}
C_{11}^0&=&C^0(m_1^\prime,m_1^{\prime\prime})=(1-y)m_1^{\prime2}+y m_1^{\prime\prime2}-y(1-y)q^2,\nonumber\\
C^0_{\Lambda\Lambda}&=&C^0(\Lambda^\prime,\Lambda^{\prime\prime}),C^0_{1\Lambda}
=C^0(m_1^\prime,\Lambda^{\prime\prime}),C^0_{\Lambda 1}=C^0(\Lambda^\prime,m_1^{\prime\prime}).\nonumber
\end{eqnarray}
The loop integral of $Z_2$ is obtained as
\begin{eqnarray}
I[Z_2]&=&\frac{i}{(2\pi)^4}\int d^4p_1^{\prime}\frac{Z_2}{N^{\prime}_\Lambda N_1^{\prime}N_2N_1^{\prime\prime}N_\Lambda^{\prime\prime}}\nonumber\\
&=&\frac{1}{16\pi^2(\Lambda^{\prime2}-m_1^{\prime2})(\Lambda^{\prime\prime2}-m_1^{\prime\prime2})}
\int_0^1du \text{ln}\frac{C^0_{11}C^0_{\Lambda\Lambda}}{C^0_{1\Lambda}C^0_{\Lambda 1}}.\nonumber\\
\label{Z2}
\end{eqnarray}
Readers can refer to Ref. \cite{Jaus:1999zv} for the definition of $u$. Thus, one obtains the replacement $N_2\rightarrow Z_2$, which is related to $C_1^{(1)}$ function.

Further, the integral of $A_1^{(2)}$ is given by
\begin{eqnarray}
I[A_1^{(2)}]&=&\frac{1}{32\pi^2(\Lambda^{\prime2}-m_1^{\prime2})(\Lambda^{\prime\prime2}-m_1^{\prime\prime2})}\nonumber\\
&&\times\int_0^1dx\int_0^1dy(1-x)\text{ln}\frac{C_{11}C_{\Lambda\Lambda}}{C_{1\Lambda}C_{\Lambda 1}},
\label{A12}
\end{eqnarray}
where
\begin{eqnarray}
C_{11}&=&C(m_1^{\prime},m_1^{\prime\prime})\nonumber\\
&=&(1-x)(1-y)m_1^{\prime2}+(1-x)y m_1^{\prime\prime2}+x m_2^{2}\nonumber\\&&-x(1-x)[(1-y)M^{\prime2}+y M^{\prime\prime2}]-(1-x)^2y(1-y)q^2,\nonumber\\
C_{\Lambda\Lambda}&=&C(\Lambda^{\prime},\Lambda^{\prime\prime}),C_{1\Lambda}=C(m_1^{\prime},\Lambda^{\prime\prime}),
C_{\Lambda 1}=C(\Lambda^{\prime},m_1^{\prime\prime}).
\end{eqnarray}
Now we take the integral $I[B_2^{(4)}]$ as an example to perform its integration. The $B_2^{(4)}$ function is similar to $B_1^{(2)}$ in Ref. \cite{Jaus:1999zv}. It contains $A_3^{(3)}N_2$ term, which is proportional to $x\delta(p_1^{\prime+})$. This means that this term vanishes. Now we only need to prove that the rest of the terms in the integral vanishes, i.e.,
\begin{eqnarray}
I[A_3^{(3)}Z_2-3A_2^{(2)}A_1^{(2)}]=0.
\end{eqnarray}
By using Eq. (\ref{Z2}) and performing a partial integration, we have
\begin{eqnarray}
I[A_{3}^{(3)}Z_2]&=&\frac{3}{128\pi^2(\Lambda^{\prime2}-m_1^{\prime2})(\Lambda^{\prime\prime2}-m_1^{\prime\prime2})}\nonumber\\
&&\times\int_0^1dx\int_0^{1}dy(1-x)x^2\text{ln}\frac{C_{11}C_{\Lambda\Lambda}}{C_{1\Lambda}C_{\Lambda 1}},
\end{eqnarray}
and by introducing Eq. (\ref{A12}), we can directly obtain
\begin{eqnarray}
I[3A_2^{(2)}A_1^{(2)}]&=&\frac{3}{128\pi^2(\Lambda^{\prime2}-m_1^{\prime2})(\Lambda^{\prime\prime2}-m_1^{\prime\prime2})}\nonumber\\
&\times&\int_0^1dx\int_0^{1}dy(1-x)x^2\text{ln}\frac{C_{11}C_{\Lambda\Lambda}}{C_{1\Lambda}C_{\Lambda 1}},
\end{eqnarray}
which menas $I[A_3^{(3)}Z_2-3A_2^{(2)}A_1^{(2)}]=0$
and hence,
\begin{eqnarray}
I[B_2^{(4)}]=0.
\end{eqnarray}
Equation $I[B_3^{(5)}]=0$ can also be obtained by using Eq. (\ref{Z2}) and Eq. (\ref{A12}).
For the loop integrals of $B^{(m)}_{n}$ and $C^{(m)}_{n}$ functions in which $I[A_{k}^{(j)}\hat{N}_2]\not= 0$, one can use the techniques in Ref. \cite{Jaus:1999zv} and
show that $I[B^{(m)}_{n}]=I[C^{(m)}_{n}]=0$.
\end{multicols}
\end{small}
\clearpage

\end{CJK*}
\end{document}